\documentclass[oneside,11pt]{article}

\usepackage[utf8]{inputenc}
\usepackage[T1]{fontenc}

\usepackage{amsthm}
\usepackage{amsmath}
\usepackage{amsfonts}
\usepackage{amssymb}

\usepackage{graphicx} 
\usepackage{graphbox}
\usepackage{float}
\usepackage{booktabs}
\usepackage{multirow}
\usepackage{rotating}
\usepackage{array}
\usepackage{xcolor}
\usepackage{subcaption}
\usepackage[ruled]{algorithm2e}

\usepackage{comment}

\newcolumntype{C}[1]{>{\centering\arraybackslash}p{#1}}
\newcolumntype{R}[1]{>{\raggedleft\let\newline\\\arraybackslash\hspace{0pt}}m{#1}}
\newcolumntype{L}[1]{>{\raggedright\let\newline\\\arraybackslash\hspace{0pt}}m{#1}}

\usepackage{breakcites}

\usepackage{nowidow}

\usepackage{enumitem}

\usepackage[top=1.5cm,bottom=2cm,right=2.5cm,left=2.5cm]{geometry}
\usepackage{titling}

\usepackage{natbib}

\usepackage{ifplatform}

\usepackage{textcomp}
\usepackage{bbm}

\ifwindows
\fi

\allowdisplaybreaks

\newcommand{\lp}{\left(}
\newcommand{\rp}{\right)}
\newcommand{\lc}{\left[}
\newcommand{\rc}{\right]}

\newcommand{\R}{\mathbb{R}}

\newcommand{\rd}{\mathrm{d}}

\newcommand{\bx}{\boldsymbol{x}}
\newcommand{\be}{\boldsymbol{e}}

\newcommand{\by}{\boldsymbol{y}}
\newcommand{\bX}{\boldsymbol{X}}
\newcommand{\bY}{\boldsymbol{Y}}

\newcommand{\bU}{\boldsymbol{U}}
\newcommand{\bR}{\boldsymbol{R}}

\newcommand{\bmu}{\boldsymbol\mu}
\newcommand{\bu}{\boldsymbol{u}}
\newcommand{\bv}{\boldsymbol{v}}
\newcommand{\bb}{\boldsymbol{b}}

\newcommand{\bs}{\boldsymbol{s}}
\newcommand{\bt}{\boldsymbol{t}}

\newcommand{\bh}{{\boldsymbol{h}}}
\newcommand{\bd}{\boldsymbol{d}}

\newcommand{\one}{\mathbf{1}}
\newcommand{\zero}{\mathbf{0}}
\newcommand{\two}{\mathbf{2}}
\newcommand{\bxi}{\boldsymbol\xi}

\newcommand{\ba}{\boldsymbol\alpha}

\newcommand{\bga}{\boldsymbol\gamma}

\newcommand{\bkappa}{\boldsymbol\kappa}

\newcommand{\Ical}{\mathcal{I}}

\newcommand{\bSigma}{\boldsymbol\Sigma}

\newcommand{\bHcal}{\boldsymbol{\mathcal{H}}}

\newcommand{\bnabla}{\boldsymbol\nabla}
\newcommand{\blambda}{\boldsymbol\lambda}
\newcommand{\bB}{\boldsymbol{B}}

\newcommand{\bA}{\boldsymbol{A}}

\newcommand{\bI}{\boldsymbol{I}}

\newcommand{\bT}{\boldsymbol{\mathcal{T}}}

\newcommand{\inlaw}{\stackrel{d}{\rightsquigarrow}}
\DeclareMathAlphabet\mathbfcal{OMS}{cmsy}{b}{n}

\newcommand{\bnab}{\boldsymbol\nabla}

\newcommand{\sigmad}{\sigma_{d}}
\newcommand{\sigmar}{\sigma_{\bd}}

\newcommand{\sigmaim}{\sigma_{d_i-1}}
\newcommand{\sigmaj}{\sigma_{d_j}}

\newcommand{\sigmak}{\sigma_{d_k}}
\newcommand{\sigmaell}{\sigma_{d_\ell}}
\newcommand{\sigmajm}{\sigma_{d_j-1}}
\newcommand{\sigmarm}{\sigma_{\bd-1}}

\newcommand{\Sdr}{(\mathbb{S}^{d})^r}
\newcommand{\Sr}{\mathbb{S}^{\bd}}
\newcommand{\Srm}{\mathbb{S}^{\bd-1}}
\newcommand{\Sd}{\mathbb{S}^{d}}
\newcommand{\Sj}{\mathbb{S}^{d_j}}
\newcommand{\Sk}{\mathbb{S}^{d_k}}
\newcommand{\Sell}{\mathbb{S}^{d_\ell}}

\newcommand{\Sjm}{\mathbb{S}^{d_j-1}}
\newcommand{\Sim}{\mathbb{S}^{d_i-1}}

\newcommand{\lrp}[1]{\left(#1\right)}
\newcommand{\lrc}[1]{\left[#1\right]}
\newcommand{\lrb}[1]{\left\{#1\right\}}
\newcommand{\lrpbig}[1]{\big(#1\big)}
\newcommand{\lrcbig}[1]{\big[#1\big]}

\newcommand{\blrp}[1]{\big(#1\big)}

\newcommand{\Es}[2]{\mathbb{E}_{#2}\lc #1\rc}

\newcommand{\diag}[1]{\mathrm{diag}\lp #1\rp}

\newcommand{\abs}[1]{\left| #1\right|}
\newcommand{\tr}[1]{\mathrm{tr}\left[#1\right]}

\newcommand{\om}[1]{\omega_{#1}}

\DeclareFontFamily{OT1}{pzc}{}
\DeclareFontShape{OT1}{pzc}{m}{it}{<-> s * [1.10] pzcmi7t}{}
\DeclareMathAlphabet{\mathpzc}{OT1}{pzc}{m}{it}

\newtheorem{theorem}{Theorem}[section]
\newtheorem{corollary}{Corollary}[section]
\newtheorem{remark}{Remark}[section]
\newtheorem{proposition}{Proposition}[section]

\newtheorem{algo}{Algorithm}[section]

\newcommand{\pct}{\%}

\newcommand{\defin}{:=}

\usepackage[pdftex,final,bookmarksnumbered,bookmarksopen=false,breaklinks,colorlinks]{hyperref}
\hypersetup{
	final,
	bookmarksnumbered=true,
	bookmarksopen=true,
	bookmarksopenlevel=0,
	unicode=false,
	pdftoolbar=true,
	pdfmenubar=true,
	pdffitwindow=false,
	pdftitle={Kernel density estimation with polyspherical data and its applications},
	pdfdisplaydoctitle=true,
	pdftoolbar=true,
	pdfmenubar=true,
	pdflang={English},
	pdfauthor={Eduardo Garcia-Portugues, Andrea Meilan-Vila},
	pdfsubject={arXiv paper},
	pdfcreator={Eduardo Garcia-Portugues},
	pdfproducer={Eduardo Garcia-Portugues},
	pdfkeywords={Circular data}{Directional data}{Projection}{Uniformity},
	pdfnewwindow=true,
	breaklinks=true,
	hidelinks,
	linkcolor=black,
	citecolor=black,
	filecolor=magenta,
	urlcolor=cyan
}

\newif\ifmain
\maintrue
\newif\ifsupplement
\supplementtrue

\newif\iffigstabs
\figstabstrue

\begin{document}

\ifmain

%-----------------------------------------------%
\title{Kernel density estimation with polyspherical data and its applications}
\setlength{\droptitle}{-1cm}
\predate{}%
\postdate{}%
\date{}
%-----------------------------------------------%

%-----------------------------------------------%
\author{Eduardo Garc\'ia-Portugu\'es$^{1,2}$ and Andrea Meil\'an-Vila$^{1}$}
\footnotetext[1]{Department of Statistics, Carlos III University of Madrid (Spain).}
\footnotetext[2]{Corresponding author. e-mail: \href{mailto:edgarcia@est-econ.uc3m.es}{edgarcia@est-econ.uc3m.es}.}
\maketitle
%-----------------------------------------------%

\begin{abstract}
A kernel density estimator for data on the polysphere $\mathbb{S}^{d_1}\times\cdots\times\mathbb{S}^{d_r}$, with $r,d_1,\ldots,d_r\geq 1$, is presented in this paper. We derive the main asymptotic properties of the estimator, including mean square error, normality, and optimal bandwidths. We address the kernel theory of the estimator beyond the von Mises--Fisher kernel, introducing new kernels that are more efficient and investigating normalizing constants, moments, and sampling methods thereof. Plug-in and cross-validated bandwidth selectors are also obtained. As a spin-off of the kernel density estimator, we propose a nonparametric $k$-sample test based on the Jensen--Shannon divergence. Numerical experiments illuminate the asymptotic theory of the kernel density estimator and demonstrate the superior performance of the $k$-sample test with respect to parametric alternatives in certain scenarios. Our smoothing methodology is applied to the analysis of the morphology of a sample of hippocampi of infants embedded on the high-dimensional polysphere $(\mathbb{S}^2)^{168}$ via skeletal representations ($s$-reps).
\end{abstract}
\begin{flushleft}
\end{flushleft}

%-------------------------------%
\section{Introduction}
\label{sec:intro}
%-------------------------------%

%
Directional data are observations recorded as directions or angles that are typically represented on the circumference, the sphere, or the Cartesian products thereof with other manifolds. In the spirit of \cite{Eltzner2015}, we use the term \emph{polyspherical data} to refer to observations on the product space of spheres $\mathbb{S}^{d_1}\times \cdots\times\mathbb{S}^{d_r}$, where $d_1,\ldots,d_r\geq 1$ and $\Sd$ denotes the (hyper)sphere of dimension $d\ge1$. The polysphere encompasses particular important cases, such as the circle ($r=d_1=1$), the sphere ($r=1$, $d_1=2$), and the torus ($d_1=\cdots=d_r=1$). Specific instances of polyspherical data include medically imaged objects \cite[e.g.,][]{Vuollo2016,piezer2022skeletons}, protein/RNA structures \cite[e.g.,][]{Boomsma2008,Eltzner2018}, and several studies within the field of astronomy, such as cosmic rays and crater locations \cite[e.g.,][]{Baldi2009,Garcia-Portugues2022}, among others. More generally, high-dimensional spherical data appear when standardizing multivariate data by Euclidean norms, as done in genetics \citep[e.g.,][]{Eisen1998} and text mining \citep[e.g.,][]{Banerjee2005}, when considering \emph{preshapes} in shape analysis \citep{Dryden2016}, or when applying embeddings of different data types to the sphere \citep[e.g.,][]{Li2023,Zoubouloglou2023}. \cite{Mardia1999a} and \cite{Ley2017a} collect the fundamental methods on directional statistics, while \cite{Pewsey2021} provides a survey of recent advances in this field. In addition, \cite{Marron2021} reviews novel methodology and different case studies involving directional data.

In the context of three-dimensional shapes, effective methods for parametrization of their structures are skeletal representations, referred to as $s$-reps \citep{Pizer2013}. Skeletal models consist of a centrally located skeleton, and spokes represented as vectors that emanate from the skeletal structure and terminate at the boundary of the object (see Figure \ref{fig:srepa}). Consequently, these models provide benefits compared to those that solely capture its boundary \citep[see, e.g.,][]{liu2021fitting}. The geometry characteristics of $s$-reps, particularly the directions represented by the spokes, imply that they reside within a space involving a Cartesian product of spheres. Skeletal models have been particularly used to analyze three-dimensional shapes derived from medical imaging. For instance, \cite{Pizer2013} provided a method for analyzing the shape of the hippocampus using nested sphere statistics of skeletal models, revealing a difference in hippocampus shapes between infants diagnosed as autistic and non-autistic. \cite{schulz2016non} introduced a permutation test to identify morphological differences between two populations based on geometric features extracted from their $s$-reps. The approach utilizes Composite Principal Nested Spheres (CPNS), a composition of Principal Nested Spheres \citep[PNS;][]{Jung2012} with Principal Component Analysis (PCA) to detect the main source of variation in the space of $s$-reps. More recently, \cite{liu2023analysis} conducted multi-object shape analysis to extract joint shape variation between two functionally related brain structures, the hippocampus and the caudate. \cite{liu2023geometric} demonstrated the effectiveness of considering the joint analysis of within- and between-object shape features, resulting in improved classification and hypothesis testing performance. From a different perspective, \cite{Garcia-Portugues2023} identified the primary source of variability in hippocampus shapes using nonparametric density ridge estimation. For further applications involving hippocampus shapes parametrized using $s$-reps, we refer to the recent survey by \cite{piezer2022skeletons}.

This work tackles the problem of nonparametric density estimation with polyspherical data. In the statistical literature, considerable research has been done on nonparametric density estimation within directional supports, mostly based on kernel smoothing approaches \citep[Section 7.1]{Pewsey2021}. The seminal works by \cite{Hall1987} and \cite{Bai1988} defined three highly related kernel density estimators on $\Sd$, $d\geq1$, deriving their asymptotic formulae of bias and variance \citep{Hall1987}, and establishing its pointwise and $L_1$ consistency \citep{Bai1988}. Somehow more explicit derivations of the mean-square error analysis, for \cite{Bai1988}'s definition, were later given in \cite{Zhao2001} and \cite{Garcia-Portugues2013b}. \cite{Klemela2000} presented an alternative estimator for the density and its derivatives on $\Sd$, $d\geq2$, using a kernel featuring an intrinsic distance. \cite{DiMarzio2011} introduced a specific class of product kernels to construct a kernel density estimator on the torus, while \cite{Garcia-Portugues2013b} constructed an estimator on $\Sd\times\R$, also using product kernels. In a broader context, the problem of nonparametric density estimation on Riemannian manifolds has been investigated by \cite{Pelletier2005} and \cite{Henry2009}, among others. Paralleling the development of these density estimators, several bandwidth selection methods have been developed for them. \cite{Taylor2008} was the first to introduce a plug-in bandwidth selector for circular data, while \cite{Oliveira2012} derived a mixture-based selector. \cite{DiMarzio2011} proposed a cross-validatory bandwidth selection method for the torus. For $\Sd$, \cite{Hall1987} provided two cross-validatory bandwidth selectors, and \cite{Garcia-Portugues2013a} gave a rule-of-thumb bandwidth selector, assuming that the underlying density follows a von Mises--Fisher distribution. Kernel density estimators on directional supports have also been used to guide nonparametric statistical inference, such as in goodness-of-fit \citep{Boente2014,Garcia-Portugues2015}, mode \citep{Ameijeiras-Alonso2019b}, or independence testing \citep{Garcia-Portugues2015}.

In this paper, we propose a kernel density estimator for data on the polysphere $\mathbb{S}^{d_1}\times\cdots\times\mathbb{S}^{d_r}$. The estimator accommodates previous estimators as particular cases and has manifold associated methodological contributions. The first set of contributions involves the main asymptotic properties of the estimator, namely, mean-square error analysis, consistency, normality, and optimal bandwidths. Unlike previous specific approaches, we highlight that these results are obtained for general non-product kernels, and that the asymptotic bias is derived with a smaller remaining order. These properties are exemplified through a numerical study that enlightens the role of the bandwidth sequence on the asymptotic normality. The second series of contributions relates to the definition and efficiency analysis of other kernels beyond the von Mises--Fisher kernel, which is omnipresent in kernel density estimation with directional data. In this regard, we introduce several more efficient kernels for both the $r=1$ and $r>1$ cases, addressing practically relevant matters such as normalizing constants, moments, and simulation methods. The third contribution consists of deriving an inexpensive rule-of-thumb plug-in bandwidth selector and formulating appropriate cross-validation selectors. The fourth contribution entails the introduction of a new $k$-sample permutation test based on the kernel density estimator and Jensen--Shannon divergence that, unlike previous smoothing tests, is applicable in a high-dimensional space. Numerical experiments show a better performance of this $k$-sample test compared to parametric alternatives in certain scenarios, as well as its applicability on high dimensions. Fifth, our methodology is applied to analyze the hippocampus shapes of 6-months-old infants \citep{liu2021fitting,liu2023geometric} that are embedded in the high-dimensional polysphere $(\mathbb{S}^2)^{168}$ through $s$-reps. In this application, we identify the most prototypical and outlying hippocampi, as well as find a mild significance of the shape difference between children who were later diagnosed as autistic and non-autistic.

The rest of this paper is organized as follows. In Section \ref{sec:est}, the nonparametric estimator of the density function is introduced. Section \ref{sec:prop} provides asymptotic results on the bias, variance, optimal bandwidths, and normality of the estimator. Section \ref{sec:ker} is devoted to kernel theory, addressing the derivation of new kernels, their efficiency, and other useful results. Section \ref{sec:bw} gives bandwidth selectors based on plug-in and cross-validation approaches. Section \ref{sec:hom} introduces a novel $k$-sample test based on the kernel density estimator. Section \ref{sec:data} presents two applications of the density estimator to the analysis of hippocampus shapes. The paper concludes with a discussion in Section \ref{sec:dis}. Proofs and numerical experiments are relegated to the Supplementary Material (SM).

%-------------------------------%
\section{Estimator}
\label{sec:est}
%-------------------------------%

%
Let $f$ be a probability density function (pdf) on $\Sr\defin\mathbb{S}^{d_1}\times \cdots\times\mathbb{S}^{d_r}$, $\bd=(d_1,\ldots,d_r)'$ and $d_1,\ldots,d_r\geq 1$, with respect to the product measure $\sigmar\defin\sigma_{d_1}\times\cdots\times\sigma_{d_r}$, where $\sigma_{d_j}$ is the surface area measure on $\Sj$. We denote $\smash{\om{d_j}\defin\sigma_{d_j}(\Sj)=2\pi^{(d_j+1)/2}/\Gamma((d_j+1)/2)}$ and $\smash{\om{\bd}\defin\prod_{j=1}^r \om{d_j}}$ to the surface areas of $\Sj$ and $\Sr$, respectively. We denote $\tilde{d}\defin\sum_{j=1}^r d_j$, and so $\Sr\subset\R^{\tilde{d}+r}$.

Let $\bX_1,\ldots,\bX_n$ be an independent and identically distributed (iid) sample from $f$. Let $\bx=(\bx_1',\ldots,\bx_r')'\in\Sr$, with $\bx_j\in\mathbb{S}^{d_j}$ for $j=1,\ldots,r$, and set $\bh\defin(h_1,\ldots,h_r)'\in\R_+^r$. Building on \cite{Bai1988}'s definition, our kernel density estimator (kde) of $f$ at $\bx$ is defined~as
\begin{align}
    \hat{f}(\bx;\bh)&\defin\frac{1}{n}\sum_{i=1}^nL_{\bh}(\bx,\bX_i),\quad L_{\bh}(\bx,\by)\defin c_{\bd,L}(\bh)L\lrp{\frac{1-\bx_1' \by_1}{h_1^2},\ldots, \frac{1-\bx_r' \by_r}{h_r^2}},\label{eq:estimator}
\end{align}
where the normalized kernel $L_{\bh}:\Sr\times\Sr\to\R_{\geq 0}$ is based on the kernel $L:\R_{\geq 0}^r\to\R_{\geq 0}$. Applying $r$ marginal tangent-normal changes of variables \citep[see, e.g., Lemma 2 in][]{Garcia-Portugues2013b}, the normalizing constant of $L$ is
\begin{align}
    c_{\bd,L}(\bh)^{-1}\defin&\int_{\Sr}L\lrp{\frac{1-\bx_1' \by_1}{h_1^2},\ldots, \frac{1-\bx_r' \by_r}{h_r^2}}\,\sigmar(\rd\bx)\nonumber\\
    =&\int_{[-1,1]^r}L\lrp{\frac{1-t_1}{h_1^2},\ldots, \frac{1-t_r}{h_r^2}}\times \prod_{j=1}^r \left(1-t_j^2\right)^{d_j/2-1}\om{d_j-1}\,\rd \bt   \nonumber \\
    =&\; \rho(\bh) \lambda_{\bh,\bd}(L)\asymp \rho(\bh)\lambda_{\bd}(L),\label{eq:equiv}
\end{align}
with $a_\ell\asymp b_\ell$ denoting $a_\ell/b_\ell\to1$ as $\ell\to\infty$ in \eqref{eq:equiv}, $I_\bh\defin [0,2h^{-2}_1]\times\stackrel{r}{\cdots}\times[0,2h^{-2}_r]$, $\lambda_{\bh,\bd}(L)\defin\om{\bd-1}\int_{I_\bh}L\lrp{\bs} \times\prod_{j=1}^r s_j^{d_j/2-1} (2-s_jh_j^2)^{d_j/2-1}\,\rd \bs$, and
\begin{align*}
    \rho(\bh)\defin \prod_{j=1}^r h_j^{d_j}\quad\text{and} \quad \lambda_{\bd}(L)\defin \om{\bd-1}\int_{\R_+^r}L\lrp{\bs} \times\prod_{j=1}^r s_j^{d_j/2-1} 2^{d_j/2-1}\,\rd \bs.
\end{align*}

The most popular kernel on the sphere is the von Mises--Fisher (vMF) kernel, defined as $L_{\mathrm{vMF}}(t)\defin e^{-t}$ for $t\geq0$. It is related with the widely known vMF distribution on $\Sd$, $d\geq 1$, whose pdf is
\begin{align*}
    f_{\mathrm{vMF}}(\bx;\bmu,\kappa)\defin c^\mathrm{vMF}_{d}(\kappa) e^{\kappa\bx'\bmu},\quad c_{d}^\mathrm{vMF}(\kappa)\defin\frac{\kappa^{(d-1)/2}}{(2\pi)^{(d+1)/2}\mathcal{I}_{(d-1)/2}(\kappa)},
\end{align*}
where $\bmu\in\Sd$ is the spherical mean, $\kappa\geq0$ is the concentration parameter about $\bmu$, and $\mathcal{I}_\nu$ is the modified Bessel function of order $\nu$. For this kernel, it easily follows that $c_{d,L_{\mathrm{vMF}}}(h)^{-1}=c_{d}^{\mathrm{vMF}}(1/h^2)^{-1}e^{-1/h^2}$. If the product kernel $L_\mathrm{vMF}^P(\bs)\defin\prod_{j=1}^r L_\mathrm{vMF}(s_j)$ is used in \eqref{eq:estimator}, the kde becomes a mixture of componentwise independent vMF pdfs:
\begin{align}
    \hat f(\bx;\bh)=\frac{1}{n}\sum_{i=1}^n f_{\mathrm{PvMF}}\blrp{\bx;\bX_i,\bh^{\odot(-2)}},\quad f_{\mathrm{PvMF}}(\bx;\bmu,\bkappa)\defin\prod_{j=1}^r f_{\mathrm{vMF}}(\bx_j;\bmu_j,\kappa_j),\label{eq:pmvf}
\end{align}
where $\odot$ denotes the Hadamard product operator, $\mathrm{PvMF}$ means \emph{Polyspherical vMF}, and $\bx,\bmu\in\Sr$ and $\bkappa\in\mathbb{R}^r_+$. In this case, $\smash{c_{\bd, L^P_{\mathrm{vMF}}}(\bh)^{-1}=c_{\bd}^{\mathrm{vMF}}\blrp{\bh^{\odot(-2)}}^{-1} \prod_{j=1}^re^{-1 / h_j^2}}$, where $\smash{c_{\bd}^{\mathrm{vMF}}(\bkappa)=\prod_{j=1}^r c_{d_j}^{\mathrm{vMF}}(\kappa_j)}$.

Another relevant kernel is the \emph{Epanechnikov} kernel $L_{\mathrm{Epa}}(t)\defin(1-t)1_{\{0\leq t\leq 1\}}$ (henceforth abbreviated as ``Epa'') defined in \cite{Hall1987}. In that paper, it was stated (p. 758) that this kernel is the most efficient for \eqref{eq:estimator}, when $r=1$ and, at least, $d=2$ (the statement allows for some degree of ambiguity about its scope in $d$). The precise quantification of this efficiency on $\Sr$ will be addressed in Section \ref{sec:ker}.

The estimator \eqref{eq:estimator} accommodates as particular cases the usual kdes by \citet[p. 1169]{Beran1979}, \citet[p. 25]{Bai1988}, and \citet[p. 52]{Hall1987} when $r=1$ and $d_1\geq1$. When $d_1=\cdots=d_r=1$, this kde is closely related to that proposed by \citet[p. 2158]{DiMarzio2011} on the torus after a reparametrization. Specifically, set $\bh=\bkappa^{\odot(-1/2)}$, where $\bkappa$ is the vector of the concentrations appearing in \cite{DiMarzio2011}. Then, after introducing polar coordinates, the kde \eqref{eq:estimator} restricted to $d_1=\cdots=d_r=1$ and the one presented by \cite{DiMarzio2011} coincide for a common vMF kernel. Within the toroidal setting, \cite{DiMarzio2011}'s kde is more general, as it also targets density derivative estimation, which we do not pursue in this article. Nonetheless, their work is restricted to kernel products. Finally, estimator \eqref{eq:estimator} also contains as a particular case the kde on $\mathbb{S}^{d_1}\times \mathbb{S}^{d_2}$ from \citet[p. 1210]{Garcia-Portugues2015}, where product kernels are also assumed.

%-------------------------------%
\section{Asymptotic properties}
\label{sec:prop}
%-------------------------------%

%-------------------------------%
\subsection{Main results}
\label{sec:mainres}
%-------------------------------%

To derive the asymptotic properties of estimator~\eqref{eq:estimator}, let us first introduce $\bar{f}$, the radial extension of $f$ from $\Sr$ to $\R^{\tilde{d}+r}\backslash\{\mathbf{0}\}$ defined by $\bar{f}(\bx)\defin f(\bx_1/\|\bx_1\|,\ldots,\bx_r/\|\bx_r\|)$. This extension allows taking derivatives of the density function $f$ (defined only on $\Sr$) in a straightforward manner, with the addition of vanishing properties in those derivatives that can be explicitly controlled. The following assumptions are required:
\begin{enumerate}[label=\textbf{A\arabic{*}}., ref=\textbf{A\arabic{*}}]
    \item $\bar{f}$ is twice continuously differentiable on $\Sr$. \label{A1}
    \item The kernel $L:\R_{\geq 0}^r\to\R_{\geq 0}$ is a bounded function such that $0<\int_{\R_+^r}L^\ell\lrp{\bs} \prod_{j=1}^r s_j^{d_j/2-1}\,\rd \bs<\infty$ for $\ell=1,2$ and all $d_j\geq 1$, $j=1,\ldots,r$. \label{A2}
    \item $\bh\equiv\bh_n$ is a sequence in $\mathbb{R}_+^r$ such that $\bh_n\to\mathbf{0}$ and $n\rho(\bh_n) \to\infty$ as $n\to\infty$. \label{A3}
\end{enumerate}

Assumption \ref{A1} is needed to enable operative Taylor expansions on $f$ via its radial extension. Note that the radial extension implies that $\bnabla \bar{f}(\bx)\bx=0$ for all $\bx\in\Sr$, i.e., that the directional derivatives of $\bar{f}$ normal to $\bx$ are null. Assumption \ref{A2} is required to ensure the existence of first- and second-order moments of the kernel $L$. The condition implies that, in particular, $L$ has to decay faster than any power function. This is trivially satisfied if $L$ has compact support, such as the Epa product kernel $L_\mathrm{Epa}^P(\bs)\defin\prod_{j=1}^r L_\mathrm{Epa}(s_j)$, or if the kernel decays exponentially, such as the vMF product kernel $L_\mathrm{vMF}^P$. Assumption \ref{A3} is standard to ensure the consistency of a kde.

The following theorem provides the asymptotic bias and variance expansions of the estimator~\eqref{eq:estimator}.

\begin{theorem}[Asymptotic bias and variance]\label{thm:biasvar}
Under \ref{A1}--\ref{A3}, for $\bx\in\Sr$,
\begin{align}
 \mathbb{E}[\hat{f}(\bx;\bh_n)]&=f(\bx)+(\bh_n^{\odot 2})'\bT\bar{f}(\bx)\bb_{\bd}(L)+o\blrp{(\bh_n^{\odot 2})'\mathbf{1}_r},\label{eq:estimator:bias}\\
	\mathbb{V}\mathrm{ar}[\hat{f}(\bx;\bh_n)]&=\frac{v_{\bd}(L)}{n\rho(\bh_n)}f(\bx) +o\blrp{\blrp{n \rho(\bh_n)}^{-1}},\label{eq:estimator:variance}
\end{align}
where
\begin{align}
        \bT\bar{f}(\bx)\defin&\;\diag{\tr{\bHcal_{11}\bar{f}(\bx)},\ldots,\tr{\bHcal_{rr}\bar{f}(\bx)}},\nonumber\\
        \bb_{\bd}(L)\defin&\; (b_{\bd,1}(L),\ldots,b_{\bd,r}(L))',\quad b_{\bd,j}(L)\defin\frac{\lambda_{\bd}(L(\bs)s_j)}{d_j \lambda_{\bd}(L)}, \quad v_{\bd}(L)\defin\, \frac{\lambda_{\bd}(L^2)}{\lambda_{\bd}(L)^2},\label{eq:moments}
\end{align}
with
$\lambda_{\bd}(L(\bs)s_j)\equiv\om{\bd-1}\int_{\R_+^r}L\lrp{\bs} s_j \prod_{j=1}^r s_j^{d_j/2-1} 2^{d_j/2-1}\,\rd \bs$ and  $\bHcal_{jj}\bar{f}(\bx)$ being the $(d_j+1)\times(d_j+1)$ marginal Hessian matrix matrix corresponding to the $j$th diagonal block of the Hessian matrix of $\bar{f}$ at $\bx$.
\end{theorem}

A lower order for the remaining term of the bias given in \eqref{eq:estimator:bias} can be obtained with the following stronger differentiability assumption, given that the third-order term vanishes due to symmetry.

\begin{enumerate}[label=\textbf{A\arabic{*}*}., ref=\textbf{A\arabic{*}*}]
    \item $\bar{f}$ is four times continuously differentiable on $\Sr$. \label{A1*}
\end{enumerate}

\begin{corollary}[Improved asymptotic bias expansion]\label{cor:bias:order}
Under \ref{A1*} and \ref{A2}--\ref{A3}, for $\bx\in\Sr$,
\begin{align*}
    \mathbb{E}[\hat{f}(\bx;\bh_n)]&=f(\bx)+(\bh_n^{\odot 2})'\bT\bar{f}(\bx)\bb_{\bd}(L)+O\blrp{(\bh_n^{\odot 4})'\mathbf{1}_r}.
\end{align*}
\end{corollary}

\begin{remark} \label{rem:prod}
In the case of a \emph{product kernel} $L^P(\bs)\defin\prod_{j=1}^r L(s_j)$, common dimensions $d\defin d_1=\cdots=d_r$, and equal bandwidths $h_n\defin h_{1,n}=\cdots=h_{r,n}$, Theorem \ref{thm:biasvar} becomes
\begin{align*}
    \mathbb{E}[\hat{f}(\bx;h_n\one_r)]&=f(\bx)+h_n^{2}b_d(L)\tr{\bT\bar{f}(\bx)}+o\left(h_n^2\right),\\
    \mathbb{V}\mathrm{ar}[\hat{f}(\bx;h_n\one_r)]&=\frac{v_{d}^r(L)}{nh_n^{dr}} f(\bx) +o\blrp{\blrp{n h_n^{dr}}^{-1}},
\end{align*}
with $b_{d}\left(L\right)=\lambda_{d}(L(s)s)/(d\lambda_{d}(L))$, $v_{d}(L)=\lambda_{d}(L^2)/\lambda_{d}(L)^2$, and $\lambda_{d}(L)=2^{d/2-1}\om{d-1}\int_{\R_+} L(s) s^{d/2-1}\,\rd s$.
\end{remark}

An immediate consequence of Theorem \ref{thm:biasvar} is the strong consistency of $\hat{f}(\bx;\bh_n)$.

\begin{corollary}[Strong pointwise consistency]\label{cor:consist}
Under \ref{A1}--\ref{A3}, for $\bx\in\Sr$, $\hat{f}(\bx;\bh_n)\stackrel{\mathrm{a.s.}}{\to} f(\bx)$ when $n\to\infty$.
\end{corollary}

The following corollary yields the Asymptotic Mean Integrated Squared Error (AMISE) optimal bandwidth and the optimal AMISE, particularized to the setting of Remark \ref{rem:prod}. %

\begin{corollary}[AMISE optimal bandwidth]\label{cor:amisebwd}
Under \ref{A1}--\ref{A3},
\begin{align}
    \mathrm{AMISE}[\hat{f}(\cdot;\bh_n)]=&\;\int_{\Sr}((\bh_n^{\odot 2})'\bT\bar{f}(\bx)\bb_{\bd}(L))^2\,\sigmar(\rd\bx)+\frac{v_{\bd}(L)}{n\rho(\bh_n)},\label{eq:amise}
\end{align}
and, therefore,
\begin{align}
    \bh_\mathrm{AMISE}\defin \arg\min_{\bh>\mathbf{0}}\mathrm{AMISE}\lrcbig{\hat{f}(\cdot;\bh)}.\label{eq:hamisedh}
\end{align}

When the kernel is $L^P$, the common dimension is $d$, and the equal bandwidth is $h_n$,
\begin{align}
    \mathrm{AMISE}\lrcbig{\hat{f}(\cdot;h_n\one_r)}=&\;h_n^4b^{2}_d(L)R\lrpbig{\nabla^2 \bar{f}}+\frac{v^r_{d}(L)}{nh_n^{dr}},\label{eq:amise1d}
\end{align}
where $R\lrpbig{\nabla^2 \bar{f}}\defin \int_{\Sdr}\lrpbig{\nabla^2 \bar{f}(\bx)}^2\,\sigma_{\bd}(\rd\bx)$. In this case,
\begin{align}
    h_\mathrm{AMISE}\defin \arg\min_{h>0}\mathrm{AMISE}\lrcbig{\hat{f}(\cdot;h\one_r)}=\lrc{\frac{v^r_d(L)}{b^{2}_d(L)}\times\frac{dr}{4R\lrpbig{\nabla^2 \bar{f}}n}}^{1/(dr+4)}\label{eq:hamise}
\end{align}
and
\begin{align}
    \mathrm{AMISE}\lrcbig{\hat{f}(\cdot;h_\mathrm{AMISE}\one_r)}=c_{d,r}C_{d,r}(L^P)R\lrpbig{\nabla^2 \bar{f}}^{dr/(dr+4)}n^{-4/(dr+4)},\label{eq:optamise}
\end{align}
with $c_{d,r}\defin (dr/4)^{4/(dr+1)}(1+4/(dr))$ and $C_{d,r}(L^P)\defin\lrc{v_d^{4r}(L) b^{2dr}_d(L)}^{1/(dr+4)}$.
\end{corollary}

We conclude the section stating the pointwise asymptotic distribution of estimator \eqref{eq:estimator}. To do so, we require an extra condition on the kernel that is slightly stronger than \ref{A3}. This assumption is satisfied by $L^P_\mathrm{vMF}$ and $L^P_\mathrm{Epa}$ kernels.

\begin{enumerate}[label=\textbf{A\arabic{*}}., ref=\textbf{A\arabic{*}}]
    \setcounter{enumi}{3}
    \item For some $\delta>0$, $\int_{\R_+^r} L^{2+\delta}(\bs) \prod_{j=1}^r s_j^{d_j/2-1}\,\rd \bs<\infty$ for all $d_j\geq 1$,  $j=1,\ldots,r$. \label{A4}
\end{enumerate}

\begin{theorem}[Pointwise asymptotic normality]\label{thm:distr}
Under \ref{A1}--\ref{A4}, for $\bx\in\Sr$,
\begin{align}
    \sqrt{n\rho(\bh_n)}\lrp{\hat{f}(\bx;\bh_n)-\mathbb{E}[\hat{f}(\bx;\bh_n)]}\rightsquigarrow\mathcal{N}(0,v_{\bd}(L)f(\bx)).\label{eq:asymp1}
\end{align}
If, in addition, $\sqrt{n\rho(\bh_n)}(\bh_n^{\odot 2})'\mathbf{1}_r=O(1)$, then
\begin{align}
    \sqrt{n\rho(\bh_n)}\lrp{\hat{f}(\bx;\bh_n)-f(\bx)-(\bh_n^{\odot 2})'\bT\bar{f}(\bx)\bb_{\bd}(L)}\rightsquigarrow\mathcal{N}(0,v_{\bd}(L)f(\bx)).\label{eq:asymp2}
\end{align}
\end{theorem}

\begin{remark} \label{rem:orders_norm1}
Within the setting of common dimensions and equal bandwidths of Corollary \ref{cor:amisebwd}, condition $\sqrt{n\rho(\bh_n)}\bh^{\odot 2'}_n\mathbf{1}_r=O(1)$ in Theorem \ref{thm:distr} reads $nh_n^{dr+4}=O(1)$. Together with \ref{A3}, this implies that the order of the sequences of allowed bandwidths for \eqref{eq:asymp2} is such that $n^{-1}\prec h_n \preceq n^{-1/(dr+4)}$, i.e., the bandwidth order needs to prioritize the reduction of bias, with the AMISE-optimal order $n^{-1/(dr+4)}$ being a limiting case (we use $a_n\prec b_n$ and $a_n\preceq b_n$ to denote $a_n=o(b_n)$ and $a_n=O(b_n)$, respectively).
\end{remark}

\begin{remark} \label{rem:orders_norm2}
Under \ref{A1*} instead of \ref{A1}, and still within the setting of common dimensions and equal bandwidths, Corollary \ref{cor:bias:order} transforms $nh_n^{dr+4}=O(1)$ into $nh_n^{dr+4}O(h_n^4)=o(1)$ and the sequence of allowed bandwidths is expanded to $n^{-1}\prec h_n \prec n^{-1/(dr+8)}$.
\end{remark}

Numerical experiments that illustrate and empirically validate Theorems \ref{thm:biasvar}--\ref{thm:distr} are provided in Section B.1 of the SM.

%-------------------------------%
\section{Kernel theory}
\label{sec:ker}
%-------------------------------%

%-------------------------------%
\subsection{Non-vMF kernels on the sphere}
\label{sec:nonvMFker}
%-------------------------------%

In Section B.2 of the SM the Epa kernel is seen to have a higher efficiency than the vMF kernel on $\Sd$, $d\geq 1$. For this reason, we provide next the explicit formulae for its normalizing constant and an effective means of simulating from it. Both tasks are well-established for the vMF kernel given its association with the vMF distribution, but have been elusive for the Epa kernel.

\begin{proposition}[Normalizing constant for the Epa kernel]\label{prop:Epa1}
Let $d\geq 1$ and $h>0$. Then,
\begin{align*}
    c_{d,L_\mathrm{Epa}}(h)^{-1}&=\om{d}(1-h^{-2})\lrc{1-F_d(m_h)}+\frac{\om{d-1}\lrp{1-m_h^2}^{d/2}}{dh^{2}},
\end{align*}
where $m_h\defin\max(-1,1-h^2)$, and $F_d$ is the cumulative distribution function (cdf) of $\bga'\bX$ for  $\bX\sim\mathrm{Unif}(\Sd)$ and an arbitrary $\bga\in\Sd$. If $h\geq\sqrt{2}$, then $c_{d,L_\mathrm{Epa}}(h)^{-1}=\om{d}(1-h^{-2})$.

For $d=1,2$ and $h>0$,
\begin{align*}
    c_{1,L_\mathrm{Epa}}(h)^{-1}&=2h^{-2}[(h^2-1)\arccos(m_h)+(1-m_h^2)^{1/2}]\quad\text{and}\\
    c_{2,L_\mathrm{Epa}}(h)^{-1}&=\pi(1-m_h)(2-h^{-2}(1-m_h)).
\end{align*}
\end{proposition}

The cdf $F_d$ is tractable for any $d\geq 1$. Explicit formulae are available for $d=1,2$, while recurrence relations allow evaluating $F_d$ for any $d>2$ \citep[see, e.g., Section 2.2 in][]{Garcia-Portugues2022}.

Simulation from the pdf $L_{h,\mathrm{Epa}}(\cdot,\bmu):=c_{d,L_\mathrm{Epa}}(h)L_{\mathrm{Epa}}((1-\bmu'\cdot)/h^2):\Sd\to \R_{\geq0}$, $\bmu\in\Sd$, can be achieved, as in any rotationally symmetric distribution \citep[see][for the vMF case]{Wood1994}, through the tangent-normal decomposition: if $\bY\defin T\bmu + \sqrt{1-T^2}\bB_{\bmu}\boldsymbol{\Xi}$, where $\boldsymbol{\Xi}\sim\mathrm{Unif}(\mathbb{S}^{d-1})$ is independent from the \emph{angular} random variable $T \sim F_{d,\mathrm{Epa},h}$ on $[-1,1]$ and $\bB_{\bmu}$ is a semiorthogonal $(d+1)\times d$ matrix with column vectors orthogonal to $\bmu$, then $\bY\sim L_{h,\mathrm{Epa}}(\cdot,\bmu)$. %

\begin{proposition}[Angular cdf of the Epa kernel]\label{prop:Epa2}
Let $d\geq 1$, $h>0$, and $\bY\sim L_{h,\mathrm{Epa}}(\cdot,\bmu)$. Then, the pdf of $T=\bY'\bmu$ is $f_{d,\mathrm{Epa},h}(t)=\om{d-1} c_{d,L_\mathrm{Epa}}(h)\{1-(1-t)/h^2\}(1-t^2)^{d/2-1}1_{\{m_h<t<1\}}$. Its cdf is \begin{align*}
    F_{d,\mathrm{Epa},h}(t)=\frac{\om{d-1} c_{d,L_\mathrm{Epa}}(h)}{h^2}\lrb{(h^2-1)\frac{\om{d}}{\om{d-1}}\lrc{F_d(t)-F_d(m_h)}-\frac{1}{d}\lrc{(1-t^2)^{d/2} - (1-m_h^2)^{d/2}}},
\end{align*}
with $t\in[m_h,1]$. In addition, when $d=2$:
\begin{align*}
    F^{-1}_{2,\mathrm{Epa},h}(u)=1-h^2\pm\sqrt{(h^2-1)^2-m_h+(m_h+u(1-m_h))(2h^2-1+m_h)},\quad u\in(0,1).
\end{align*}
\end{proposition}

We next consider the ``softplus'' function $\mathrm{sfp}(t)\defin\log(1+e^t)$ to build a smooth approximator of the Epa kernel that combines the higher efficiency of the latter kernel with the smoothness and positiveness of the vMF kernel. Indeed, the kernel $L_{\mathrm{sfp}_{\upsilon}}(t)\defin \mathrm{sfp}(\upsilon(1-t)) / \mathrm{sfp}(\upsilon)$ for $\upsilon>0$ is such that $\lim_{\upsilon\to\infty}L_{\mathrm{sfp}_\upsilon}(t)=L_{\mathrm{Epa}}(t)$ (see Figure \ref{fig:effic:a}). The normalizing constant of $L_{\mathrm{sfp}_\upsilon}$ and the angular cdf $F_{d,\mathrm{sfp}_\upsilon,h}$ are challenging for general $d\geq1$, but they admit tractable forms for $d=2$.

\begin{proposition}[Normalizing constant and angular cdf of the sfp kernel]\label{prop:sp}
Let $d=2$, $h>0$, $\upsilon>0$, and $\bY\sim L_{h,\mathrm{sfp}_\upsilon}(\cdot,\bmu)$. Then, the pdf of $T=\bY'\bmu$ is $f_{d,\mathrm{sfp}_\upsilon,h}(t)=\om{d-1} c_{d,L_{\mathrm{sfp}_\upsilon}}(h)L_{\mathrm{sfp}_\upsilon}((1-t)/h^2) (1-t^2)^{d/2-1}$,
\begin{align*}
	c_{2,L_{\mathrm{sfp}_\upsilon}}(h)^{-1}&=\frac{2\pi}{\mathrm{sfp}(\upsilon)\upsilon}\Big[\mathrm{Li}_{2}\big(-e^{\upsilon(1-2h^{-2})}\big)-\mathrm{Li}_{2}(-e^{\upsilon})\Big]h^2,\quad \text{and}\\
	F_{2,\mathrm{sfp}_\upsilon,h}(t)&=\frac{\mathrm{Li}_2\big(-e^{\upsilon(1-2h^{-2})}\big)-\mathrm{Li}_2\big(-e^{\upsilon(1-(1-t)h^{-2})}\big)}{  \mathrm{Li}_2\big(-e^{\upsilon(1-2h^{-2})}\big)-\mathrm{Li}_2\big(-e^\upsilon\big)}
\end{align*}
for $t\in[-1,1]$, where $\mathrm{Li}_s$ is the polylogarithm function with integral representation $\mathrm{Li}_s(z)=\Gamma(s)^{-1}\allowbreak \int_{0}^{\infty} t^{s-1}/(e^t/z-1)\,\rd t$ for $s>0$ and $z<1$.
\end{proposition}

The following result collects the moments of the presented kernels. These appear in the results of Section \ref{sec:mainres} regarding product kernels and are key in establishing kernel efficiency in Section \ref{sec:effiker}.

\begin{proposition}[Moments of the vMF, Epa, and sfp kernels]\label{prop:moments}
Let $d\geq1$. Then,
\begin{align*}
    b_d(L_{\mathrm{vMF}})&=\frac{1}{2},&
    v_d(L_{\mathrm{vMF}})&=\frac{1}{(2\pi^{1/2})^d},\\
    b_d(L_{\mathrm{Epa}})&=\frac{1}{d+4},&
    v_d(L_{\mathrm{Epa}})&=\frac{4\Gamma(d/2+2)}{(2\pi)^{d/2}(d+4)},\\
    b_d(L_{\mathrm{sfp}_\upsilon})&=\frac{\mathrm{Li}_{d/2+2}(-e^\upsilon)}{2\upsilon\,\mathrm{Li}_{d/2+1}(-e^\upsilon)},&
    v_d(L_{\mathrm{sfp}_\upsilon})&=\frac{\upsilon^d J_d(\upsilon)}{(2\pi)^{d/2}\Gamma(d/2) \,\mathrm{Li}^2_{d/2+1}(-e^\upsilon)},
\end{align*}
where $J_d(\upsilon)\defin\int_0^\infty \log(1+e^{\upsilon(1-t)})^2 t^{d/2-1} \,\rd t$.
\end{proposition}

\begin{remark}\label{rem:equiv}
    For all $d\geq 1$ and $h>0$, $\lim_{\upsilon\to\infty}c_{2,L_{\mathrm{sfp}_\upsilon}}(h)=c_{2,L_\mathrm{Epa}}(h)$. Similarly, $\lim_{\upsilon\to\infty}b_d(L_{\mathrm{sfp}_\upsilon})=b_d(L_\mathrm{Epa})$ and $\lim_{\upsilon\to\infty}v_d(L_{\mathrm{sfp}_\upsilon})=v_d(L_\mathrm{Epa})$. These statements can be checked using the asymptotic equivalence $\mathrm{Li}_{s}\left(\pm e^w\right) \asymp-w^s/\Gamma(s+1)$ as $w\to\infty$, for $s\in\mathbb{N}$ \cite[Equation (11.3)]{Wood1992}.
\end{remark}

Simulation from $f_{d,\mathrm{Epa},h}$, for general $d\geq1$, can also be approached by acceptance-rejection: sample $T\sim f_{d,\mathrm{vMF},h}$ and accept it if $U<L_{\mathrm{Epa}}((1-T)/h^2)/[M L_{\mathrm{vMF}}((1-T)/h^2)]$, where $U\sim\mathrm{Unif}[0,1]$ and $M=c_{d,L_\mathrm{Epa}}(h)/c_{d,\mathrm{vMF}}(h)$ since $L_{\mathrm{vMF}}\geq L_{\mathrm{Epa}}$. A analogous approach holds for $f_{d,\mathrm{sfp}_\nu,h}$, with $M=c_{d,\mathrm{sfp}_\nu}(h)/c_{d,\mathrm{vMF}}(h)$ for sufficiently large $\nu>0$.

%-------------------------------%
\subsection{Spherically symmetric kernels on the polysphere}
\label{sec:sphker}
%-------------------------------%

The product kernel $\smash{L^{P}(\bs)=\prod_{j=1}^r L(s_j)}$ from Remark \ref{rem:prod} is an example of a kernel on $\Sr$ induced by a kernel $L:\R_{\geq 0} \to\R_{\geq 0}$ on $\Sd$. Another instance of such construction is the \emph{spherically symmetric kernel} $L^{S}(\bs)\defin L(\|\bs\|_1)$. In the Euclidean setting, disparities are commonly observed between product and spherically symmetric kernels \citep[Section 4.5]{Wand1994}, with the Gaussian kernel being a notable exception. Analogously, the vMF kernel verifies that $L_\mathrm{vMF}^{P}=L_\mathrm{vMF}^{S}$. From the Epa and sfp kernels on $\Sd$, we define their spherically symmetric counterparts on $\Sr$ $\smash{L_\mathrm{Epa}^{S}(\bs)}\defin L_\mathrm{Epa}(\|\bs\|_1)=(1-\|\bs\|_1)1_{\{\|\bs\|_1\leq 1\}}$ and $\smash{L_{\mathrm{sfp}_\upsilon}^{S}(\bs)}\defin L_{\mathrm{sfp}_\upsilon}(\|\bs\|_1)=\mathrm{sfp}(\upsilon(1-\|\bs\|_1)) / \mathrm{sfp}(\upsilon)$, respectively.

First, we establish a useful direct connection between the moments of any spherically symmetric kernel $L^S$ on $\Sr$ and those for $L$ on $\mathbb{S}^{\tilde{d}}$, where $\tilde{d}=\sum_{j=1}^r d_j$.

\begin{proposition}[Moments of spherically symmetric kernels] \label{prop:moments2} If $d_1,\ldots,d_r\geq1$ and $L^{S}(\bs)= L(\|\bs\|_1)$, then $\lambda_{\bd}(L^S)=\lambda_{\tilde{d}}(L)$, $\bb_{\bd}(L^S)= b_{\tilde{d}}(L)\one_r$, and $v_{\bd}(L^S)=v_{\tilde{d}}(L)$.
\end{proposition}

General asymptotic formulae for the normalizing constants of $L_{\mathrm{Epa}}^S$ and $L_{\mathrm{sfp}_\upsilon}^S$ are given next. For the specific $\smash{(\mathbb{S}^2)^r}$ case, exact expressions are derivable.

\begin{proposition}[Normalizing constants for the spherically symmetric Epa and sfp kernels]\label{prop:sp2}
For $d_1,\ldots,d_r\geq 1$, $\upsilon>0$, $\bh\in\mathbb{R}_+^r$, and as $\bh\to\zero$,
\begin{align}
    c_{\bd,L_{\mathrm{Epa}}^S}(\bh)^{-1}\asymp\rho(\bh)\frac{4(2\pi)^{\tilde{d}/2}}{\tilde{d}(\tilde{d}+2)\Gamma(\tilde{d}/2)}, \quad c_{\bd,L^S_{\mathrm{sfp}_\upsilon}}(\bh)^{-1}\asymp-\rho(\bh)\left(\frac{2\pi}{\upsilon}\right)^{\tilde{d}/2} \frac{\mathrm{Li}_{\tilde{d}/2+1}(-e^\upsilon)}{\mathrm{sfp}(\upsilon)}. \label{eq:asymp-c-d2}
\end{align}

Under common dimension $d=2$ and equal bandwidth $h$, then
\begin{align*}
   c_{\two,L_\mathrm{Epa}^{S}}(h\one_r)^{-1}&=\frac{(2\pi h^2)^r(-1)^{r}}{(r+1)!}\sum_{\ell=\lceil r-h^2/2 \rceil}^{r}\binom{r}{\ell} (-1)^\ell\, \lrp{1-2h^{-2}(r-\ell)}^{r+1},\\
   c_{\two,L_{\mathrm{sfp}_\upsilon}^{S}}(h\one_r)^{-1}&=\frac{(2\pi h^2)^r}{\mathrm{sfp}(\upsilon)}\frac{(-1)^{r-1}}{\upsilon^r}\sum_{\ell=0}^{r}\binom{r}{\ell} (-1)^\ell\, \mathrm{Li}_{(r+1)}\big(-e^{\upsilon(1-rh^{-2})+\upsilon h^{-2}(2\ell-r)}\big).
\end{align*}
If $h<\sqrt{2}$, then $c_{\two,L_\mathrm{Epa}^{S}}(h\one_r)^{-1}=(2\pi h^2)^r/(r+1)!$.
\end{proposition}

\begin{remark}\label{rem:optamiseLS}
Using Proposition \ref{prop:moments2}, we can readily derive analogous results to \eqref{eq:amise1d}--\eqref{eq:optamise} when the considered kernel is $L^S$, the common dimension is $d$, and the equal bandwidth is $h_n$. In this case, \eqref{eq:amise1d}--\eqref{eq:hamise} hold replacing $b_d^2(L)$ and $v_{d}^r(L)$ with $b_{dr}^2(L)$ and $v_{dr}(L)$, while \eqref{eq:optamise} becomes
\begin{align}
    \mathrm{AMISE}\lrcbig{\hat{f}(\cdot;h_\mathrm{AMISE}\one_r)}=c_{d,r}C_{d,r}(L^S)R\lrpbig{\nabla^2 \bar{f}}^{dr/(dr+4)}n^{-4/(dr+4)},\label{eq:optamiseLS}
\end{align}
with $C_{d,r}(L^S)\defin\lrc{v_{dr}^4(L) b^{2dr}_{dr}(L)}^{1/(dr+4)}$.
\end{remark}

Similarly to Section \ref{sec:nonvMFker}, simulation from the pdfs $L_{\bh,\mathrm{Epa}}^S(\cdot,\bmu):=c_{\bd,L_\mathrm{Epa}^S}(\bh)L_{\mathrm{Epa}}(\sum_{j=1}^r(1-\bmu_j'\cdot_j)/h_j^2):\Sr\to \R_{\geq0}$ and $L_{\bh,\mathrm{sfp}_\nu}^S(\cdot,\bmu)$, $\bmu\in\Sr$, can be performed through acceptance-rejection from $L_{\bh,\mathrm{vMF}}^P(\cdot,\bmu)$, which admits straightforward sampling.

%-------------------------------%
\subsection{Kernel efficiency}
\label{sec:effiker}
%-------------------------------%

We restrict to the special case of common dimension $d$ and equal bandwidth $h_n$ throughout this subsection. This simplification is motivated by $\Sr$ being the most practically relevant polyspherical space, and on it the components of $\bh_{\mathrm{AMISE}}$ are of equal order. Given that the kernel functionals $C_{d,r}(L^P)$ and $C_{d,r}(L^S)$ determine the optimal AMISE for a product kernel $L^P$ and a spherically symmetric kernel $L^S$ (see \eqref{eq:optamise} and \eqref{eq:optamiseLS}, respectively), we define the \emph{efficiency} of the product or spherically symmetric kernel $L$ relative to $L_{\mathrm{Epa}}^S$ as
\begin{align*}
    \mathrm{eff}_{d,r}(L)\defin\lrc{\frac{C_{d,r}(L_{\mathrm{Epa}}^S)}{C_{d,r}(L)}}^{(dr+4)/4}.
\end{align*}
The rationale of this definition is standard in kernel smoothing. In view of \eqref{eq:optamise} and \eqref{eq:optamiseLS}, $\mathrm{eff}_{d,r}(L)$ represents the fraction of the sample that the kernel $L^S_{\mathrm{Epa}}$ requires to achieve the minimum AMISE of the kernel $L$ when both are used with their AMISE bandwidths. The kernel $L_{\mathrm{Epa}}^S$ is taken as the reference since, as Table 1 in the SM corroborates, it is the most efficient among the explored kernels.

\begin{proposition}[Efficiencies of the $L^S_{\mathrm{vMF}}$, $L^S_{\mathrm{sfp}_\upsilon}$, $L^P_{\mathrm{sfp}_\upsilon}$, and $L^P_{\mathrm{Epa}}$ kernels]\label{cor:effic2}
Let $r,d\geq 1$ and $\upsilon>0$. Then:
\begin{align}
  \mathrm{eff}_{d,r}(L^S_{\mathrm{vMF}})=&\;\frac{2^{dr+2} \Gamma(dr/2+2)}{(dr+4)^{dr/2+1}},\label{eq:eff:vmf}\\
  \mathrm{eff}_{d,r}(L^S_{\mathrm{sfp}_\upsilon})=&\;\frac{2^{dr/2+2}\Gamma(dr/2+2)\Gamma(dr/2)|\mathrm{Li}|^{dr/2+2}_{dr/2+1}(-e^\upsilon)}{(dr+4)^{dr/2+1} \upsilon^{dr/2}J_{dr}(\upsilon) \,|\mathrm{Li}|_{dr/2+2}^{dr/2}(-e^\upsilon)},
  \nonumber\\
  \mathrm{eff}_{d,r}(L^P_{\mathrm{sfp}_\upsilon})=&\;\mathrm{eff}_{d,r}(L_{\mathrm{sfp}_\upsilon}^S) \times \frac{\Gamma(d/2)^rJ_{dr}(\upsilon)|\mathrm{Li}|_{dr/2+2}^{dr/2}(-e^\upsilon)|\mathrm{Li}|_{d/2+1}^{r(d/2+2)}(-e^\upsilon)}{\Gamma(dr/2)J_d^r(\upsilon)|\mathrm{Li}|_{dr/2+1}^{dr/2+2}(-e^\upsilon)|\mathrm{Li}|_{d/2+2}^{dr/2}(-e^\upsilon)},\label{eq:eff:sfpprod}\\
  \mathrm{eff}_{d,r}(L^P_{\mathrm{Epa}})=&\;\frac{4^{1-r}\Gamma(dr/2+2)(d+4)^{r(d/2+1)}}{\Gamma(d/2+2)^r(dr+4)^{dr/2+1}},\label{eq:eff:epa}
\end{align}
where $J_d(\upsilon)$ was defined in Proposition \ref{prop:moments} and $|\mathrm{Li}|_s$ denotes the absolute value of the $\mathrm{Li}_s$ function.
\end{proposition}

Several comments on these efficiencies follow. First, the efficiency of the vMF kernel with respect to $L^S_{\mathrm{Epa}}$ on $\Sdr$ is exactly that of the Gaussian kernel $K^S_\mathrm{Gauss}$ with respect to the spherically symmetric Epa kernel
on $\R^p$ with $p=dr$. Indeed, from Theorem 2 of \cite{Duong2015}, it follows that
\begin{align*}
  \mathrm{eff}_p(K_\mathrm{Gauss}^S)%
  =\frac{[4\Gamma(p/2+2)/(\pi^{p/2}(p+4))](p+4)^{-p/2}}{(4\pi)^{-p/2}}
  =\frac{2^{p+2} \Gamma(p/2+2)}{(p+4)^{p/2+1}},
\end{align*}
which coincides with \eqref{eq:eff:vmf}. Second, \eqref{eq:eff:epa} also coincides with the efficiency of the Epa kernel $\smash{K_{\mathrm{Epa},d}^r}$ on $\smash{(\R^d)^r}$ formed by the $r$-product of spherically symmetric Epa kernels on $\R^d$. This efficiency is also obtainable from Theorem 2 of \cite{Duong2015}:
\begin{align*}
  \mathrm{eff}_{dr}(K_{\mathrm{Epa},d}^r)%
  =\frac{[4\Gamma(dr/2+2)/(\pi^{dr/2}(dr+4))] (dr+4)^{-dr/2}}{[4\Gamma(d/2+2)/(\pi^{d/2}(d+4))]^r (d+4)^{-dr/2}}
  =\frac{4^{1-r}\Gamma(dr/2+2) (d+4)^{r(d/2+1)}}{\Gamma(d/2+2)^r (dr+4)^{dr/2+1}}.
\end{align*}
Finally, the effect of $(d,r)$ is symmetric for spherically symmetric kernels. For these kernels, the effect of adding dimensions on the spherical or product space is equivalent in terms of efficiency.

The efficiencies of the vMF, sfp, and Epa kernels are plotted in Figure \ref{fig:effic}, for fixed $r=1$ and $d=2$. Numerical results are listed in Section B.2 of the SM.

\begin{figure}[htpb!]
    \centering
    \hspace{-0.5cm}
    \begin{subfigure}{0.33\textwidth}
        \centering
        \includegraphics[width=1.1\textwidth,clip,trim={0cm 0.5cm 0cm 1.5cm}]{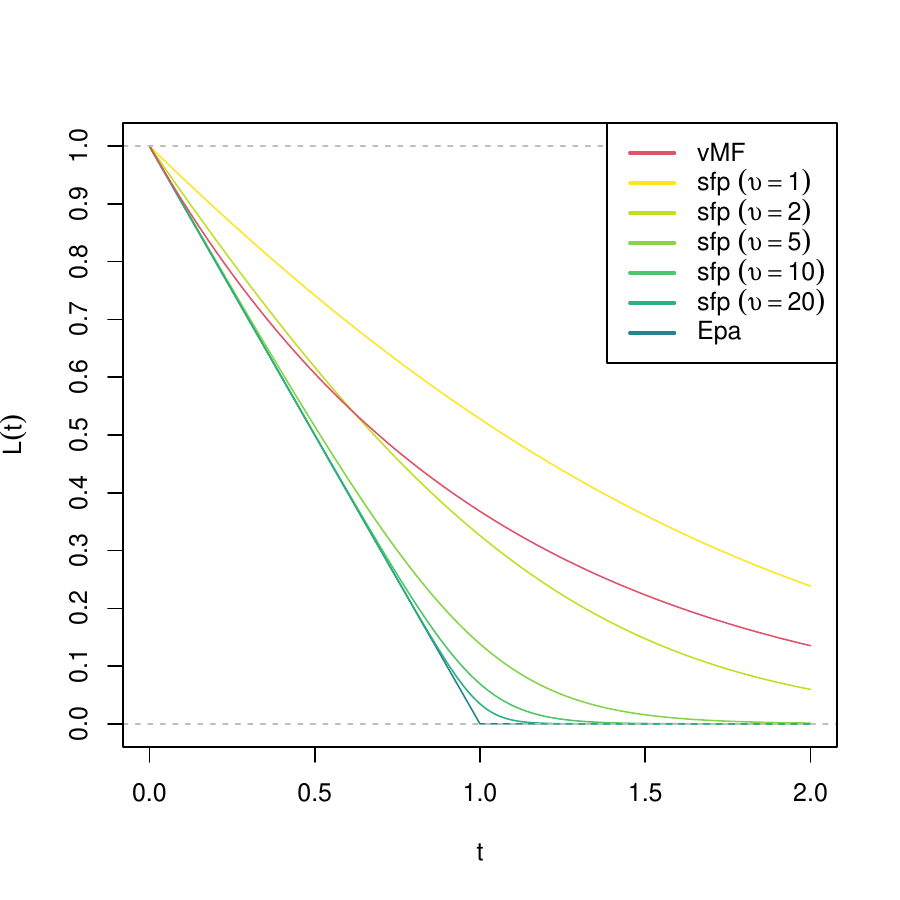}
        \caption{vMF, Epa, and sfp kernels} \label{fig:effic:a}
    \end{subfigure}%
    \
    \begin{subfigure}{0.33\textwidth}
        \centering
        \includegraphics[width=1.1\textwidth,clip,trim={0cm 0.5cm 0cm 1.5cm}]{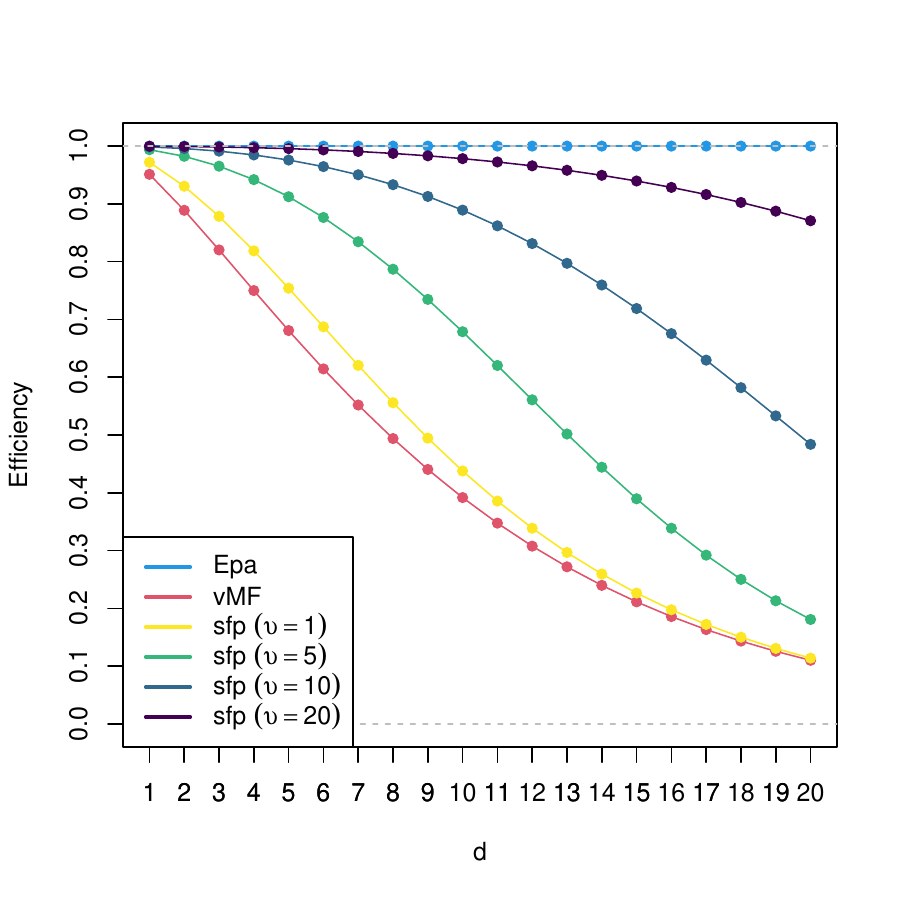}
        \caption{Efficiencies for $d$ and $r=1$} \label{fig:effic:b}
    \end{subfigure}%
    \
    \begin{subfigure}{0.33\textwidth}
        \centering
        \includegraphics[width=1.1\textwidth,clip,trim={0cm 0.5cm 0cm 1.5cm}]{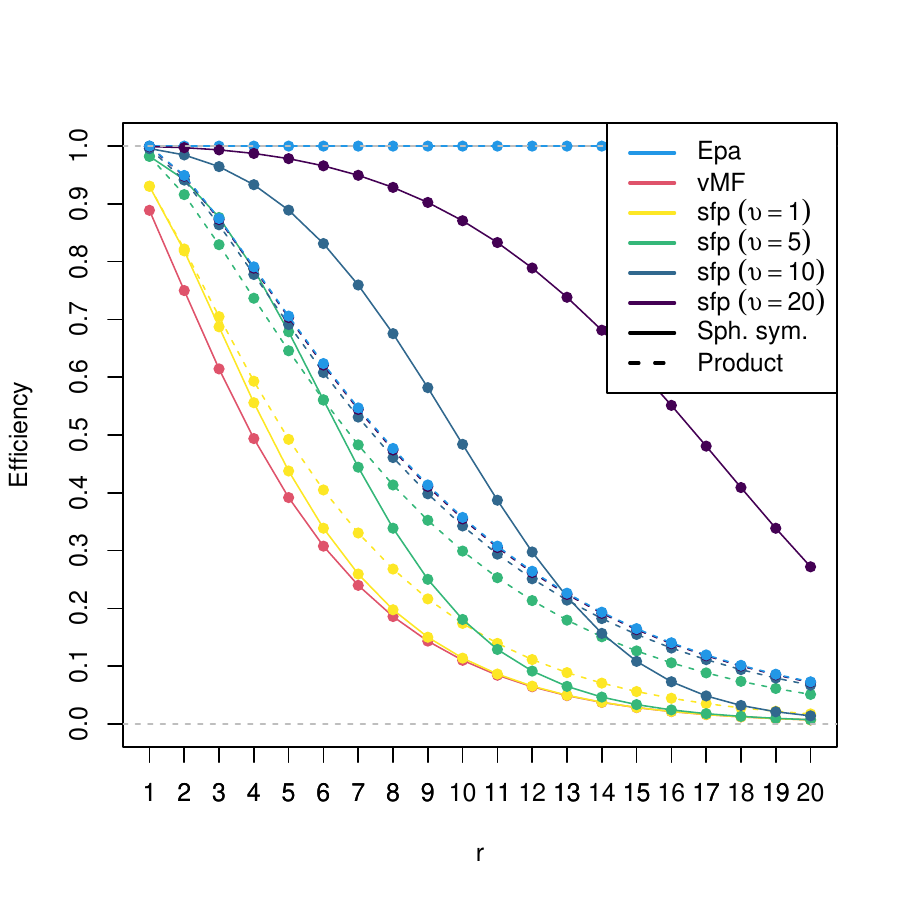}
        \caption{Efficiencies for $r$ and $d=2$} \label{fig:effic:c}
    \end{subfigure}%
    \caption{\small Kernels and their efficiencies. Figure \ref{fig:effic:a} shows the approximation of the Epa kernel by the sfp kernel as $\upsilon$ increases. Figure \ref{fig:effic:b} displays the efficiencies of the vMF and sfp kernels with respect to the $\mathrm{Epa}^S$ kernel for a fixed $r=1$. Figure \ref{fig:effic:c} does the same, but now for a fixed $d=2$.}
    \label{fig:effic}
\end{figure}

%-------------------------------%
\section{Bandwidth selection}
\label{sec:bw}
%-------------------------------%

%-------------------------------%
\subsection{Plug-in bandwidths}
\label{sec:bw_pi}
%-------------------------------%

Using the properties of the PvMF density, it is possible to compute the curvature term in the AMISE optimal bandwidth for this pdf, and then derive a (zero-stage) plug-in bandwidth selector.

\begin{proposition}[PvMF blockwise Hessian traces]\label{prop:nabla2}
For $\bx,\bmu\in\Sr$ and $\bkappa\in\R_{\geq 0}^r$,
\begin{align*}
    \bT_{\bar{f}_{\mathrm{PvMF}}}(\bx)=f_{\mathrm{PvMF}}(\bx;\bmu,\bkappa)\ \diag{\psi_1(\bx_1'\bmu_1,\kappa_1),\ldots,\psi_r(\bx_r'\bmu_r,\kappa_r)},
\end{align*}
with $\psi_j(t,\kappa)\defin\;\kappa[\kappa(1-t^2)-d_jt]$.
\end{proposition}

\begin{proposition}[AMISE optimal bandwidth for PvMF]\label{prop:plugin}
Assume that $f$ is a $f_\mathrm{PvMF}(\cdot; \bmu,\bkappa)$. Then, under \ref{A1}--\ref{A3},
\begin{align*}
    \mathrm{AMISE}[\hat{f}(\cdot;\bh_n)]=&\;(\bh_n^{\odot 2}\odot\bb_{\bd}(L))'\bR(\bkappa)(\bh_n^{\odot 2}\odot\bb_{\bd}(L))+\frac{v_{\bd}(L)}{n\rho(\bh_n)},
\end{align*}
where $\bR(\bkappa)$ is the $r\times r$ matrix of curvature terms given by
\begin{align*}
\bR(\bkappa)=\frac{1}{4}\lrb{\frac{1}{2}\mathrm{diag}(\bv(\bkappa))+(\bu(\bkappa)\bu(\bkappa)')^\circ}\times\prod_{j=1}^{r} R_0(\kappa_j),\quad R_0(\kappa_j)=\frac{\kappa^{(d_j-1)/2}\mathcal{I}_{(d_j-1)/2}(2\kappa_j)}{2^{d_j} \pi^{(d_j+1)/2}\mathcal{I}_{(d_j-1)/2}(\kappa_j)^2},
\end{align*}
being $\bv(\bkappa)= \bd\odot\bkappa\odot\lrc{2(2+\bd)\odot\bkappa-(\bd^{\odot 2}-\bd+2)\odot\mathbfcal{I}}$ and $\bu(\bkappa)=\bd\odot\bkappa\odot\mathbfcal{I}$, with $\mathbfcal{I}=\big(\mathcal{I}_{(d_1+1)/2}(2\kappa_1)\big/\mathcal{I}_{(d_1-1)/2}(2\kappa_1),\ldots,\mathcal{I}_{(d_r+1)/2}(2\kappa_r)\big/\mathcal{I}_{(d_r-1)/2}(2\kappa_r)\big)'$, and $\bA^\circ$ standing for a matrix $\bA$ whose diagonal is set to zero. Therefore, $\bh_{\mathrm{AMISE}}$ is the solution to the system
\begin{align}
 4 \bR(\bkappa)({\bh}_{\mathrm{AMISE}}^{\odot 2}\odot\bb_{\bd}(L))\odot {\bh}_{\mathrm{AMISE}}\odot\bb_{\bd}(L)-\frac{v_{\bd}(L)}{n\rho(\bh_{\mathrm{AMISE}})}\big(\bd\odot{\bh}_{\mathrm{AMISE}}^{\odot(-1)}\big)=\zero_r.\label{eq:amisevmf2}
\end{align}

For the special case of equal bandwidth $h_n$,
\begin{align}
        h_\mathrm{AMISE}=\lrc{\frac{\tilde{d}\, v_{\bd}(L)}{4\bb_{\bd}(L)'\bR(\bkappa)\bb_{\bd}(L)n}}^{1/(\tilde{d}+4)}. \label{eq:hamisevmf}
\end{align}
If, in addition, the common dimension is $d$, then $R\lrpbig{\nabla^2 \bar{f}_\mathrm{PvMF}(\cdot; \boldsymbol{\mu},\bkappa)}=\one_r'\bR(\bkappa)\one_r$.
\end{proposition}

If $\hat{\bkappa}$ is a suitable estimator for $\bkappa$,
the Rule-Of-Thumb (ROT) bandwidth selector $\hat{\bh}_{\mathrm{ROT}}$ for the kde \eqref{eq:estimator} with a general $L$ kernel is the solution to the system
\begin{align}
   4 \bR(\hat\bkappa)(\hat{\bh}_{\mathrm{ROT}}^{\odot 2}\odot\bb_{\bd}(L))\odot {\hat\bh}_{\mathrm{ROT}}\odot\bb_{\bd}(L)-\frac{v_{\bd}(L)}{n\rho(\hat{\bh}_{\mathrm{ROT}})}\big(\bd\odot\hat{\bh}_{\mathrm{ROT}}^{\odot(-1)}\big)=\zero_r.\label{eq:rotgeneral}
\end{align}

\begin{remark}
If $r=1$, the bandwidth selector $\hat{h}_\mathrm{ROT}$ reduces to that proposed by \citet[Proposition 2]{Garcia-Portugues2013a}, taking into account that $R\lrpbig{\nabla^2 \bar{f}_\mathrm{PvMF}(\cdot ; \boldsymbol{\mu},\hat{\kappa})}=\allowbreak d^2 R(\Psi(f_{\mathrm{vMF}}(\cdot ;\bmu, \hat\kappa), \cdot))$ (see the proof of Proposition \ref{prop:plugin} of the present paper).
\end{remark}

The next algorithm defines the steps involved in the computation of the ROT bandwidth selector.

\begin{algo}[Rule-of-thumb bandwidth selector] \label{algo:pi}
Given a sample $\bX_1,\ldots,\bX_n$ on $\Sr$:
\begin{enumerate}
    \item Compute the maximum likelihood estimator $\hat{\boldsymbol{\kappa}}$, where $\hat{\kappa}_j=A_{d}^{-1}(\|\frac{1}{n}\sum_{i=1}^n\bX_{ij}\|)$ with $A_d(r)\defin\Ical_{(d+1)/2}(r)/\Ical_{(d-1)/2}(r)$.
    \item Determine the marginal ROT bandwidths for each $\mathbb{S}^{d_j}$,  $j=1,\ldots, r$:
    \begin{align*}
        \hat{h}_{j,\mathrm{ROT}}\defin\left[\frac{4 \pi^{1/2} \mathcal{I}_{(d_j-1)/2}(\hat{\kappa}_j)^2}{\hat{\kappa}_j^{(d_j+1)/2}\big[2 d_j \Ical_{(d_j+1)/2}(2 \hat{\kappa}_j)+(2+d_j) \hat{\kappa}_j \Ical_{(d_j+3)/2}(2 \hat{\kappa}_j)\big] n}\right]^{1/(4+d_j)}.
    \end{align*}
    \item Solve the nonlinear equation \eqref{eq:rotgeneral} using $(\hat{h}_{1,\mathrm{ROT}},\ldots,\hat{h}_{r,\mathrm{ROT}})'$ as initial values.
\end{enumerate}
\end{algo}

The rule-of-thumb selector assumes that the underlying density curvature can be approximated by that of a PvMF. The selector is therefore expected to perform as well as this assumption is practically met. In particular, when the underlying distribution is centro-symmetric, the selector in Algorithm \ref{algo:pi} is expected to perform worse than cross-validation approaches.

%-------------------------------%
\subsection{Cross-validation bandwidths}
\label{sec:bw_cv}
%-------------------------------%

The selection of the bandwidth can also be achieved with cross-validation techniques. Specifically, one such method is the Least Squares Cross-Validation (LSCV) technique. %
The procedure aims to select the bandwidth by minimizing
\begin{align}\label{eq:LSCV}
    \mathrm{LSCV}(\bh)\defin\int_{\Sr}\hat{f}(\bx;\bh)^2 \,\sigmar (\rd\bx)-2n^{-1}\sum_{i=1}^n\hat{f}_{-i}(\bX_i;\bh),
\end{align}
that is, the bandwidth-dependent terms of an unbiased estimate of $\mathrm{ISE}[\hat{f}(\cdot;\bh_n)]$. In \eqref{eq:LSCV}, $\hat{f}_{-i}(\bx;\bh)\defin\allowbreak \frac{1}{n-1}\sum_{j=1,\,i\neq j}^nL_{\bh}(\bx,\bX_j)$ represents the leave-one-out kde.

The following result provides a closed form of \eqref{eq:LSCV} when using a vMF kernel.

\begin{proposition}[Explicit LSCV loss for vMF kernel]\label{prop:LSCV}
If the considered kernel in \eqref{eq:estimator} is $L_\mathrm{vMF}^P$, then
\begin{align*}
\mathrm{LSCV}(\bh)=&\;\frac{c_{\bd}^{\mathrm{vMF}}\big(\bh^{\odot(-2)}\big)^2}{n c_{\bd}^{\mathrm{vMF}}\big(2 \bh^{\odot(-2)}\big)}-\frac{2 c_{\bd}^{\mathrm{vMF}}\big(\bh^{\odot(-2)}\big)}{n}\\
                &\times \sum_{i=1}^n \sum_{j>i}^n\lrc{\frac{2}{n-1} \exp\bigg\{\sum_{\ell=1}^r \bX_{i\ell}'\bX_{j\ell}/h_\ell^2\bigg\}-\frac{c_{\bd}^{\mathrm{vMF}}\big(\bh^{\odot(-2)}\big)}{n c_{\bd}^{\mathrm{vMF}}\big(\|\bX_i+\bX_j\|  \bh^{\odot(-2)}\big)}}.
\end{align*}
\end{proposition}

An alternative cross-validation approach is Likelihood Cross-Validation (LCV). In this case, the pseudo log-likelihood function
\begin{align*}
    \mathrm{LCV}(\bh)\defin\sum_{i=1}^n\log\hat{f}_{-i}(\bX_i;\bh)
\end{align*}
is maximized. Care is needed when computing the LCV bandwidth for compactly-supported kernels, as the LCV loss is not smooth and can be minus infinite for small bandwidths. The next result is useful when performing numerical optimization with the $\mathrm{Epa}^P$ kernel.

\begin{proposition}[Critical LCV bandwidth for Epa kernel] \label{prop:hmin}
Consider the kernel $L_\mathrm{Epa}^P$ and a common bandwidth $\bh=h\one_r$. Then, $\mathrm{LCV}(\bh) > -\infty$ if and only if $h > h_{\min,\mathrm{Epa}}$, where
\begin{align*}
    h_{\min,\mathrm{Epa}}^2\defin \max_{i=1,\ldots,n} \min_{j\neq i} \max_{k=1,\ldots,r}\{1-\bX_{ik}'\bX_{jk}\}.
\end{align*}
In the non-common bandwidth case $\bh$, $\min_{k=1,\ldots,r}h_k > h_{\min,\mathrm{Epa}}$ implies $\mathrm{LCV}(\bh) > -\infty$.
\end{proposition}

%-------------------------------%
\section{Kernel-based homogeneity testing}
\label{sec:hom}
%-------------------------------%

The kde \eqref{eq:estimator} can be used to construct a nonparametric test of the homogeneity of two classes, which will be applied later in Section \ref{sec:classes}. Let $f_1$ and $f_2$ denote two pdfs on $\Sr$. The two sample problem is formalized as
\begin{align}
    \mathcal{H}_0:f_1=f_2 \quad\text{vs.} \quad \mathcal{H}_1:f_1\neq f_2.\label{eq:twosamp}
\end{align}
Different test statistics can be constructed for this testing problem considering distinct distances between $f_1$ and $f_2$. The one with the longest tradition is the $L_p$-distance
\begin{align}
    d_p(f_1,f_2)^p\defin \int_{\mathbb{S}^{\bd}}|f_1(\bx)-f_2(\bx)|^p\,\sigmar(\mathrm{d}\bx)\label{eq:dp}
\end{align}
with $p=2$; see \citet[Chapter 3]{Mammen1992}, \cite{Anderson1994a}, and \cite{Li1996} for tests based on $d_2(f_1,f_2)^2$ on $\R^d$ and, e.g., \cite{Boente2014} for the use of $p=1,2$ in the context of goodness-of-fit testing on $\Sd$. However, \eqref{eq:dp} is remarkably unpractical in high-dimensional settings due to the vanishing of the densities $f_1$ and $f_2$, which is amplified by the power $p$. On the high-dimensional polysphere $(\mathbb{S}^2)^{168}$, the setting in the real data application of Section \ref{sec:classes}, this vanishing can easily yield numerically null distances when working with machine double precision. The Hellinger distance $d_\mathrm{H}(f_1,f_2)^2\defin 1-\int_{\mathbb{S}^{\bd}}\sqrt{f_1(\bx) f_2(\bx)}\,\sigmar(\mathrm{d}\bx)$ slightly alleviates the issue of vanishing densities in \eqref{eq:dp}, but it is still not enough for high-dimensional setups, where working on the logarithmic scale of the density is imperative. Furthermore, these two distances involve computing numerically an integral on $\Sr$, which is inexact and expensive, even if adapting exact formulae for \eqref{eq:dp} when $p=2$ and $f_1$ and $f_2$ are mixtures of vMFs \citep[Proposition~4]{Garcia-Portugues2013b}.

We utilize the Jensen--Shannon Divergence (JSD) introduced by \cite{Lin1991} to measure the closeness between the density functions $f_1$ and $f_2$ on $\Sr$:
\begin{align*}
    \mathrm{JSD}(f_1,f_2)\defin&\;\pi_1\mathrm{D}_\mathrm{KL}(f_1\| f_0)+\pi_2\mathrm{D}_\mathrm{KL}(f_2\| f_0),\quad \mathrm{D}_\mathrm{KL}(f_j\| f_0)\defin\int_{\Sr}\log\left(\frac{f_j(\bx)}{f_0(\bx)}\right)f_j(\bx)\,\sigmar(\mathrm{d}\bx)
\end{align*}
where $f_0(\bx)\defin\pi_1 f_1(\bx)+\pi_2 f_2(\bx)$ is the pooled density for both classes, $\pi_1$ and $\pi_2$ denote the prior probability of each class, and $\mathrm{D}_\mathrm{KL}(f_j\| f_0)$ is the Kullback--Leibler divergence of $f_j$ from $f_0$. The JSD is a symmetrized version of the Kullback--Leibler divergence that satisfies $0\leq \mathrm{JSD}(f_1,f_2)\leq \log(2)$, with $\mathrm{JSD}(f_1,f_2)=0$ if and only if $f_1=f_2$ almost everywhere \citep{Lin1991}. Its variant $\sqrt{\mathrm{JSD}(f_1,f_2)}$ also satisfies the triangular inequality and is thus a proper distance \citep{Endres2003}. However, this transformation is unnecessary when using $\mathrm{JSD}(f_1,f_2)$ as a test statistic.

\cite{Lin1991} also introduced the generalized JSD to compare $k\geq2$ distributions:
\begin{align}
    \mathrm{JSD}(f_1,\ldots,f_k)\defin&\;\sum_{j=1}^k \pi_j\mathrm{D}_\mathrm{KL}(f_j\| f_0)=\mathrm{H}(f_0)-\sum_{j=1}^k \pi_j \mathrm{H}(f_j),\label{eq:jsd}\\
    \mathrm{H}(f_j)\defin& -\int_{\Sr} \log(f_j(\bx))f_j(\bx)\,\sigmar(\mathrm{d}\bx)=-\Es{\log(f_j(\bX_j))}{f_j},\nonumber
\end{align}
with $f_0(\bx)\defin\sum_{j=1}^r \pi_j f_j(\bx)$ and where $\mathrm{H}(f_j)$ is the Shannon entropy of $\bX_j\sim f_j$. The generalized JSD satisfies $0\leq \mathrm{JSD}(f_1,\ldots,f_k)\leq \log(k)$, with $\mathrm{JSD}(f_1,\ldots,f_k)=0$ if and only if $f_1=\cdots=f_k$ almost everywhere. This generalized JSD also allows tackling the generalized version of \eqref{eq:twosamp}:
\begin{align}
    \mathcal{H}_0:f_1=\cdots=f_k \quad\text{vs.} \quad \mathcal{H}_1: \exists\, i,j\in\{1,\ldots,k\}:f_i\neq f_j.\label{eq:ksamp}
\end{align}

Let $\{\bX_{i,1}\}_{i=1}^{n_1},\ldots,\{\bX_{i,k}\}_{i=1}^{n_k}$ be $k$ independent samples, each of them respectively iid with the pdfs $f_1,\ldots,f_k$. The sample sizes are such $n_1+\cdots+n_k=n$. We define a test statistic for \eqref{eq:ksamp} as the following kernel-based estimator of \eqref{eq:jsd}:
\begin{align}
    T_{n,\mathrm{JSD}}(\bh)\defin&\; \hat{\mathrm{H}}(\hat{f}_0(\cdot;\bh))-\sum_{j=1}^k \hat{\pi}_j \hat{\mathrm{H}}(\hat{f}_j(\cdot;\bh)), \label{eq:TnJSD}
\end{align}
where $\hat{\pi}_j\defin n_j/n$, $j=1,\ldots,k$, are the relative frequencies of each class, and the Shannon entropies are estimated with
\begin{align}
\begin{aligned}
    \hat{\mathrm{H}}(\hat{f}_j(\cdot;\bh))\defin -\frac{1}{n_j}\sum_{i=1}^{n_j} \log(\hat{f}_j^{-i}(\bX_{i,j};\bh))\quad\text{and}\\
    \hat{\mathrm{H}}(\hat{f}_0(\cdot;\bh))\defin -\frac{1}{n}\sum_{j=1}^k\sum_{i=1}^{n_j} \log(\hat{f}_0^{-(i,j)}(\bX_{i,j};\bh))
        \end{aligned}
        \label{eq:Hs}
\end{align}
for the leave-one-out cross-validated kdes

\begin{align}
\begin{aligned}
    \hat{f}_j^{-i}(\bX_{i,j};\bh)&\defin\frac{1}{n_j-1}\sum_{\substack{\ell=1\\\ell\neq i}}^{n_j} L_\bh(\bX_{i,j},\bX_{\ell,j})\quad\text{and}\\
    \hat{f}_0^{-(i,j)}(\bX_{i,j};\bh)&\defin
    \frac{1}{n-1}\Bigg\{\sum_{\substack{m=1\\m\neq j}}^k \pi_m n_m \hat{f}_m(\bX_{i,j};\bh)+\pi_j(n_j-1)\hat{f}_j^{-i}(\bX_{i,j};\bh)\Bigg\}
    \end{aligned}
    \label{eq:cvkdes}
\end{align}

Several comments on the construction of \eqref{eq:TnJSD} are in order. First, the expectations in the Shannon entropies are estimated by their empirical versions. This approach is much more stable than carrying out a Monte Carlo approximation of $\mathrm{H}(f_j)$ %
with a simulated sample from %
$\hat{f}_j(\cdot;\bh)$, especially since in high dimensions the kde is sparse and can spike close to the sample observations, hence bringing variability to the final estimate. Second, the terms $L_\bh(\bX_{i,j},\bX_{i,j})$ are excluded to remove unnecessary bias and avoid numerical overflows. Third, the same bandwidths are used for the $k$ kdes $\hat{f}_j(\cdot;\bh)$, $j=1\ldots,k$, to balance their contributions in the test statistic and to avoid different smoothing biases, this being a common practice  \cite[see, e.g.,][]{Mammen1992,Li1996}.

Our test for \eqref{eq:ksamp} based on the test statistic $T_{n,\mathrm{JSD}}(\bh)$ rejects $\mathcal{H}_0$ for large values of this statistic. The $p$-value of the test can be easily approximated with permutations, by shuffling the class labels under $\mathcal{H}_0$. For each $b=1,\ldots,B$, a permutation resample of $T_{n,\mathrm{JSD}}(\bh)$ is
\begin{align*}
    T_{n,\mathrm{JSD}}^{*b}(\bh)\defin \hat{\mathrm{H}}^{*b}(\hat{f}_0^{*b}(\cdot;\bh))-\sum_{j=1}^k \hat{\pi}_j \hat{\mathrm{H}}^{*b}(\hat{f}^{*b}_j(\cdot;\bh)),
\end{align*}
where $\hat{\mathrm{H}}^{*b}(\hat{f}_0^{*b}(\cdot;\bh))$ and $\hat{\mathrm{H}}^{*b}(\hat{f}^{*b}_j(\cdot;\bh))$ are defined as in \eqref{eq:Hs}--\eqref{eq:cvkdes}, but with the sample $\{\bX_{i,j}\}_{j,i=1}^{k,n_j}$ replaced with the permuted sample $\smash{\{\bX_{i,j}^{*b}\}_{j,i=1}^{k,n_j}}$ that is randomly extracted without replacement from $\smash{\{\bX_{i,j}\}_{j,i=1}^{k,n_j}}$. The $p$-value of the permutation test is approximated as $ B^{-1}\sum_{b=1}^B 1_{\{T_{n,\mathrm{JSD}}^{*b}(\bh)>T_{n,\mathrm{JSD}}(\bh)\}}$.

The proposed test is nonparametric, and thus its main advantage is the absence of strong distributional assumptions. As a consequence of its construction via the JSD, it is expected to be consistent against any kind of alternative to $\mathcal{H}_0$. With the aim of critically assessing the performance of this test in Section B.3 of the SM, we next construct two possible competing parametric tests. These tests evaluate the equality of population location and scatter measures for the two-sample problem. For the random vectors  $\bX_{j}=(\bX_{1,j}',\ldots, \bX_{r,j}')'$, $j=1,2$, on the sphere $\Sr$, define the location vectors $\boldsymbol{\mu}_j\defin\mathbb{E}[\bX_j]/\|\mathbb{E}[\bX_j]\|$ and scatter matrices $\boldsymbol{\Sigma}_j\defin\mathbb{E}[\bX_j\bX_j']$, and the associated testing problems $\mathcal{H}_0:\bmu_1=\bmu_2$ vs. $\mathcal{H}_1:\bmu_1\neq\bmu_2$ and $\mathcal{H}_0:\bSigma_1=\bSigma_2$ vs. $\mathcal{H}_1:\bSigma_1\neq\bSigma_2$. The empirical counterparts are $\smash{\hat{\boldsymbol{\mu}}_j\defin\bar{\bX}_j/\|\bar{\bX}_j\|}$ and $\smash{\hat{\boldsymbol{\Sigma}}_j\defin n_j^{-1}\sum_{i=1}^{n_j}\bX_{i,j}\bX_{i,j}'}$. %
For each spherical component $\mathbb{S}^{d_\ell}$ in $\Sd$, $\ell=1,\ldots,r$, consider the marginal versions of the above estimators: $\hat{\boldsymbol{\mu}}_{\ell,j}$ and $\hat{\boldsymbol{\Sigma}}_{\ell,j}$. We define the location and scatter test statistics as the maximum discrepancy along the $r$ spheres between $\hat{\boldsymbol{\mu}}_{\ell,1}$ and $\hat{\boldsymbol{\mu}}_{\ell,2}$, and $\hat{\boldsymbol{\Sigma}}_{\ell,1}$ and $\hat{\boldsymbol{\Sigma}}_{\ell,2}$, respectively:
\begin{align*}
T_{n,\mathrm{loc}}\defin\max_{\ell=1,\ldots,r} \|\hat{\boldsymbol{\mu}}_{\ell,1}-\hat{\boldsymbol{\mu}}_{\ell,2}\|_2,\quad T_{n,\mathrm{sc}}\defin \max_{\ell=1,\ldots,r} d_{\mathcal{P}(d_\ell+1)}(\hat{\boldsymbol{\Sigma}}_{\ell,1},\hat{\boldsymbol{\Sigma}}_{\ell,2}),
\end{align*}
where $d_{\mathcal{P}(d_\ell+1)}(\bSigma_{\ell,1},\bSigma_{\ell,2})=\|\log(\blambda_\ell)\|_2$, with $\blambda_\ell=(\lambda_{1,\ell},\ldots,\lambda_{(d+1),\ell})'$ the eigenvalues of $\bSigma_{\ell,1}^{-1}\bSigma_{\ell,2}$, is the affine-invariant/Fisher--Rao metric in the space $\mathcal{P}(d_\ell+1)$ of positive definite matrices of size $d_\ell+1$ \citep{Pennec2006}. The tests based on $T_{n,\mathrm{loc}}$ and $T_{n,\mathrm{sc}}$ can be calibrated by permutations. %

Simulations showing the satisfactory performance of the $k$-sample test can be found in Section B.3 of the SM.

%-------------------------------%
\section{Real data application}
\label{sec:data}
%-------------------------------%

We analyze the hippocampi dataset briefly presented in Section \ref{sec:intro}. The data comprises shapes of hippocampi from 177 $6$-months-old infants, which have been parametrized as $s$-reps \citep{liu2021fitting}.  %
Figure \ref{fig:srepa} depicts the process of constructing the $s$-rep, where the hippocampus surface is parametrized as a collection of inner skeletal points (red) and a set of 168 spokes (green), which are vectors originating from the skeleton and extending to the boundary points (blue).

The analyzed dataset reveals variations in the shapes of the hippocampus, but the internal skeletal structures exhibit somewhat similar configurations. As a result, it is justifiable to use the average internal skeletal configuration as a common reference point, and analyze the vectors that emanate from this reference point and extend to the boundary. To focus specifically on the shape of the hippocampus, we standardize the vectors by normalizing their radii to match the sample means. This standardization removes size information and results in a simplified representation observed within the polysphere $(\mathbb{S}^2)^{168}$, capturing the hippocampus shape as an $s$-rep. Figure \ref{fig:srepb} shows the $\mathbb{S}^2$-directions of the $n=177$ subjects at four boundary points.

\begin{figure}[htpb!]
    \centering
     \begin{subfigure}[b]{0.5\textwidth}
        \includegraphics[width=\textwidth,trim={2cm 8.425cm 0cm 8.65cm},clip]{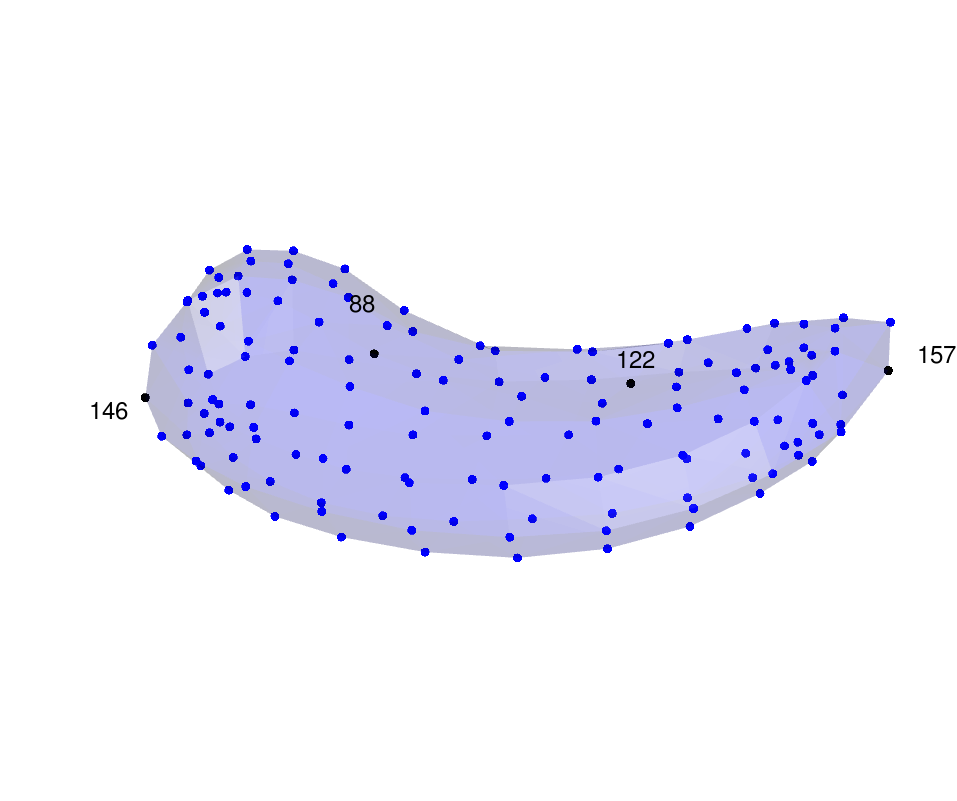}\\
        \includegraphics[width=\textwidth,trim={2cm 8.425cm 0cm 8.65cm},
        clip]{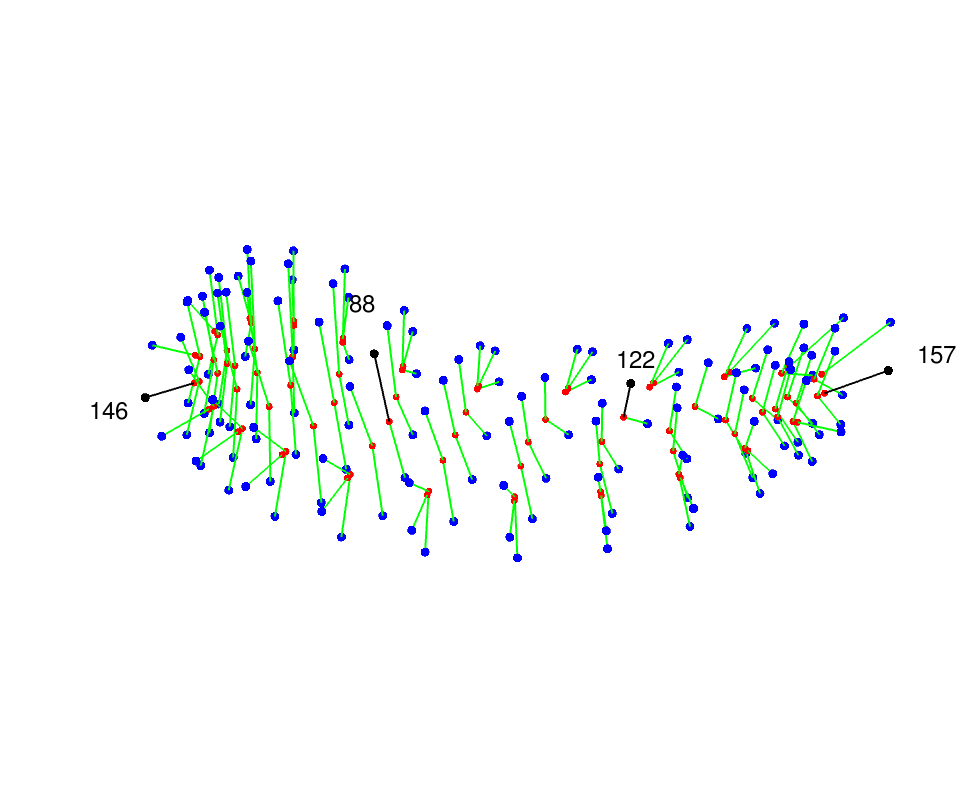}
        \caption{\small Construction process of an $s$-rep}
        \label{fig:srepa}
    \end{subfigure}%
    \hspace*{0.25cm}%
    \begin{subfigure}[b]{0.5\textwidth}
        \centering
        \includegraphics[width=0.125\textheight]{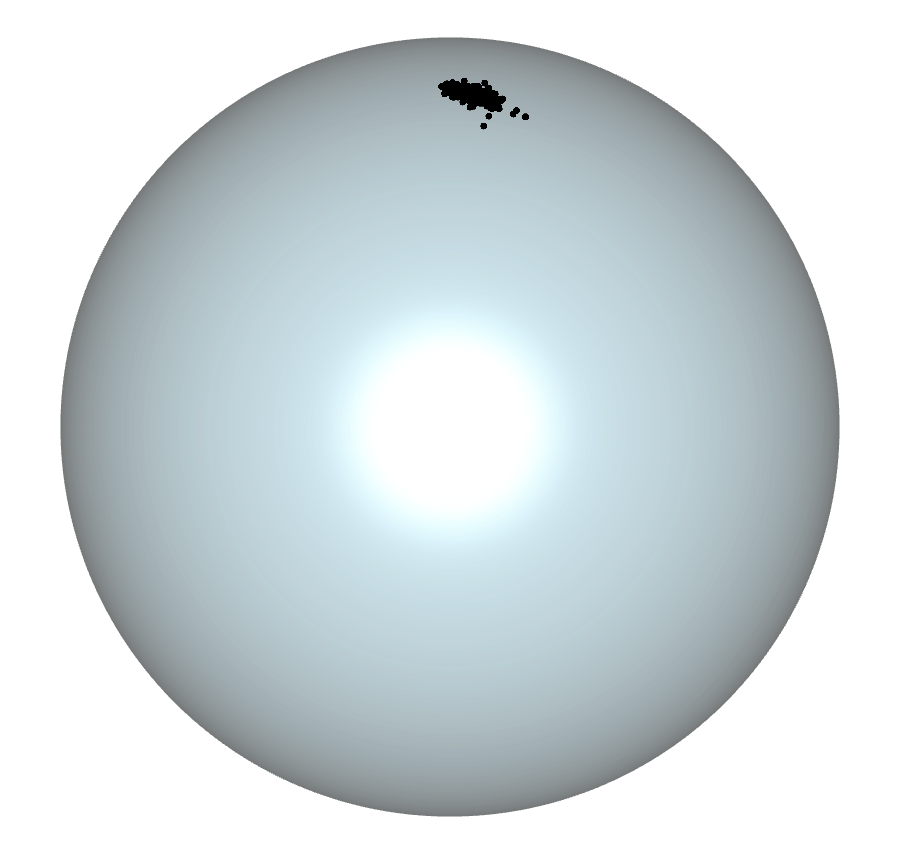}%
        \includegraphics[height=0.125\textheight]{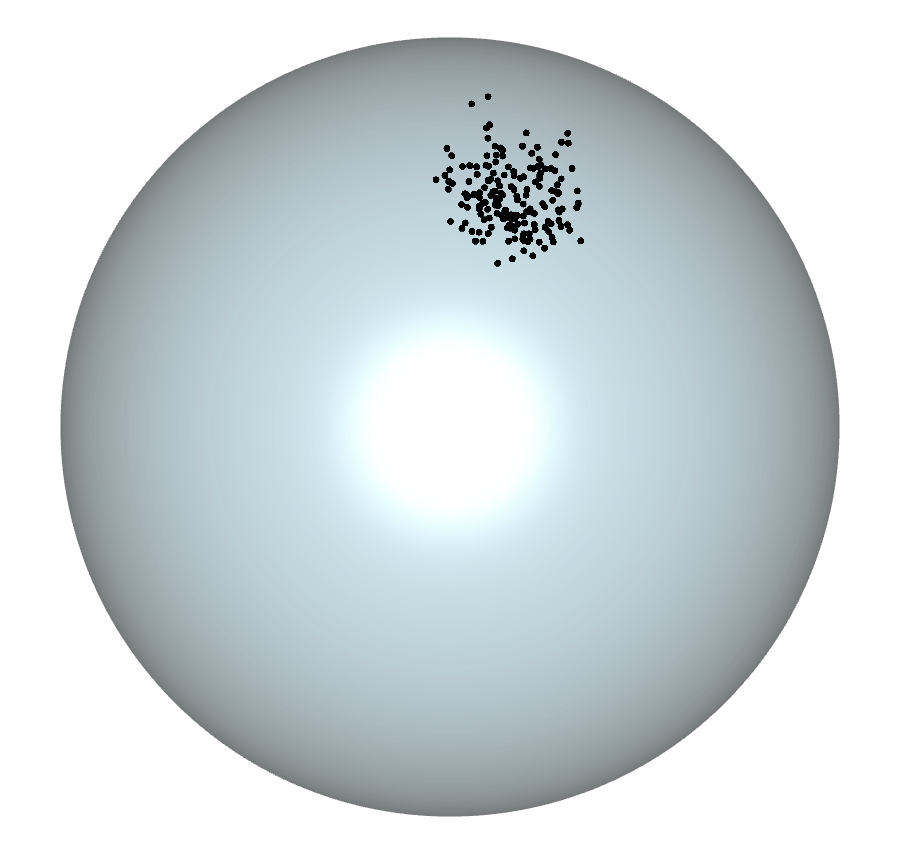}\\
        \includegraphics[height=0.125\textheight]{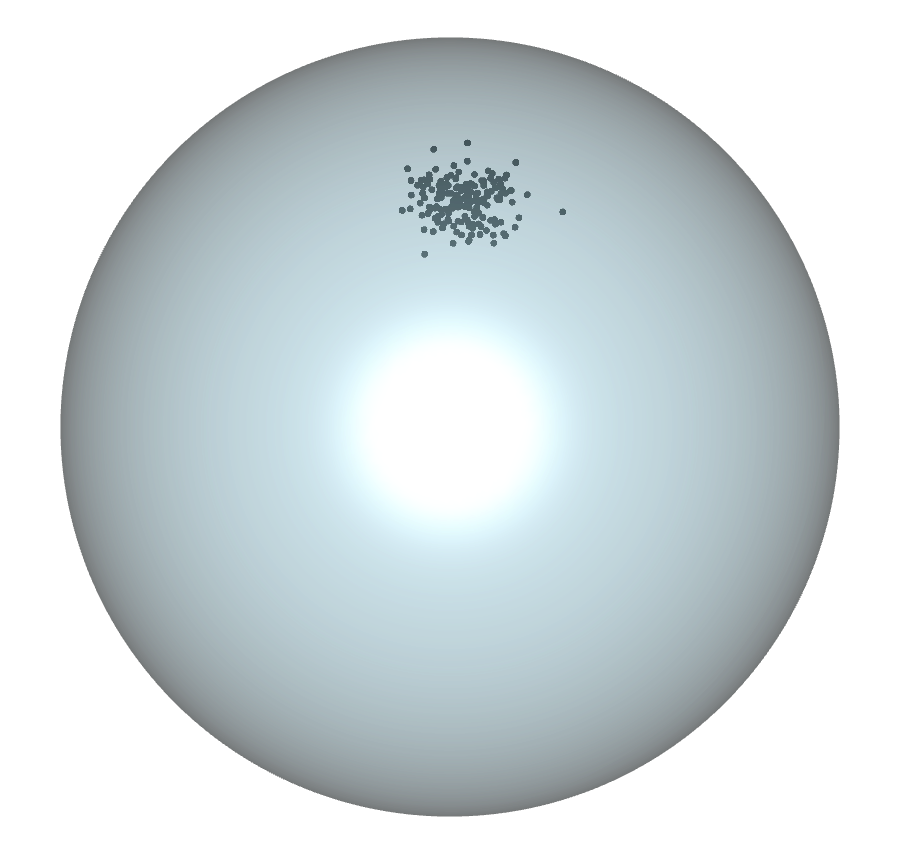}%
        \includegraphics[height=0.125\textheight]{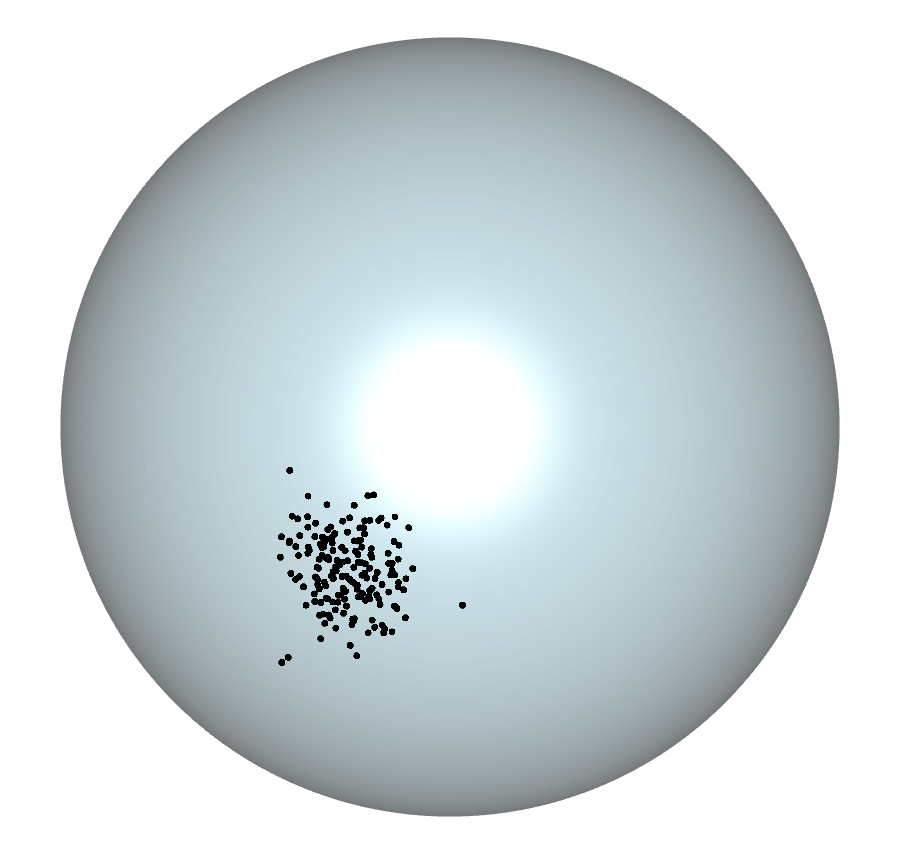}
    \caption{\small Directions on $\mathbb{S}^2$ of specific spokes}
    \label{fig:srepb}
    \end{subfigure}%
    \caption{\small Construction of an $s$-rep and scatterplots of directions of different spokes. Figure \ref{fig:srepa} sketches the $s$-rep fitting process \citep{liu2021fitting}, where a hippocampus surface (blue, top) is parametrized as a set of base skeletal points (red) and a collection of $168$ spokes (green segments) connecting with the boundary (blue points). Figure \ref{fig:srepb} shows the $\mathbb{S}^2$-directions of the $n=177$ spokes located at boundary points $88$, $122$, $146$, and $157$ (from left to right and top to bottom). The locations of these spokes are shown in Figure \ref{fig:srepa}.}
    \label{fig:srep}
\end{figure}

The space $(\mathbb{S}^2)^{168}$ is high-dimensional, at least according to smoothing standards. Consequently, there are many computational challenges involved in evaluating \eqref{eq:estimator} and \eqref{eq:TnJSD}. The first and most pressing issue is the density vanishing, resulting in numerical underflows. We countermeasured it  by (\textit{i}) evaluating the log-kde and (\textit{ii}) computing the density with respect to the uniform measure $1/\omega_{\bd}$ %
(this is only beneficial if $\omega_{\bd}>1$, which for a common dimension $d$ happens if $d\leq17$). Second, numerical stability through safe computations is also important. Applying (\textit{i}) involves a recurrent use of the ``LogSumExp trick'' to safely evaluate the logarithms of sums. Additionally, Bessel functions appearing in the vMF kernel normalizing constant can easily overflow for small $h$'s, so it is crucial to exponentially scale them, use log-forms, and exploit specific cases. For $d=2$, useful numerically-stable normalizing constants for small bandwidths are $\log(c_{2,L_{\mathrm{vMF}}}(h))=-[\log(2\pi)+2\log(h) + \operatorname{log1p}(-\exp\{-2/h^2\})]$ and expressions \eqref{eq:asymp-c-d2}. %
Finally, we addressed the slowness of repeatedly evaluating \eqref{eq:estimator} by (\textit{i}) implementing in C++ the kde and (\textit{ii}) designing computationally-tractable statistics such as $\hat{\bh}_\mathrm{ROT}$ and $T_{n,\mathrm{JSD}}(\bh)$.

%-------------------------------%
\subsection{Inward--outward ranking of hippocampi}
\label{sec:inout}
%-------------------------------%

The first application of the kde \eqref{eq:estimator} involves establishing a nonparametric inward--outward ranking of the hippocampi based on the density of their shapes. Denoting by $\mathbb{X}=\{\bX_1,\ldots,\bX_n\}\subset(\mathbb{S}^2)^{168}$ the $n=177$ hippocampus shapes, we establish the rank of each subject $\bX_i$ within the sample through the ranking function
\begin{align}
    \mathrm{rank}(\bX_i;\mathbb{X})\defin\mathrm{rank}(\ell_{(i)}(\bh);\ell_{1}(\bh),\ldots,\ell_{n}(\bh)),\quad \ell_i(\bh)\defin\log \hat{f}^{-i}(\bX_i;\bh), \label{eq:rank}
\end{align}
where $\mathrm{rank}(x_i;x_1,\ldots,x_n)$ %
is the standard ranking function of the univariate sample $\{x_1,\ldots,\allowbreak x_n\}$. %
In \eqref{eq:rank}, the logarithm does not affect the ranking, but it is fundamental for numerical stability. Similarly, removing the contribution of the $i$th point in $\hat{f}^{-i}(\bX_i;\bh)$ does not alter the ranking, but has the advantage of excluding the easily-overflowing addend $c_{\bd,L}(\bh)L(\zero)$ from within the logarithm. %

\begin{figure}[htpb!]
    \centering
    \includegraphics[width=\textwidth]{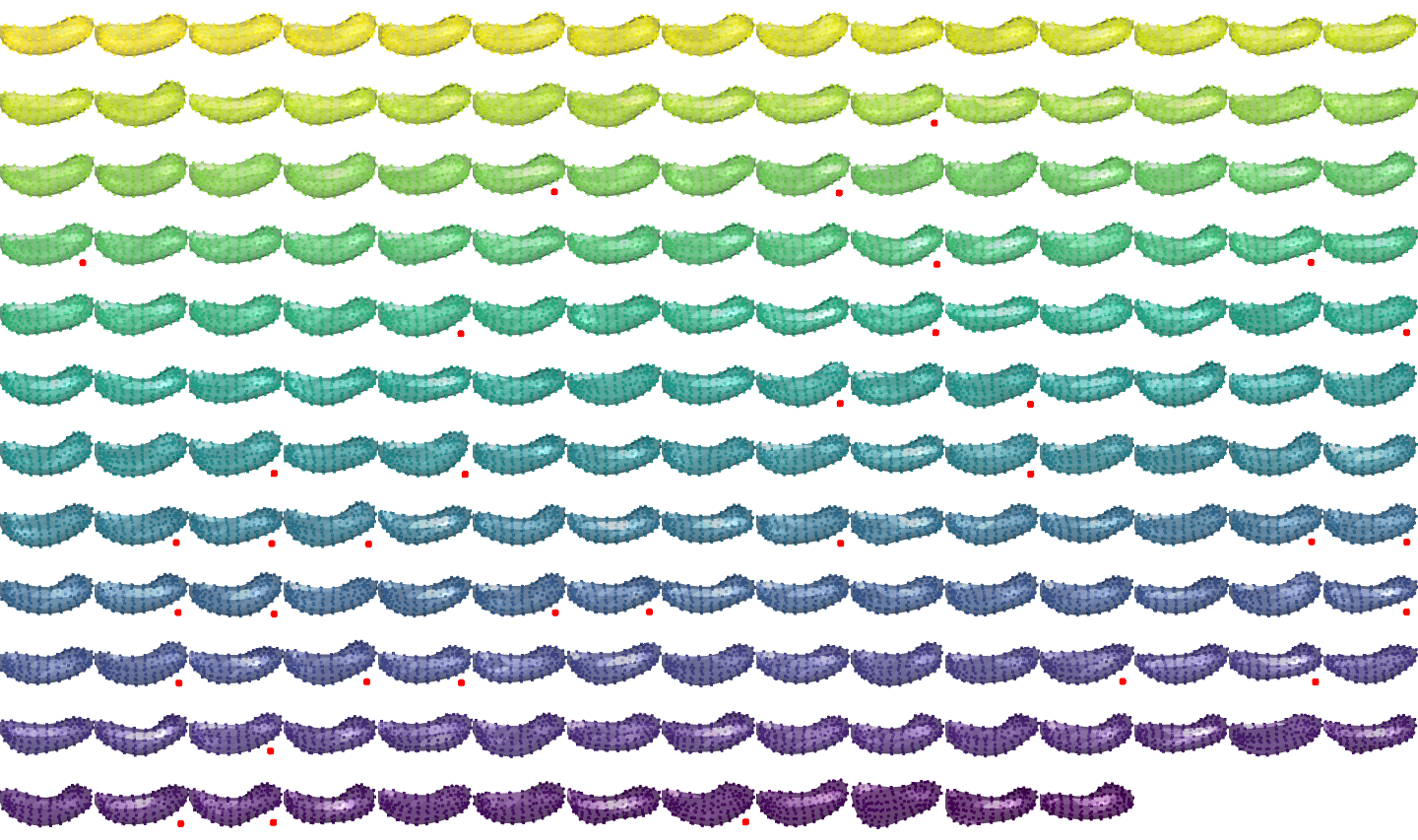}%
    \caption{\small Dataset of $n=177$ hippocampi ranked by the density of their shapes, as determined by \eqref{eq:rank} using $\hat{\bh}_{\mathrm{ROT}}$ and the sfp kernel with $\upsilon=100$. The yellow--violet color gradient indicates the inward--outward ranking. The red dots indicate whether the 6-month-old infant later developed autism (see Section \ref{sec:classes}).}
    \label{fig:subjects_sorted}
\end{figure}

The ranking \eqref{eq:rank} was implemented with the sfp kernel with $\upsilon=100$ due to its positivity and higher efficiency than the vMF kernel. The bandwidth vector was selected using $\hat{\bh}_{\mathrm{ROT}}$ as described in Algorithm \ref{algo:pi}. This choice is justified by the unimodality of the distributions of the spokes directions (see Figure~\ref{fig:srepb}).

The resulting ranking sorts the hippocampi shapes in terms of their density within the sample: the first subjects represent prototypical hippocampi in high-density regions, while the last ones represent outlying hippocampi from low-density regions. Figure \ref{fig:subjects_sorted} shows the subjects sorted according to their density rank, with a yellow--violet color gradient indicating the inward--outward ranking. The subjects shown in yellow/violet correspond to the most central/extreme ones, and can be regarded as the most prototypical/outlying, respectively.

%-------------------------------%
\subsection{Homogeneity of classes}
\label{sec:classes}
%-------------------------------%

The JSD two-sample test from Section \ref{sec:hom} represents the second application of the kde \eqref{eq:estimator}. This test has an attractive application in the hippocampi dataset: determining whether there exist significant shape differences, as measured by the polyspherical parameterization of $s$-reps, between the autistic and non-autistic classes. A preliminary analysis on the separability of these two groups was conducted with the ranking induced in Figure \ref{fig:subjects_sorted}, which revealed no distinguishable distribution pattern of the subjects in the autistic class. In particular, the autistic class does not have the most inward or outward subjects in terms of density. However, this analysis is not conclusive of the existence of significant differences between the distributions of both classes.

To test the homogeneity of the two classes, we used $B=5,\!000$ permutations and a range of bandwidths $c\times \hat{\bh}_\mathrm{ROT}$, $c=2^\ell$ with $\ell\in\{-3,-2.5,\ldots,5\}$. The selection of this bandwidth range is informed by the experience with \textbf{E2} in Section B.3 of the SM. Figure \ref{fig:hippotest} indicates a mild significant difference between the two classes, with $p$-values slightly below the $5\pct$ significance level for at least $c\geq 3$ and $p$-values on the $10\pct$ significance level for $c=2$. Observe that in \textbf{E2} the largest power gains occurred for $c\in\{2,4,8,16,32\}$. The mild significant difference found is consistent with that reported in \citet[Section 7.4]{liu2023analysis}, where, among others, a two-sample test based on DiProPerm \citep{Wei2016} and PNS scores was applied ($p\text{-value}=0.028$). In summary, their test works on the PNS scores of the hippocampus preshapes. %
Therefore, as ours, their analysis considers a framework where the size information of the hippocampi is removed.

\begin{figure}[htpb!]
    \centering
        \includegraphics[width=0.8\textwidth,clip,trim={0cm 0.5cm 0cm 1.5cm}]{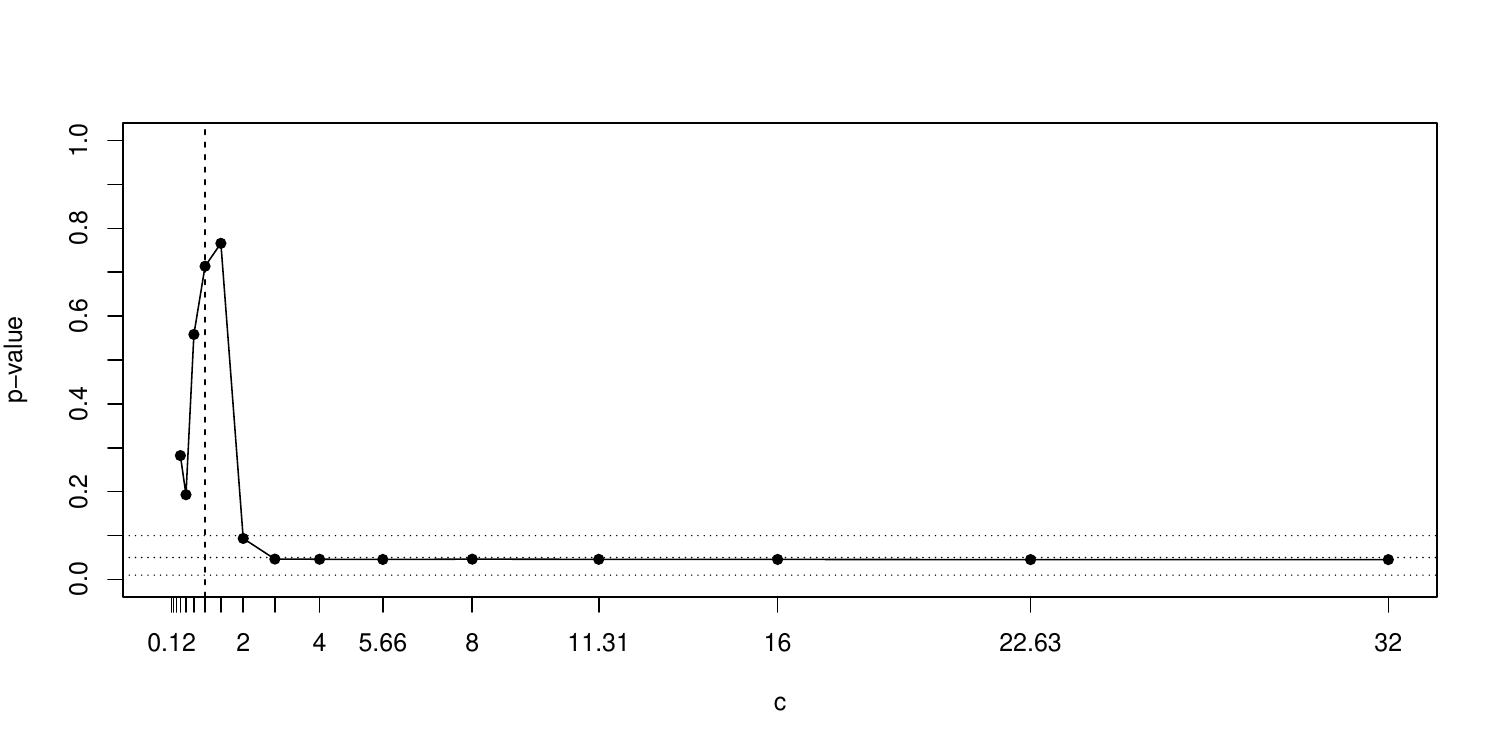}%
    \caption{\small $p$-values of the test based on $T_{n,\mathrm{JSD}}(c\times\hat{\bh}_\mathrm{ROT})$ as a function of the factor $c$. The sfp kernel with $\upsilon=100$ is used in the test.}
    \label{fig:hippotest}
\end{figure}

The location and scatter tests gave $p$-values $0.2524$ and $0.3372$, using as well $B=5,\!000$ permutations, thus failing to reject. The proposed JSD test is therefore adding a clear value to this particular application over parametric approaches. These tests also fail to reject when inspecting the differences between groups in each of the $168$ marginal $\mathbb{S}^2$-distributions and correcting the resulting $p$-values of each test to control the False Discovery Rate \citep[FDR;][]{Benjamini2001}. The FDR-corrected $p$-values, omitted, are exactly one. The rejection of the null hypothesis also does not happen for the JSD test run on each spoke, for the range of $c$'s considered for $(\mathbb{S}^2)^{168}$ and with FDR-corrected $p$-values across spokes (virtually all the FDR-corrected $p$-values are one). Hence, the significant differences between classes are subtle and appear in the joint variation of the spokes on $(\mathbb{S}^2)^{168}$, not in the marginal distributions on $\mathbb{S}^2$.

%-------------------------------%
\section{Discussion}
\label{sec:dis}
%-------------------------------%

%
This paper introduces a kernel density estimator for data on the polysphere $\Sr$ and, as a spin-off, a $k$-sample test that is tailored to be applicable in relatively high dimensions. The mean square error analysis of the estimator provides an inexpensive bandwidth selector that is particularly useful for estimating unimodal densities. The (non-product) kernel theory allows quantifying kernel efficiencies for the first time in directional supports. The estimator, bandwidth selector, and $k$-sample test are shown to be practically relevant in analyzing $s$-reps of hippocampi: the density-based ranking serves to identify prototypical and outlying subjects, and the homogeneity test reveals mild significant differences between classes that cannot be captured by exploring the marginal variation of spokes.

This paper opens some possible future research avenues, which are outlined. First, there is room for improvement in bandwidth selection, either by using plug-in selectors based on mixtures (e.g., mixtures of vMFs) or bootstrap selectors, or by developing asymptotic theory for cross-validatory selectors. Second, a natural extension of the kernel efficiency analysis will be the more general framework of different bandwidths and spherical dimensions. A final research endeavor is the derivation of the asymptotic distribution of the proposed Jensen--Shannon divergence test statistic and the exploration of its power under local alternatives.

%-----------------------------------------------%
\section*{Supplementary materials}
%-----------------------------------------------%

%-----------------------------------------------%
\section*{Acknowledgments}
%-----------------------------------------------%

The authors are supported by grant PID2021-124051NB-I00, funded by MCIN/\-AEI/\-10.13039/\-501100011033 and by ``ERDF A way of making Europe''. The first author acknowledges support from ``Convocatoria de la Universidad Carlos III de Madrid de Ayudas para la recualificación del sistema universitario español para 2021--2023'', funded by Spain's Ministerio de Ciencia, Innovación y Universidades. The authors greatly acknowledge Prof. Stephen M. Pizer (University of North Carolina at Chapel Hill) and Dr. Zhiyuan Liu (Meta) for kindly providing the analyzed $s$-reps hippocampi data, as well as for insightful discussions with the latter during the ``Object Oriented Data Analysis in Health Sciences'' workshop held on the Institute for Mathematical and Statistical Innovation (IMSI). The source for the raw hippocampi data is the Infant Brain Imaging Study (IBIS) network. Part of the simulations were conducted at the Centro de Supercomputación de Galicia (CESGA). The comments by two anonymous referees are greatly acknowledged.

\fi

\ifsupplement

\newpage
%-----------------------------------------------%
\title{Supplementary materials for ``Kernel density estimation with polyspherical data and its applications''}
\setlength{\droptitle}{-1cm}
\predate{}%
\postdate{}%
\date{}
%-----------------------------------------------%

%-----------------------------------------------%
\author{Eduardo Garc\'ia-Portugu\'es$^{1,2}$ and Andrea Meil\'an-Vila$^{1}$}
\footnotetext[1]{Department of Statistics, Carlos III University of Madrid (Spain).}
\footnotetext[2]{Corresponding author. e-mail: \href{mailto:edgarcia@est-econ.uc3m.es}{edgarcia@est-econ.uc3m.es}.}
\maketitle
%-----------------------------------------------%

%-----------------------------------------------%
\begin{abstract}
The supplementary material consists of three parts. Section \ref{sec:proof} contains the proofs of the presented results. Section \ref{sec:num} gives additional numerical results not included in the main text: the validity of the asymptotic normality, the kernel efficiency, and the relevance and validity of the $k$-sample test. Section \ref{sec:extremehippo} better shows the most central and extreme hippocampi.
\end{abstract}
\begin{flushleft}
	\small\textbf{Keywords:} Directional data; Nonparametric statistics; Skeletal representation; Smoothing.
\end{flushleft}
%-----------------------------------------------%

\appendix

%-------------------------------%
\section{Proofs}
\label{sec:proof}
%-------------------------------%

\begin{proof}[Proof of Theorem \ref{thm:biasvar}]
We drop the dependence of $\bh_n$ on $n$ in this proof to alleviate the notation. Along the proof, we use $\bar{f}$ to highlight the requirement of the extension of $f$ when taking derivatives. The proof is split into two parts, for the bias and variance, and follows the steps of the proofs of Lemma 5 and Proposition 1 in \cite{Garcia-Portugues2013b}.

\textit{Bias}. We first address the bias of $\hat{f}(\bx;\bh)$:
\begin{align}
	\mathbb{E}[\hat{f}(\bx;\bh)]-f(\bx)
    =&\;
	\mathbb{E}\lrc{L_{\bh}(\bx,\bX)}-f(\bx)\nonumber\\
	=&\;\int_{\Sr}(f(\by)-f(\bx))c_{\bd,L}(\bh)L\lrp{\frac{1-\bx_1' \by_1}{h_1^2},\ldots, \frac{1-\bx_r' \by_r}{h_r^2}}\,\sigmar(\rd\by).\label{eq:bias0}
\end{align}

Let $\bB_{\bx_j}$ be a $(d_j+1)\times d_j$ semi-orthonormal matrix such that $\bB_{\bx_j}\bB_{\bx_j}'=\bI_{d_j+1}-\bx_j\bx_j'$ and $\bB_{\bx_j}'\bB_{\bx_j}=\bI_{d_j}$ for $j=1,\ldots,r$. Lemma 2 in \cite{Garcia-Portugues2013b} readily entails that
\begin{align*}
    \int_{\Sr}f(\by)\,\sigmar(\rd\by)=&\;\int_{[-1,1]^r}\int_{\mathbb{S}^{\bd-1}} f\lrp{t_1 \bx_1+\left(1-t_1^2\right)^{1/2} \bB_{\bx_1} \bxi_1,\ldots,t_r \bx_r+\left(1-t_r^2\right)^{1/2} \bB_{\bx_r} \bxi_r}\\
    &\times \prod_{j=1}^r \left(1-t_j^2\right)^{d_j/2-1}\,\sigmarm(\rd\bxi)\,\rd \bt,
\end{align*}
where $(\bB_1\bxi_1,\ldots,\bB_r\bxi_r)'\in\Sr$. Therefore,
\begin{align}
	\eqref{eq:bias0}=&\;\int_{[-1,1]^r}\int_{\mathbb{S}^{\bd-1}}\left(f\lrp{t_1 \bx_1+\left(1-t_1^2\right)^{1/2} \bB_{\bx_1} \bxi_1,\ldots,t_r \bx_r+\left(1-t_r^2\right)^{1/2} \bB_{\bx_r} \bxi_r}-f(\bx)\right)\nonumber\\
	&\times c_{\bd,L}(\bh)L\lrp{\frac{1-t_1}{h_1^2},\ldots, \frac{1-t_r}{h_r^2}}  \prod_{j=1}^r \left(1-t_j^2\right)^{d_j/2-1}\,\sigmarm(\rd\bxi)\,\rd \bt.\label{eq:bias00}
\end{align}
Applying the change of variables $s_j=\frac{1-t_j}{h_j^2}$, for $j=1,\ldots,r$, one gets that
\begin{align}
    \eqref{eq:bias00}=&\; c_{\bd,L}(\bh)\int_{I_\bh} L(\bs)\prod_{j=1}^r h_j^{d_j}s_j^{d_j/2-1} (2-h_j^2s_j)^{d_j/2-1}\nonumber\\
    & \times \int_{\mathbb{S}^{\bd-1}}\left(f\left(\bx+\ba_{\bx,\bxi}\right)-f(\bx)\right)\sigmarm(\rd\bxi)\,\rd \bs,\label{eq:bias1}
\end{align}
where $I_\bh=[0,2h^{-2}_1]\times\stackrel{r}{\cdots}\times[0,2h^{-2}_r]$, $\ba_{\bx,\bxi}=(\ba_{\bx_1,\bxi_1}',\ldots,\ba_{\bx_r,\bxi_r}')'$ with
$
\ba_{\bx_j,\bxi_j}=-s_jh_j^2\bx_j+h_j[s_j(2-h_j^2s_j)]^{1/2}\bB_{\bx_j}\bxi_j.
$
Note that  $\ba_{\bx_j,\bxi_j}\defin \ba_{\bx_j,\bxi_j,1}+\ba_{\bx_j,\bxi_j,2}$, $\ba_{\bx_j,\bxi_j,1}\defin -s_jh_j^2\bx_j$, $\ba_{\bx_j,\bxi_j,2}\defin h_j[s_j(2-h_j^2s_j)]^{1/2}\bB_{\bx_j}\bxi_j$. Under \ref{A1}, the second-order Taylor expansion of $f$ at $\bx$ is
\begin{align}
    f\left(\bx+\ba_{\bx,\bxi}\right)-f(\bx)=&\;\ba'_{\bx,\bxi}\bnabla \bar{f}(\bx)+\frac{1}{2}\ba'_{\bx,\bxi}\bHcal \bar{f}(\bx)\ba_{\bx,\bxi}+o(\|\ba_{\bx,\bxi}\|^2)\nonumber\\
    =&\;\ba'_{\bx,\bxi,2}\bnabla \bar{f}(\bx)+\ba'_{\bx,\bxi,1}\bHcal \bar{f}(\bx)\ba_{\bx,\bxi,2}\nonumber\\
    &+\frac{1}{2}\ba'_{\bx,\bxi,2}\bHcal \bar{f}(\bx)\ba_{\bx,\bxi,2}+o(\|\ba_{\bx,\bxi}\|^2),\label{eq:Taylor}
\end{align}
where the second equality follows by realizing that $\bnabla \bar{f}(\bx) \bx=0$ and $\bx'\bHcal \bar{f}(\bx) \bx=0$ (but $\bHcal \bar{f}(\bx) \bx\neq\zero$ in general). Decomposition \eqref{eq:Taylor} induces a splitting of \eqref{eq:bias1} into four terms.

The first and the second of these terms vanish immediately: $\int_{\Sjm}\bxi_j\,\sigmajm(\rd\bxi_j)=\zero$ \cite[see, e.g.,][Lemma 3]{Garcia-Portugues2013b}.
For the third term, split $\bHcal\bar{f}(\bx)$ in blocks as
\begin{align*}
    \bHcal\bar{f}(\bx)=\left(\begin{array}{@{}c|c|c@{}}
        \bHcal_{11}\bar{f}(\bx) & \cdots & \bHcal_{1r}\bar{f}(\bx) \\\hline
        \vdots & \ddots & \vdots \\\hline
        \bHcal'_{1r}\bar{f}(\bx) & \cdots & \bHcal_{rr}\bar{f}(\bx)
    \end{array}\right),
\end{align*}
where subscripts in Hessian matrices denote the entries associated with the derivatives with respect to $\bx_j$, $j=1,\ldots,r$. From the previous observation, it becomes evident that, when $i\neq j$,
\begin{align*}
    \int_{\Srm}&(\bB_{\bx_i}\bxi_i)'\bHcal_{ij} \bar{f}(\bx)\bB_{\bx_j}\bxi_j\,\sigmarm (\rd\bxi)\\
    =&\;\lrc{\int_{\Sim}\bxi_i'\,\sigmaim(\rd\bxi_i)} \frac{\omega_{\bd-1}}{\omega_{d_i-1,d_j-1}}\bB_{\bx_i}'\bHcal_{ij} \bar{f}(\bx)\bB_{\bx_j}\lrc{\int_{\Sjm}\bxi_j\,\sigmajm(\rd\bxi_j)}=0.
\end{align*}
Therefore, the off-diagonal blocks of $\bHcal\bar{f}(\bx)$ result in null integrals in \eqref{eq:bias1}. For the diagonal blocks, it can be obtained that
\begin{align*}
    \int_{\Srm}(\bB_{\bx_j}\bxi_j)'\bHcal_{jj} \bar{f}(\bx)\bB_{\bx_j}\bxi_j\,\sigmarm (\rd\bxi)
    =&\;\tr{\bB_{\bx_j}'\bHcal_{jj}\bar{f}(\bx)\bB_{\bx_j}\lrc{\int_{\Sjm}\bxi_j\bxi_j'\,\sigmajm(\rd\bxi_j)} \frac{\omega_{\bd-1}}{\omega_{d_j-1}}}\\
    =&\;\omega_{\bd-1}\tr{\bHcal_{jj}\bar{f}(\bx)} \, d_j^{-1}.
\end{align*}

In the last step of the above equation, it is used that $\bB_{\bx_j}$ is a semi-orthonormal matrix and Lemma 3 of \cite{Garcia-Portugues2013b}.
Taking into account the previous comments, the third term can be written as
\begin{align}\label{eq:bias_int22}
	\int_{\mathbb{S}^{\bd-1}}&\ba'_{\bx,\bxi,2}\bHcal \bar{f}(\bx)\ba_{\bx,\bxi,2}\,\sigmarm (\rd\bxi)\nonumber\\
    =&\int_{\mathbb{S}^{\bd-1}}\left(\sum_{j=1}^rh_j^2s_j(2-h_j^2s_j)\bxi'_{j}\bB_{\bx_j}'\bHcal_{jj}\bar{f}(\bx)\bB_{\bx_j}\bxi_j\right)\sigmarm (\rd\bxi)\nonumber\\
    &+\int_{\mathbb{S}^{\bd-1}}\left(\sum_{i\neq j}^rh_ih_js_i^{1/2}s_j^{1/2}(2-h_i^2s_i)^{1/2}(2-h_j^2s_j)^    {1/2}\bxi'_{i}\bB_{\bx_i}'\bHcal_{ij}\bar{f}(\bx)\bB_{\bx_j}\bxi_j\right)\nonumber\\
    &\times\,\sigmarm (\rd\bxi)\nonumber\\
	=&\;\omega_{\bd-1}\sum_{j=1}^rh_j^2s_j(2-h_j^2s_j)d_j^{-1}\tr{\bHcal_{jj}\bar{f}(\bx)}.
\end{align}

Regarding the fourth term of \eqref{eq:Taylor}, it can be easily obtained that
\begin{align}\label{eq:bias_int3}
	o(\|\ba_{\bx,\bxi}\|^2)=o\left(\sum_{j=1}^rh_j^4s_j^2+h_j^2s_j(2-h_j^2s_j)\right)=\sum_{j=1}^rs_jo(h_j^2).
\end{align}

Using that the first and second terms of \eqref{eq:Taylor} vanish, and using \eqref{eq:bias_int22} and \eqref{eq:bias_int3}, it follows that
\begin{align*}
    \eqref{eq:bias1}=&\;\omega_{\bd-1}\, c_{\bd,L}(\bh) \int_{I_\bh}L(\bs)\prod_{j=1}^r h_j^{d_j}s_j^{d_j/2-1} (2-h_j^2s_j)^{d_j/2-1}\,\nonumber\\
    &\times\sum_{k=1}^r\lrp{\frac{1}{2}h_k^2s_k(2-h_k^2s_k)d_k^{-1}\tr{\bHcal_{kk}\bar{f}(\bx)}+s_ko(h_k^2)}\,\rd \bs\\
    =&\;\frac{1}{2} \,\omega_{\bd-1}\,  c_{\bd,L}(\bh) \sum_{j=1}^r\int_{I_\bh}L(\bs)s_j^{d_j/2}h_j^{d_j+2} (2-h_j^2s_j)^{d_j/2}d_j^{-1}\tr{\bHcal_{jj}\bar{f}(\bx)}\,\nonumber\\
    &\times\prod_{k=1,k\neq j}^r h_k^{d_k}s_k^{d_k/2-1} (2-h_k^2s_k)^{d_k/2-1} \,\rd \bs\\
    &+\omega_{\bd-1}\, c_{\bd,L}(\bh) \sum_{j=1}^r\int_{I_\bh}L(\bs)h_j^{d_j}s_j^{d_j/2} (2-h_j^2s_j)^{d_j/2-1}o(h_j^2)\,\nonumber\\
    &\times\prod_{k=1,k\neq j}^r h_k^{d_k}s_k^{d_k/2-1} (2-h_k^2s_k)^{d_k/2-1} \,\rd \bs\\
    =&\;\frac{1}{2} \, \omega_{\bd-1}\, c_{\bd,L}(\bh) \rho(\bh)\sum_{j=1}^r\int_{I_\bh}L(\bs)h_j^2\tr{\bHcal_{jj}\bar{f}(\bx)}d_j^{-1}s_j (2-h_j^2s_j)\,\;\prod_{\ell=1}^r s_\ell^{d_\ell/2-1} (2-h_\ell^2s_\ell)^{d_\ell/2-1} \,\rd \bs\\
    &+\omega_{\bd-1}\,  c_{\bd,L}(\bh) \rho(\bh)\sum_{j=1}^r\int_{I_\bh}L(\bs)o(h_j^2)s_j\prod_{\ell=1}^r s_\ell^{d_\ell/2-1} (2-h_\ell^2s_\ell)^{d_\ell/2-1}  \,\rd \bs.
\end{align*}

Now consider the following functions for the indices $i=0,1$:
\begin{align*}
    \phi_{\bh,i}(\bs)\defin&\; c_{\bd,L}(\bh)\rho(\bh) L(\bs) s_j (2-h_j^2s_j)^i\,\;\prod_{\ell=1}^r s_\ell^{d_\ell/2-1} (2-h_\ell^2s_\ell)^{d_\ell/2-1} 1_{[0,2h^{-2}_\ell]}(s_\ell).
\end{align*}
Using Remark 5 in \cite{Garcia-Portugues2013b}, it can be obtained that, when $n\to\infty$ and $\bh\to \zero$,
\begin{align*}
    \phi_i(\bs)\defin\lim_{\bh\to \zero}\phi_{{\bh,i}}(\bs)\,&=\lambda_{\bd}(L)^{-1}L(\bs)s_j 2^i\,\;\prod_{\ell=1}^r s_\ell^{d_\ell/2-1} 2^{d_\ell/2-1}.
\end{align*}
Therefore, applying the Dominated Convergence Theorem (DCT) and using Lemma 1 in \cite{Garcia-Portugues2013b}, it follows that
\begin{align*}
    \lim_{\bh\to \zero}\int_{I_\bh}\phi_{{\bh,i}}(\bs)\,\rd \bs=&\;\int_{\R_+^r} \phi_i(\bs)\,\rd \bs%
    =
    \frac{2^{i}}{\omega_{\bd-1}}\frac{\int_{\R_+^r}L(\bs)s_j\prod_{\ell=1}^r s_\ell^{d_\ell/2-1}\,\rd \bs}{\int_{\R_+^r}L(\bs)\prod_{\ell=1}^rs_\ell^{d_\ell/2-1}\,\rd \bs}.
\end{align*}

Moreover, using $\int_{\R_+^r}\phi_{{\bh}}(\bs)\,\rd \bs=\int_{\R_+^r}\phi{{}}(\bs)\,\rd \bs\{1+o(1)\}$, it can be obtained that
\begin{align}\label{eq:bias_res}
	\mathbb{E}[\hat{f}(\bx;\bh)]-f(\bx)=
	&\; \sum_{j=1}^rh_j^2\tr{\bHcal_{jj}\bar{f}(\bx)}d_j^{-1}\frac{\int_{\R_+^r}L(\bs)s_j\prod_{\ell=1}^r s_\ell^{d_\ell/2-1}\,\rd \bs}{\int_{\R_+^r}L(\bs)\prod_{\ell=1}^rs_\ell^{d_\ell/2-1}\,\rd \bs}+o\left(\sum_{j=1}^rh_j^2\right).
\end{align}

\textit{Variance}. The variance of the estimator can be decomposed as
\begin{align}\label{eq:var}
	\mathbb{V}\mathrm{ar}[\hat{f}(\bx;\bh)]=&\;\frac{1}{n}\mathbb{E}\lrc{L^2_{\bh}(\bx,\bX)}-\frac{1}{n}\mathbb{E}[\hat{f}(\bx;\bh)]^2.
\end{align}
Using similar arguments to those employed for computing the bias term, the first addend of the variance \eqref{eq:var} can expressed as:
\begin{align}\label{eq:var_term1}
	\mathbb{E}\lrc{L^2_{\bh}(\bx,\bX)}
        =&\;\int_{\Sr}f(\by)c^2_{\bd,L}(\bh)L^2\lrp{\frac{1-\bx_1' \by_1}{h_1^2},\ldots, \frac{1-\bx_r' \by_r}{h_r^2}}\,\sigmar(\rd\by)\nonumber\\
        =&\int_{[-1,1]^r}\int_{\mathbb{S}^{\bd-1}}f\lrp{t_1 \bx_1+\left(1-t_1^2\right)^{1/2} \bB_{\bx_1} \bxi_1,\ldots,t_r \bx_r+\left(1-t_r^2\right)^{1/2} \bB_{\bx_r} \bxi_r}\nonumber\\
	&\times c^2_{\bd,L}(\bh)L^2\lrp{\frac{1-t_1}{h_1^2},\ldots, \frac{1-t_r}{h_r^2}} \prod_{j=1}^r \left(1-t_j^2\right)^{d_j/2-1}\,\sigmarm(\rd\bxi)\,\rd \bt\nonumber\\
        =&\; c^2_{\bd,L}(\bh)\int_{I_\bh} L^2(\bs)\prod_{j=1}^r h_j^{d_j}s_j^{d_j/2-1} (2-h_j^2s_j)^{d_j/2-1}\nonumber\\
        & \times \int_{\mathbb{S}^{\bd-1}}f\left(\bx+\ba_{\bx,\bxi}\right)\sigmarm(\rd\bxi)\,\rd \bs\nonumber\\
        =&\;\omega_{\bd-1}\, c^2_{\bd,L}(\bh) \int_{I_\bh}L^2(\bs)\prod_{j=1}^r h_j^{d_j}s_j^{d_j/2-1} (2-h_j^2s_j)^{d_j/2-1}\,\nonumber\\
	&\times\lrc{f(\bx)+\sum_{k=1}^r\lrp{\frac{1}{2}h_k^2s_k(2-h_k^2s_k)d_k^{-1}\tr{\bHcal_{kk}\bar{f}(\bx)}+s_ko(h_k^2)}}\,\rd \bs\nonumber\\
         =&\;\frac{1}{2} \, \omega_{\bd-1}\, c^2_{\bd,L}(\bh)\rho(\bh) \sum_{j=1}^r\int_{I_\bh}L^2(\bs)h_j^2\tr{\bHcal_{jj}\bar{f}(\bx)}d_j^{-1}s_j (2-h_j^2s_j)\,\nonumber\\
         &\times \prod_{\ell=1}^r s_\ell^{d_\ell/2-1} (2-h_\ell^2s_\ell)^{d_\ell/2-1} \,\rd \bs\nonumber\\
         &+\omega_{\bd-1}\,  c^2_{\bd,L}(\bh) \rho(\bh)f(\bx)\int_{I_\bh}L(\bs)\prod_{\ell=1}^r s_\ell^{d_\ell/2-1} (2-h_\ell^2s_\ell)^{d_\ell/2-1}  \,\rd \bs\nonumber\\
         &+\omega_{\bd-1}\,  c^2_{\bd,L}(\bh) \rho(\bh)\sum_{j=1}^r\int_{I_\bh}L^2(\bs)o(h_j^2)s_j\prod_{\ell=1}^r s_\ell^{d_\ell/2-1} (2-h_\ell^2s_\ell)^{d_\ell/2-1}  \,\rd \bs.
\end{align}

Analogously to the proof of the bias, we consider the functions
\begin{align*}
    \phi_{\bh,i,k}(\bs)\defin&\; c_{\bd,L}(\bh) \rho(\bh) L^2(\bs) s_j^k  (2-h_j^2s_j)^i\,\;\prod_{\ell=1}^r s_\ell^{d_\ell/2-1} (2-h_\ell^2s_\ell)^{d_\ell/2-1} 1_{[0,2h^{-2}_\ell]}(s_\ell).
\end{align*}
for the indices $i=0,1$ and $k=0,1$, and their limits
\begin{align*}
    \phi_{{i,k}}(\bs)\defin\lim_{\bh\to \zero}\phi_{{\bh,i,k}}(\bs)\,&=\lambda_{\bd}(L)^{-1}L^2(\bs)s_j^k 2^i\,\;\prod_{\ell=1}^r s_\ell^{d_\ell/2-1} 2^{d_\ell/2-1}.
\end{align*}
as $n\to\infty$ and $\bh\to \zero$. By the %
DCT and using Lemma 1 in \cite{Garcia-Portugues2013b}, it follows that
\begin{align*}
    \lim_{\bh\to \zero}\int_{I_\bh}\phi_{{\bh,i,k}}(\bs)\,\rd \bs=\int_{\R_+^r}\phi_{i,k}(\bs)\,\rd \bs%
    =
    \frac{2^{i}}{\omega_{\bd-1}}\frac{\int_{\R_+^r}L^2(\bs)s_j^k\prod_{\ell=1}^r s_\ell^{d_\ell/2-1}\,\rd \bs}{\int_{\R_+^r}L(\bs)\prod_{\ell=1}^rs_\ell^{d_\ell/2-1}\,\rd \bs}.
\end{align*}

Moreover, using $\int_{\R_+^r}\phi_{{\bh}}(\bs)\,\rd \bs=\int_{\R_+^r}\phi{{}}(\bs)\,\rd \bs\{1+o(1)\}$, it can be obtained that
\begin{align}\label{var_calc2}
	\eqref{eq:var_term1}=
	&\; c_{\bd,L}(\bh)\Bigg[\sum_{j=1}^rh_j^2\tr{\bHcal_{jj}\bar{f}(\bx)}d_j^{-1}\frac{\int_{\R_+^r}L^2(\bs)s_j\prod_{\ell=1}^r s_\ell^{d_\ell/2-1}\,\rd \bs}{\int_{\R_+^r}L(\bs)\prod_{\ell=1}^rs_\ell^{d_\ell/2-1}\,\rd \bs}\nonumber\\
        &+f(\bx)\frac{\int_{\R_+^r}L^2(\bs)\prod_{\ell=1}^r s_\ell^{d_\ell/2-1}\,\rd \bs}{\int_{\R_+^r}L(\bs)\prod_{\ell=1}^rs_\ell^{d_\ell/2-1}\,\rd \bs}\Bigg]+o\bigg(\prod_{j=1}^r h_j^{-d_j}\bigg).
\end{align}

Using \eqref{eq:bias_res}, the second term in \eqref{eq:var} is given by
\begin{align}
    \mathbb{E}[\hat{f}(\bx;\bh)]^2&
    =\Bigg[f(\bx)+\sum_{j=1}^rh_j^2\tr{\bHcal_{jj}\bar{f}(\bx)}d_j^{-1}\frac{\int_{\R_+^r}L(\bs)s_j\prod_{\ell=1}^r s_\ell^{d_\ell/2-1}\,\rd \bs}{\int_{\R_+^r}L(\bs)\prod_{\ell=1}^rs_\ell^{d_\ell/2-1}\,\rd \bs}\Bigg]^2+o\bigg(\sum_{j=1}^rh_j^2\bigg).\label{eq:biassq}
\end{align}

Then, from \eqref{var_calc2} and \eqref{eq:biassq}, the variance \eqref{eq:var} is given by
\begin{align*}
	\mathbb{V}\mathrm{ar}[\hat{f}(\bx;\bh)]=&\;\frac{c_{\bd,L}(\bh)}{n}
	\Bigg[\sum_{j=1}^rh_j^2\tr{\bHcal_{jj}\bar{f}(\bx)}d_j^{-1}\frac{\int_{\R_+^r}L^2(\bs)s_j\prod_{\ell=1}^r s_\ell^{d_\ell/2-1}\,\rd \bs}{\int_{\R_+^r}L(\bs)\prod_{\ell=1}^rs_\ell^{d_\ell/2-1}\,\rd \bs}\nonumber\\
        &+ f(\bx)\frac{\int_{\R_+^r}L^2(\bs)\prod_{\ell=1}^r s_\ell^{d_\ell/2-1}\,\rd \bs}{\int_{\R_+^r}L(\bs)\prod_{\ell=1}^rs_\ell^{d_\ell/2-1}\,\rd \bs}\Bigg]%
        +O(n^{-1})+o\bigg(n^{-1}\prod_{j=1}^r h_j^{-d_j}\bigg)\nonumber\\
	=&\;\frac{c_{\bd,L}(\bh)}{n} \frac{\int_{\R_+^r}L^2(\bs)\prod_{\ell=1}^r s_\ell^{d_\ell/2-1}\,\rd \bs}{\int_{\R_+^r}L(\bs)\prod_{\ell=1}^rs_\ell^{d_\ell/2-1}\,\rd \bs}f(\bx)+o\bigg(n^{-1}\prod_{j=1}^r h_j^{-d_j}\bigg).
\end{align*}
\end{proof}

\begin{proof}[Proof of Corollary \ref{cor:bias:order}]
The proof of this result is analogous to the derivation of the bias in Theorem \ref{thm:biasvar}. Due to \ref{A1*}, instead of considering the second-order Taylor expansion in \eqref{eq:Taylor}, we use a fourth-order Taylor expansion of $f$ at~$\bx$:
\begin{align}\label{eq:Taylor:four}
f\left(\bx+\ba_{\bx,\bxi}\right)=\sum_{\ell=0}^4\frac{1}{\ell!}(\mathsf{D}^{\otimes \ell}f(\bx))'\ba_{\bx,\bxi}^{\otimes \ell}+o(\|\ba_{\bx,\bxi}\|^4),
\end{align}
where the $\ell$th order differential operator $\mathsf{D}^{\otimes \ell}$ is defined as the formal $\ell$-fold Kronecker product of $\mathsf{D}$ with itself \citep[Section 2.6]{Chacon2018}. From the proof of Theorem \ref{thm:biasvar}, the first-order term of \eqref{eq:Taylor:four} vanishes, whereas the second-order term remains. The calculation of $\int_{\mathbb{S}^{\bd-1}}(\mathsf{D}^{\otimes 2}f(\bx))'\ba_{\bx,\bxi}^{\otimes 2}\,\sigmarm(\rd\bxi)$ can be seen in the aforementioned proof. The third-order term $\int_{\mathbb{S}^{\bd-1}}(\mathsf{D}^{\otimes 3}f(\bx))'\ba_{\bx,\bxi}^{\otimes 3}\,\sigmarm(\rd\bxi)=0$ also vanishes using Remark \ref{rem:int:prod} below and similar arguments to those employed for computing $\int_{\mathbb{S}^{\bd-1}}(\mathsf{D}^{\otimes 2}f(\bx))'\ba_{\bx,\bxi}^{\otimes 2}\,\sigmarm(\rd\bxi)$. Regarding the fourth-order term, it follows that $\int_{\mathbb{S}^{\bd-1}}(\mathsf{D}^{\otimes 4}f(\bx))'\ba_{\bx,\bxi}^{\otimes 4}\,\sigmarm(\rd\bxi)=O(\sum_{j=1}^r h_j^4)$ also using Remark \ref{rem:int:prod}. Moreover, $o(\|\ba_{\bx,\bxi}\|^4)=o(\sum_{j=1}^r h_j^4)$. The rest of the proof is analogous to that performed for Theorem \ref{thm:biasvar}.
\end{proof}

\begin{remark}\label{rem:int:prod}
Consider $\bx \in \Sd$ with entries $\left(x_1, \ldots, x_{d+1}\right)$. For all $i, j,k,\ell=1, \ldots, d+1$, using similar arguments to those in Lemma 3 of \cite{Garcia-Portugues2013b}, it can be obtained that
\begin{align*}
\int_{\Sd} x_i x_j x_k\,\sigma_d(\rd \bx)&=0,\\ %
\int_{\Sd} x_i x_j x_k x_\ell\,\sigma_d(\rd \bx)&= \begin{cases} 0& \text{all different indices }i, j, k, \ell, \\0  & i=j,i\neq k, i\neq \ell, \\ \frac{\omega_{d}}{(d+1)(d+3)} & i=j,k=\ell,i\neq k,\\ 0 & i=j=k,i\neq \ell,\\
\frac{3\omega_d}{(d+1)(d+3)} & i=j=k=\ell.\end{cases}
\end{align*}
\end{remark}

\begin{proof}[Proof of Corollary \ref{cor:consist}]
From Theorem \ref{thm:biasvar} it trivially follows that, for each $\bx\in\Sr$, $\mathbb{E}[(\hat{f}(\bx;\bh_n)- f(\bx))^2]\to0$ as $n\to\infty$ and under \ref{A1}--\ref{A3}. Consequently, $\hat{f}(\bx;\bh_n)$ converges in probability to $f(\bx)$. Since $\hat{f}(\bx;\bh_n)$ is a sum of independent variables, this convergence also holds almost surely.
\end{proof}

\begin{proof}[Proof of Corollary \ref{cor:amisebwd}]
The expression of the AMISE is trivial after integration of the squared bias and variance from Theorem \ref{thm:biasvar}.

Using $L^P$, $d=d_1=\cdots=d_r$, and $h_n=h_{1,n}=\cdots=h_{r,n}$ in \eqref{eq:amise} readily gives
\begin{align*}
\int_{\Sdr}((\bh_n^{\odot 2})'\bT\bar{f}(\bx)\bb_{\bd}(L))^2\,\sigmar(\rd\bx)
=h_n^4b_d^2(L)\int_{(\Sd)^r}\tr{\bHcal\bar{f}(\bx)}^2 \sigmar(\rd\bx)%
=h_n^4b_d^2(L)R\lrpbig{\nabla^2 \bar{f}}.
\end{align*}

Moreover, $\rho(\bh_n)=h_n^{rd}$, and therefore
\begin{align}
    \mathrm{AMISE}\lrcbig{\hat{f}(\cdot;h_n\one_r)}=&\;h_n^4b^{2}_d(L)R\lrpbig{\nabla^2 \bar{f}}+\frac{v^r_{d}(L)}{nh_n^{dr}}\label{eq:amise1d_proof}.
\end{align}

Differentiation of \eqref{eq:amise1d_proof} yields
\begin{align*}
\frac{\rd }{\rd h}\mathrm{AMISE}\lrcbig{\hat{f}(\cdot;h\one_r)}&=4b^{2}_d(L)h^3R\lrpbig{\nabla^2 \bar{f}}-dr\frac{v^r_d(L)}{nh^{dr+1}},%
\end{align*}
so solving $\frac{\rd }{\rd h}\mathrm{AMISE}\lrcbig{\hat{f}(\cdot;h\one_r)}=0$ readily gives $h_\mathrm{AMISE}$, the global minimum of the convex function $h\in\R_+\mapsto \mathrm{AMISE}\lrcbig{\hat{f}(\cdot;h\one_r)}$. Replacing $h_\mathrm{AMISE}$ in \eqref{eq:amise1d_proof} produces:
\begin{align*}
    \mathrm{AMISE}\lrcbig{\hat{f}(\cdot;h_\mathrm{AMISE}\one_r)}=&\;b^{2}_d(L)\lrc{\frac{dr\,v^r_d(L)}{4b^{2}_d(L)R\lrpbig{\nabla^2 \bar{f}}n}}^{4/(dr+4)}R\lrpbig{\nabla^2 \bar{f}}\\
    &+\frac{dr\,v^r_d(L)}{4b^{2}_d(L)R\lrpbig{\nabla^2 \bar{f}}n}\lrc{\frac{dr\,v^r_d(L)}{4b^{2}_d(L)R\lrpbig{\nabla^2 \bar{f}}n}}^{-dr/(dr+4)}\frac{4b^{2}_d(L)R\lrpbig{\nabla^2 \bar{f}}}{dr}\\
    =&\;\frac{(dr/4)^{4/(dr+1)}(dr+4)}{dr} \lrc{v_d(L)}^{4r/(dr+4)} \lrc{b^{2}_d(L)R\lrpbig{\nabla^2 \bar{f}}}^{dr/(dr+4)} n^{-4/(dr+4)}.
\end{align*}
\end{proof}

\begin{proof}[Proof of Theorem \ref{thm:distr}]
We apply Lyapunov's CLT %
to prove \eqref{eq:asymp1}, since $\hat{f}(\bx;\bh_n)$ equals $\bar{V}_n\defin\frac{1}{n}\sum_{i=1}^nV_{n,i}$ for the triangular array $\{V_{n,i}\}_{i=1}^n$ that is formed by the rowwise iid random variables $V_{n,i}\defin L_{\bh_n}(\bx,\bX_i)$. Then, if $\mathbb{E}\big[\abs{V_{n,i}}^{2+\delta}\big]<\infty$ and the Lyapunov's condition
\begin{align*}
    \lim_{n\to\infty}\frac{\mathbb{E}\big[|V_{n,i}-\mathbb{E}[V_{n,i}]|^{2+\delta}\big]}{n^{\delta/2}\mathbb{V}\mathrm{ar}[V_{n,i}]^{1+\delta/2}}=0
\end{align*}
is satisfied for some $\delta>0$, it follows that
\begin{align*}
    \sqrt{n}\frac{\bar V_{n}-\mathbb{E}[V_{n,i}]}{\sqrt{\mathbb{V}\mathrm{ar}[V_{n,i}]}}\rightsquigarrow\mathcal{N}(0,1).
\end{align*}

First, the condition $\mathbb{E}[\abs{V_{n,i}}^{2+\delta}]<\infty$ can be readily addressed:
\begin{align}
    \mathbb{E}[\abs{V_{n,i}}^{2+\delta}]=&\;\int_{\Sr}L^{2+\delta}_{\bh_n}(\bx,\by)f(\by)\,\sigmar(\rd \by)\nonumber\\
    =&\;c_{\bd,L}(\bh_n)^{2+\delta}\rho(\bh_n)\int_{I_\bh} L^{2+\delta}(\bs)\prod_{j=1}^rs_j^{d_j/2-1} (2-h_j^2s_j)^{d_j/2-1}\nonumber\\
        & \times \int_{\mathbb{S}^{\bd-1}}f\left(\bx+\ba_{\bx,\bxi}\right)\sigmarm(\rd\bxi)\,\rd \bs\nonumber\\
    \asymp&\;\lrp{\rho(\bh_n)^{-1}\lambda_{\bd}(L)^{-1}}^{2+\delta}\rho(\bh_n)\,\omega_{\bd-1}\int_{\R_+^r} L^{2+\delta}(\bs)\prod_{j=1}^rs_j^{d_j/2-1} 2^{d_j/2-1}\,\rd\bs\times f(\bx)  \nonumber\\
    =&\;\lrp{\rho(\bh_n)}^{-(1+\delta)}\frac{\int_{\R_+^r} L^{2+\delta}(\bs)\prod_{j=1}^rs_j^{d_j/2-1} \,\rd\bs}{\lrp{\prod_{j=1}^r 2^{d_j/2-1}\omega_{\bd-1}}^{1+\delta}\lrp{\int_{\R_+^r}L\lrp{\bs} \prod_{j=1}^r s_j^{d_j/2-1} \,\rd \bs}^{2+\delta}} \times f(\bx)\nonumber\\
    =&\;O\blrp{\rho(\bh_n)^{-(1+\delta)}}.\label{eq:Vnd}
\end{align}

Second, the $C_p$ and Jensen's inequalities yield that $\mathbb{E}\big[\abs{V_{n,i}-\mathbb{E}[V_{n,i}]}^{2+\delta}\big]=O\lrpbig{\mathbb{E}\big[\abs{V_{n,i}}^{2+\delta}\big]}$. Moreover, by Theorem \ref{thm:biasvar}, it can be deduced that the variance of $V_n$ has order
\begin{align}
    \mathbb{V}\mathrm{ar}[V_{n,i}]=&\;\frac{v_{\bd}(L)}{\rho(\bh_n)}f(\bx)(1+o(1))\asymp \rho(\bh_n)^{-1}.\label{eq:Vnvar}
\end{align}

Consequently, using \eqref{eq:Vnvar} and \eqref{eq:Vnd}, and from assumption \ref{A3}, it can be obtained that when $n\to\infty$,
\begin{align*}
    \frac{\mathbb{E}\big[\abs{V_{n,i}-\mathbb{E}[V_{n,i}]}^{2+\delta}\big]}{n^{\delta/2}\mathbb{V}\mathrm{ar}[V_{n,i}]^{1+\delta/2}}=O\lrp{\frac{\rho(\bh_n)^{-(1+\delta)}}{n^{\delta/2}\rho(\bh_n)^{-(1+\delta/2)}}}= O\lrpbig{\lrpbig{n\rho(\bh_n)}^{-\delta/2}}\to 0.
\end{align*}

Therefore,
\begin{align*}
    N_{n,1}\defin\frac{\hat f(\bx;\bh_n)-\mathbb{E}[\hat f(\bx;\bh_n)]}{\sqrt{\mathbb{V}\mathrm{ar}[\hat f(\bx;\bh_n)]}}\inlaw \mathcal{N}(0,1),
\end{align*}
pointwise for $\bx\in\Sr$. Considering $R_n\defin (n\rho(\bh_n))^{-1}v_{\bd}(L)f(\bx)/\mathbb{V}\mathrm{ar}[\hat f(\bx;\bh_n)]=1+o(1)$, the application of Slutsky's theorem ensures $R_nN_{n,1}\rightsquigarrow \mathcal{N}(0,1)$, thereby proving \eqref{eq:asymp1}.

For the proof of \eqref{eq:asymp2}, let us define $N_{n,2}\defin\sqrt{n\rho(\bh_n)}\lrpbig{\hat{f}(\bx;\bh_n)-\mathbb{E}[\hat{f}(\bx;\bh_n)]}$ and $N_{n,3}\defin\sqrt{n\rho(\bh_n)}\allowbreak\lrpbig{\hat{f}(\bx;\bh_n)-f(\bx)-(\bh_n^{\odot 2})'\bT\bar{f}(\bx)\bb_{\bd}(L)}$. Then, we have $N_{n,3}=N_{n,2}+o(\sqrt{n\rho(\bh_n)}(\bh^{\odot 2}_n)'\mathbf{1}_r)$. By employing Slutsky's theorem again, \eqref{eq:asymp2} follows.
\end{proof}

\begin{proof}[Proof of Proposition \ref{prop:Epa1}]
Applying the tangent-normal decomposition gives
\begin{align}
    c_{d,L_\mathrm{Epa}}(h)^{-1}%
    &=\frac{\om{d-1}}{h^2}\int_{-1}^1\lrp{h^2-1+t}1_{\{1-h^2<t<1\}}(1-t^2)^{d/2-1}\,\rd t\nonumber\\
    &=\om{d-1}\lrc{(1-h^{-2})\int_{m_h}^1(1-t^2)^{d/2-1}\,\rd t+h^{-2}\int_{m_h}^1t(1-t^2)^{d/2-1}\,\rd t},\label{eq:int0}
\end{align}
where $m_h=\max(-1,1-h^2)$. The first integral in \eqref{eq:int0} follows as
\begin{align}
    \int_{m_h}^1(1-t^2)^{d/2-1}\,\rd t%
    =\frac{\om{d}}{\om{d-1}}\lrc{1-F_d(m_h)},\label{eq:int1}
\end{align}
where $F_d$ is the cdf of the projected uniform distribution (see, e.g., Section 2.2 in \cite{Garcia-Portugues2022}). The second integral is immediate:
\begin{align}
    \int_{m_h}^1t(1-t^2)^{d/2-1}\,\rd t%
    =\frac{\lrp{1-m_h^2}^{d/2}}{d}.\label{eq:int2}
\end{align}
Joining \eqref{eq:int1} and \eqref{eq:int2} in \eqref{eq:int0} gives
\begin{align*}
    c_{d,L_\mathrm{Epa}}(h)^{-1}&=\om{d}(1-h^{-2})\lrc{1-F_d(m_h)}+\frac{\om{d-1}\lrp{1-m_h^2}^{d/2}}{dh^{2}}.
\end{align*}
If $h\geq\sqrt{2}$, then $m_h=-1$ and $c_{d,L_\mathrm{Epa}}(h)^{-1}=\om{d}(1-h^{-2})$. When $d=1$, then $F_1(x)=1/2+\arcsin(x)/\pi$ for $x\in(-1,1)$, and
\begin{align*}
    c_{1,L_\mathrm{Epa}}(h)^{-1}
    =2(1 - h^{-2}) \arccos(m_h)+2h^{-2}(1 - m_h^2)^{1/2}
    =2h^{-2}[(h^2-1)\arccos(m_h)+(1-m_h^2)^{1/2}].
\end{align*}

When $d=2$, since $F_2(x)=(x+1)/2$ for $x\in(-1,1)$, straightforward computations give
\begin{align*}
    c_{2,L_\mathrm{Epa}}(h)^{-1}%
    =2\pi(1-h^{-2})(1-m_h)+\frac{\pi(1-m_h^2)}{h^{2}}
    =\pi(1-m_h)(2-h^{-2}(1-m_h)).
\end{align*}
\end{proof}

\begin{proof}[Proof of Proposition \ref{prop:Epa2}]
The form of $f_{d,\mathrm{Epa},h}$ follows by the tangent-normal decomposition. Let $x\in[m_h,1]$, then the cdf associated to $f_{d,\mathrm{Epa},h}$ is
\begin{align*}
    F_{d,\mathrm{Epa},h}(x)=\int_{m_h}^x f_{d,\mathrm{Epa},h}(t)\,\rd t
    =\frac{\om{d-1} c_{d,L_\mathrm{Epa}}(h)}{h^2}\int_{m_h}^x \lrb{h^2-1+t}(1-t^2)^{d/2-1}\,\rd t.
\end{align*}

From the arguments given in the proof of Proposition \ref{prop:Epa1}, it follows that
\begin{align*}
    F_{d,\mathrm{Epa},h}(t)%
    =&\;\frac{\om{d-1} c_{d,L_\mathrm{Epa}}(h)}{h^2}\lrb{(h^2-1)\frac{\om{d}}{\om{d-1}}\lrc{F_d(x)-F_d(m_h)}-\frac{1}{d}\lrc{(1-x^2)^{d/2} - (1-m_h^2)^{d/2}}}.
\end{align*}

When $d=2$,
\begin{align*}
    F_{2,\mathrm{Epa},h}(t)%
    =&\;\frac{2\pi c_{2,L_\mathrm{Epa}}(h)}{h^2}\lrb{(h^2-1)(x-m_h)+\frac{1}{2}\lrc{x^2-m_h^2}}\\
    =&\;\frac{1}{(1-m_h)(2h^2-1+m_h)}\lrb{x^2+2(h^2-1)x-m_h(2(h^2-1)+m_h)}.
\end{align*}

Therefore, given $u\in(0,1)$, the roots of $F_{2,\mathrm{Epa},h}(t)=u$ are
\begin{align*}
    t%
    =&\;1-h^2\pm\sqrt{(h^2-1)^2-m_h+(m_h+u(1-m_h))(2h^2-1+m_h)}.
\end{align*}
Only the largest root satisfies that $1-h^2<t$.
\end{proof}

\begin{proof}[Proof of Proposition \ref{prop:sp}]
For any $d\geq1$, the tangent-normal decomposition gives
\begin{align*}
    c_{d,L_{\mathrm{sfp}_\upsilon}}(h)^{-1}&=\om{d-1}\int_{-1}^1L_{\mathrm{sfp}_\upsilon}((1-t)/h^2)(1-t^2)^{d/2-1}\,\rd t\nonumber\\
    &=\frac{\om{d-1}}{\mathrm{sfp}(\upsilon)}\int_{-1}^1\log\Big(1+e^{\upsilon(1-h^{-2})+(\upsilon h^{-2})t}\Big) (1-t^2)^{d/2-1}\,\rd t.
\end{align*}

In the particular case that $d=2$, using that
\begin{align}\label{eq:intlog}
    \int_{-1}^1 \log \big(1+e^{a+b t}\big) \,\rd t
    =\frac{1}{b}\int_{-e^{a-b}}^{-e^{a+b}} \frac{\log (1-u)}{u} \,\rd u
    =\frac{1}{b}\Big[\mathrm{Li}_{2}\big(-e^{a-b}\big)-\mathrm{Li}_{2}\big(-e^{a+b}\big)\Big],
\end{align}
it follows that
\begin{align*}
    c_{2,L_{\mathrm{sfp}_\upsilon}}(h)^{-1}&=\frac{2\pi}{\mathrm{sfp}(\upsilon)}\int_{-1}^1\log\Big(1+e^{\upsilon(1-h^{-2})+(\upsilon h^{-2})t}\Big)\,\rd t\\
    &=\frac{2\pi}{\mathrm{sfp}(\upsilon)\upsilon}\Big[\mathrm{Li}_{2}\big(-e^{\upsilon(1-2h^{-2})}\big)-\mathrm{Li}_{2}(-e^{\upsilon})\Big]h^2,
\end{align*}
and $F_{2,\mathrm{sfp}_\upsilon,h}(t)$ follows similarly.
\end{proof}

\begin{proof}[Proof of Proposition \ref{prop:moments}]
Standard computations give
\begin{align*}
    b_{d}(L_\mathrm{vMF}) &= \frac{\int_0^\infty L_\mathrm{vMF}(t) t^{d/2}\,\rd t}{d\int_0^\infty L_\mathrm{vMF}(t) t^{d/2-1}\,\rd t}
    =\frac{\Gamma((d+2)/2)}{d\,\Gamma(d/2)}
    =\frac{1}{2},\\
    v_{d}(L_\mathrm{vMF})&=\frac{\int_0^\infty L_\mathrm{vMF}^2(t) t^{d/2-1}\,\rd t}{2^{d/2-1}\om{d-1}(\int_0^\infty L_\mathrm{vMF}(t) t^{d/2-1}\,\rd t)^2}
    =\frac{2^{-d/2}\Gamma(d/2)}{\Gamma(d/2)^2}=\frac{1}{(2\pi^{1/2})^d},\\
    b_{d}(L_\mathrm{Epa}) &%
    =\frac{(d/2+1)^{-1}-(d/2+2)^{-1}}{d\,[(d/2)^{-1}-(d/2+1)^{-1}]}
    =\frac{1}{d+4},\\
    v_{d}(L_\mathrm{Epa})&%
    = \frac{(d/2+2)^{-1}-2(d/2+1)^{-1}+(d/2)^{-1}}{2^{d/2-1}\om{d-1}[(d/2)^{-1}-(d/2+1)^{-1}]^2}
    = \frac{d (d + 2)}{2^{d/2-1}\om{d-1}(d+4)}\\
    &=\frac{d (d + 2)\Gamma(d/2)}{(2\pi)^{d/2}(d+4)}=\frac{4\Gamma(d/2+2)}{(2\pi)^{d/2}(d+4)}.
\end{align*}

For the sfp kernel, define $I_a(\upsilon)\defin\int_0^\infty s^a \log \left(1+e^{\upsilon(1-s)}\right) \,\rd s$, $a\in\mathbb{N}_0$, and consider the integral representation of the polylogarithm function given in Proposition \ref{prop:sp}. Integrating by parts we obtain
\begin{align*}
    I_a(\upsilon)&=\frac{\upsilon}{a + 1} \int_0^\infty s^{a+1} \frac{e^{\upsilon(1-s)}}{1+e^{\upsilon(1-s)}} \,\rd s\\
    &=-\frac{1}{(a+1)\upsilon^{a+1}}\int_0^\infty u^{(a+2)-1} \frac{1}{e^u /(-e^\upsilon)-1} \,\rd u\\
    &=-\frac{\Gamma(a+1)}{\upsilon^{a+1}} \mathrm{Li}_{a+2}(-e^\upsilon).
\end{align*}
Therefore,
\begin{align*}
    b_d(L_{\mathrm{sfp}_\upsilon}) &= \frac{I_{d/2}(\upsilon)}{dI_{d/2-1}(\upsilon)}
    =\frac{\Gamma(d/2)/\Gamma(d/2-1)}{\upsilon^{d/2+1}/\upsilon^{d/2}}\frac{\mathrm{Li}_{d/2+2}(-e^\upsilon)}{d\,\mathrm{Li}_{d/2+1}(-e^\upsilon)}=\frac{\mathrm{Li}_{d/2+2}(-e^\upsilon)}{2\upsilon\,\mathrm{Li}_{d/2+1}(-e^\upsilon)},\\
    v_d(L_{\mathrm{sfp}_\upsilon})&= \frac{1}{2^{d/2-1}\omega_{d-1}}\frac{\int_0^\infty \log(1+e^{\upsilon(1-t)})^2t^{d/2-1} \,\rd t}{I_{d/2-1}(\upsilon)^2}= \frac{\upsilon^d\int_0^\infty \log(1+e^{\upsilon(1-t)})^2 t^{d/2-1} \,\rd t}{2^{d/2-1}\omega_{d-1}\Gamma(d/2)^2\mathrm{Li}^2_{d/2+1}(-e^\upsilon)}\\
    &= \frac{\upsilon^d\int_0^\infty \log(1+e^{\upsilon(1-t)})^2 t^{d/2-1} \,\rd t}{(2\pi)^{d/2}\Gamma(d/2)\mathrm{Li}^2_{d/2+1}(-e^\upsilon)}.
\end{align*}
\end{proof}

\begin{proof}[Proof of Proposition \ref{prop:moments2}]
For the spherically symmetric kernel $L^S(\bs)=L(\|\bs\|_1)$, define
\begin{align}
    I_{\bd, b,i}(L)&\defin\int_{\R_+^r} L\bigg(\sum_{j=1}^{r} s_j\bigg) \prod_{j=1}^{r} s_j^{d_j / 2-1} s_i^{b}\,\rd s_1 \cdots \,\rd s_r.\label{eq:Idb}
\end{align}
From this notation, $\lambda_{\bd}(L^S)=\om{\bd-1}2^{\tilde{d}/2-r}I_{\bd,0,1}(L)$ and the moments in \eqref{eq:moments} are
\begin{align}
    b_{\bd,i}(L^S)=\frac{I_{\bd,1,i}(L)}{d_iI_{\bd, 0,i}(L)},\, i=1,\ldots,r, \text{ and } v_{\bd}(L^S)=\frac{I_{\bd,0,1}(L^2)}{2^{\tilde{d}/2-r}\omega_{\bd-1}(I_{\bd,0,1}(L))^2}. \label{eq:momentsI}
\end{align}

We compute \eqref{eq:Idb} using the change of variables $(u_1,\ldots,u_{r-1},s)\defin(s_1/s,\ldots,s_{r-1}/s,\sum_{j=1}^rs_j)$ with %
\begin{align*}
    I_{\bd, b,i}(L)&=\int_{0}^{\infty} \int_{\Delta^{r-1}} L(s) u_i^{b}\prod_{j=1}^{r} u_{j}^{d_j / 2-1} \, s^{d_j / 2-1+b}s^{r-1} \,\rd u_{1} \cdots \,\rd u_{r}\,\rd s\\
    &=\frac{1}{\Gamma(r)} \int_{0}^{\infty} L(s) s^{\tilde{d} / 2-1+b}\,\rd s \times \int_{\Delta^{r-1}} u_{i}^b\prod_{j=1}^{r} u_{j}^{d_j / 2-1}  \,\Gamma(r) \,\rd u_{1} \cdots \,\rd u_{r} \\
    &=\frac{1}{\Gamma(r)} \int_{0}^{\infty} L(s) s^{\tilde{d} / 2-1+b}\,\rd s \times \mathbb{E}\Bigg[U_{i}^{b}\prod_{j=1}^{r} U_{j}^{d_j / 2-1} \Bigg],
\end{align*}
where $\bU=(U_1,\ldots,U_r)'$ is uniformly distributed on $\Delta^{r-1}$, i.e., it follows a Dirichlet distribution $\mathrm{Dir}(1,\ldots,1)$. Using the well-known formulae for expectations of powers of $U_j$, $\bU\sim\mathrm{Dir}(\alpha_1,\ldots,\alpha_r)$,
\begin{align*}
    \mathbb{E}\Bigg[\prod_{j=1}^{r} U_j^{\beta_j}\Bigg]=\frac{\Gamma(\sum_{j=1}^r \alpha_j)}{\Gamma(\sum_{j=1}^r(\alpha_j+\beta_j))} \prod_{j=1}^r \frac{\Gamma(\alpha_j+\beta_j)}{\Gamma(\alpha_j)},
\end{align*}
with $\alpha_1,\ldots,\alpha_r>0$ we have that
\begin{align*}
    \mathbb{E}\Bigg[U_{i}^{b}\prod_{j=1}^{r} U_{j}^{d_j / 2-1} \Bigg]=\frac{\Gamma(r)\prod_{j=1}^r \Gamma(d_j/2)\Gamma(d_i/2+b)}{\Gamma(\tilde{d}/2+b)\Gamma(d_i/2)}
\end{align*}
and therefore
\begin{align}
    I_{\bd, b,i}(L)%
    &=\frac{\Gamma(d_i/2+b)\prod_{j=1}^r \Gamma(d_j/2)}{\Gamma(d_i/2)\Gamma(\tilde{d}/2+b)}\times \left[\int_{0}^{\infty} L(s) s^{\tilde{d} / 2-1+b} \,\rd s\right]. \label{eq:momentsI2}
\end{align}

Now, using \eqref{eq:momentsI} and \eqref{eq:momentsI2}, we have
\begin{align*}
    b_{\bd,i}(L^S)%
    &= \frac{\Gamma(d_i/2+1)/\Gamma(d_i/2)}{d_i\Gamma(\tilde{d}/2+1)/\Gamma(\tilde{d}/2)} \times\left[\frac{\int_{0}^{\infty} L(s) s^{\tilde{d} / 2} \,\rd s}{\int_{0}^{\infty} L(s) s^{\tilde{d} / 2-1} \,\rd s}\right]
    =\frac{1}{\tilde{d}}\times \left[\frac{\int_{0}^{\infty} L(s) s^{\tilde{d} / 2} \,\rd s}{\int_{0}^{\infty} L(s) s^{\tilde{d} / 2-1} \,\rd s}\right],\\
    v_{\bd}(L^S)%
    &=\frac{1}{2^{\tilde{d} / 2-r} \omega_{\bd-1}} \frac{\Gamma(\tilde{d}/2)}{\prod_{j=1}^r \Gamma(d_j/2)}\times\left[\frac{\int_{0}^{\infty} L^2(s)s^{\tilde{d} / 2-1} \,\rd s}{\big(\int_{0}^{\infty} L(s) s^{\tilde{d} / 2-1} \,\rd s\big)^{2}}\right]\\
    &=\frac{\Gamma(\tilde{d}/2)}{(2\pi)^{\tilde{d}/2}}\times \left[\frac{\int_{0}^{\infty} L^2(s)s^{\tilde{d} / 2-1} \,\rd s}{\big(\int_{0}^{\infty} L(s) s^{\tilde{d} / 2-1} \,\rd s\big)^{2}}\right]%
    =\frac{1}{2^{\tilde{d}/2-1}\omega_{\tilde{d}-1}}\times \left[\frac{\int_{0}^{\infty} L^2(s)s^{\tilde{d} / 2-1} \,\rd s}{\big(\int_{0}^{\infty} L(s) s^{\tilde{d} / 2-1} \,\rd s\big)^{2}}\right],\\pct\\
    \lambda_{\bd}(L^S)%
    &=\om{\bd-1}2^{\tilde{d}/2-r}  \frac{\prod_{j=1}^r \Gamma(d_j/2)}{\Gamma(\tilde{d}/2)}\times \left[\int_{0}^{\infty} L(s) s^{\tilde{d} / 2-1} \,\rd s\right]%
    =\omega_{\tilde{d}-1}2^{\tilde{d}/2-1}  \int_{0}^{\infty} L(s) s^{\tilde{d} / 2-1} \,\rd s.%
\end{align*}
Therefore, $\lambda_{\bd}(L^S)=\lambda_{\tilde{d}}(L)$, $\bb_{\bd}(L^S)= b_{\tilde{d}}(L)\one$, and $v_{\bd}(L^S)=v_{\tilde{d}}(L)$.
\end{proof}

\begin{proof}[Proof of Proposition \ref{prop:sp2}]
From Proposition \ref{prop:moments2} and the proof of Proposition \ref{prop:moments}, it follows that
\begin{align*}
    \lambda_{\bd}(L_\mathrm{Epa}^S)&=\lambda_{\tilde{d}}(L_\mathrm{Epa})%
    =\frac{(2\pi)^{\tilde{d}/2}}{\Gamma(\tilde{d}/2)}[(d/2)^{-1}-(d/2+1)^{-1}]%
    =\frac{4(2\pi)^{\tilde{d}/2+1}}{d(d+2)\Gamma(\tilde{d}/2)},\\
    \lambda_{\bd}(L_{\mathrm{sfp}_\upsilon}^S)&=\lambda_{\tilde{d}}(L_{\mathrm{sfp}_\upsilon})=-\frac{\omega_{\tilde{d}-1}2^{\tilde{d}/2-1}}{\mathrm{sfp}(\upsilon)} \frac{\Gamma(\tilde{d}/2)}{\upsilon^{\tilde{d}/2}} \mathrm{Li}_{\tilde{d}/2+1}(-e^\upsilon)=-\left(\frac{2\pi}{\upsilon}\right)^{\tilde{d}/2} \frac{\mathrm{Li}_{\tilde{d}/2+1}(-e^\upsilon)}{\mathrm{sfp}(\upsilon)}.
\end{align*}
The asymptotic approximations of the normalizing constants are obtained with \eqref{eq:equiv}.

For $d=d_1=\cdots=d_r$, applying the tangent-normal decomposition gives
\begin{align*}
    c_{\bd,L_{\mathrm{sfp}_\upsilon}^{S}}(\bh)^{-1}=&\;\frac{1}{\mathrm{sfp}(\upsilon)}\int_{\Sr}\log\lrp{1+e^{\upsilon\lrp{1-\sum_{j=1}^r\frac{1-\bx_j' \by_j}{h^2}}}} \,\sigmar(\rd\bx)\\
    =&\;\frac{1}{\mathrm{sfp}(\upsilon)}\int_{[-1,1]^r}\log\lrp{1+e^{\upsilon\lrp{\frac{h^2-r+\sum_{j=1}^rt_j}{h^2}}}} \prod_{j=1}^r \left(1-t_j^2\right)^{d/2-1}\om{d-1}\,\rd \bt.
\end{align*}

In the particular case that $d=2$, the normalizing constant is simplified to
\begin{align*}
    c_{\two,L_{\mathrm{sfp}_\upsilon}^{S}}(\bh)^{-1}
    =\frac{(2\pi)^r}{\mathrm{sfp}(\upsilon)}\int_{[-1,1]^r}\log\lrp{1+e^{\upsilon(1-rh^{-2})+\upsilon h^{-2}\lrp{\sum_{j=1}^rt_j}}} \,\rd \bt.
\end{align*}

Now, denoting $a=\upsilon(1-rh^{-2})$ and $b=\upsilon h^{-2}$, it follows that
\begin{align}\label{eq:polylog}
    \int_{[-1,1]^r} \log \big(1+e^{a+b \sum_{j=1}^rt_j}\big) \,\rd \bt
    =\frac{(-1)^{r-1}}{b^r}\sum_{\ell=0}^{r}\binom{r}{\ell} (-1)^\ell\, \mathrm{Li}_{(r+1)}(-e^{a+b(2\ell-r)}).
\end{align}

The proof of \eqref{eq:polylog} can be carried out by induction. In the case where $r=1$, the result has already been proved in \eqref{eq:intlog}. We now assume the statement holds for $r=k> 1$,
\begin{align*}
	\int_{[-1,1]^{k}} \log \big(1+e^{a+b \sum_{j=1}^{k}t_j}\big) \,\rd \bt
	=\frac{(-1)^{k-1}}{b^{k}}\sum_{\ell=0}^{k}\binom{k}{\ell} (-1)^\ell\, \mathrm{Li}_{(k+1)}(-e^{a+b(2\ell-k)}),
\end{align*}
and we readily show its validity for $r=k+1$:
\begin{align*}
	\int_{[-1,1]^{k+1}} \log \big(1+e^{a+b \sum_{j=1}^{k+1}t_j}\big) \,\rd \bt
	=&\int_{-1}^1\lrp{\int_{[-1,1]^{k}} \log \big(1+e^{a+bt_{k+1}+b \sum_{j=1}^{k}t_j}\big)} \,\rd \bt\\
    =&\,\frac{(-1)^{k-1}}{b^{k}}\sum_{\ell=0}^{k}\binom{k}{\ell} (-1)^\ell\int_{-1}^1 \mathrm{Li}_{(k+1)}\big(-e^{a+bt_{k+1}+b(2\ell-r)}\big) \,\rd t_{k+1}\\
    =&\, \frac{(-1)^{k-1}}{b^{k+1}}\sum_{\ell=0}^{k}\binom{k}{\ell} (-1)^\ell\int_{-e^{a+b(2\ell-r-1)}}^{-e^{a+b(2\ell-r+1)}}\frac{\mathrm{Li}_{(k+1)}(u)}{u} \,\rd u\\
    =&\,\frac{(-1)^{k}}{b^{k+1}}\sum_{\ell=0}^{k+1}\binom{k+1}{\ell} (-1)^\ell\, \mathrm{Li}_{(k+2)}\big(-e^{a+b(2\ell-k-1)}\big).
\end{align*}

Therefore, from \eqref{eq:polylog}, it follows that
\begin{align*}
    c_{\two,L_{\mathrm{sfp}_\upsilon}^{S}}(\bh)^{-1}
    =&\;\frac{(2\pi h^2)^r}{\mathrm{sfp}(\upsilon)}\frac{(-1)^{r-1}}{\upsilon^r}\sum_{\ell=0}^{r}\binom{r}{\ell} (-1)^\ell\, \mathrm{Li}_{(r+1)}\big(-e^{\upsilon(1-rh^{-2})+\upsilon h^{-2}(2\ell-r)}\big).
\end{align*}

The $L_\mathrm{Epa}^{S}$ kernel follows from the asymptotic equivalence provided in Remark \ref{rem:equiv} and the fact that the $\ell$th term of the above sum does not vanish as $\upsilon\to\infty$ if and only if $\ell\geq \lceil r-h^2/2 \rceil$:
\begin{align*}
   c_{\two,L_\mathrm{Epa}^{S}}(\bh)^{-1}=&\;\lim_{\upsilon\to\infty}c_{\two,L_{\mathrm{sfp}_\upsilon}^{S}}(\bh)^{-1}\\
    =&\;\frac{(2\pi h^2)^r(-1)^{r}}{(r+1)!}\sum_{\ell=\lceil r-h^2/2 \rceil}^{r}\binom{r}{\ell} (-1)^\ell\, \lrp{(1-rh^{-2})+h^{-2}(2\ell-r)}^{r+1}\\
    =&\;\frac{(2\pi h^2)^r(-1)^{r}}{(r+1)!}\sum_{\ell=\lceil r-h^2/2 \rceil}^{r}\binom{r}{\ell} (-1)^\ell\, \lrp{1-2h^{-2}(r-\ell)}^{r+1}.
\end{align*}
If $h<\sqrt{2}$, then $\lceil r-h^2/2 \rceil=r$ and
\begin{align*}
   c_{\two,L_\mathrm{Epa}^{S}}(\bh)^{-1}=\frac{(2\pi h^2)^r(-1)^{r}}{(r+1)!} (-1)^r=\frac{(2\pi h^2)^r}{(r+1)!}.
\end{align*}
\end{proof}

\begin{proof}[Proof of Proposition \ref{cor:effic2}]
We first compute the AMISE constants
$C_{d,r}(L^P)=%
\big[v^r_d(L)b^{dr/2}_d(L)\big]^{4/(dr+4)}$ and $C_{d,r}(L^S)=%
\big[v_{dr}(L) b_{dr}^{dr/2}(L)\big]^{4/(dr+4)}$ using Propositions \ref{prop:moments} and \ref{prop:moments2}. For the $L_\mathrm{vMF}^S=L_\mathrm{vMF}^P$ kernel, we readily have
\begin{align*}
    C_{d,r}(L_\mathrm{vMF}^S)=%
    \lrc{\frac{1}{(2\pi^{1/2})^{dr}} \times \frac{1}{2^{dr/2}}}^{4/(dr+4)}%
    =\big[(8\pi)^{-dr/2}\big]^{4/(dr+4)}.
\end{align*}
For the $L_\mathrm{Epa}^P$ and $L_\mathrm{Epa}^S$ kernels the constants are:
\begin{align*}
    C_{d,r}(L_\mathrm{Epa}^S)%
    =&\,\lrc{\frac{4\Gamma(dr/2+2)}{(2\pi)^{dr/2}(dr+4)} \times \frac{1}{(dr+4)^{dr/2}}}^{4/(dr+4)}
    =\lrc{\frac{4\Gamma(dr/2+2)}{(2\pi)^{dr/2}(dr+4)^{dr/2+1}}}^{4/(dr+4)},\\
    C_{d,r}(L_\mathrm{Epa}^P)%
    &=\lrc{\frac{4^r\Gamma(d/2+2)^r}{(2\pi)^{dr/2}(d+4)^{r}} \times \frac{1}{(d+4)^{dr/2}}}^{4/(dr+4)}
    =\lrc{\frac{4^r \Gamma(d/2+2)^r}{(2\pi)^{dr/2}(d+4)^{r(d/2+1)}}}^{4/(dr+4)}.
\end{align*}
Define the function $J_d(\upsilon)\defin\int_0^\infty \log(1+e^{\upsilon(1-t)})^2 t^{d/2-1} \,\rd t$. Then, for the $L_{\mathrm{sfp}_\upsilon}^S$ and $L_{\mathrm{sfp}_\upsilon}^P$ kernels, the constants are
\begin{align*}
    C_{d,r}(L_{\mathrm{sfp}_\upsilon}^S)%
    &=\lrc{\frac{\upsilon^{dr}J_{dr}(\upsilon)}{(2\pi)^{dr/2}\Gamma(dr/2)\mathrm{Li}^2_{dr/2+1}(-e^\upsilon)} \times \frac{|\mathrm{Li}|^{dr/2}_{dr/2+2}(-e^\upsilon)}{(2\upsilon)^{dr/2}\,|\mathrm{Li}|^{dr/2}_{dr/2+1}(-e^\upsilon)}}^{4/(dr+4)}\\
    &=\lrc{\frac{\upsilon^{dr/2}J_{dr}(\upsilon) \,|\mathrm{Li}|_{dr/2+2}^{dr/2}(-e^\upsilon)}{(4\pi)^{dr/2}\Gamma(dr/2) \,|\mathrm{Li}|^{dr/2+2}_{dr/2+1}(-e^\upsilon)}}^{4/(dr+4)},\\
    C_{d,r}(L_{\mathrm{sfp}_\upsilon}^P)%
    &=\lrc{\frac{\upsilon^{dr}J_d^r(\upsilon)}{(2\pi)^{dr/2}\Gamma(d/2)^r\,\mathrm{Li}^{2r}_{d/2+1}(-e^\upsilon)}\times \frac{|\mathrm{Li}|_{d/2+2}^{dr/2}(-e^\upsilon)}{(2\upsilon)^{dr/2}\,|\mathrm{Li}|_{d/2+1}^{dr/2}(-e^\upsilon)}}^{4/(dr+4)}\\
    &=C_{d,r}(L_{\mathrm{sfp}_\upsilon}^S)\lrc{\frac{\Gamma(dr/2)J_d^r(\upsilon)|\mathrm{Li}|_{dr/2+1}^{dr/2+2}(-e^\upsilon)|\mathrm{Li}|_{d/2+2}^{dr/2}(-e^\upsilon)}{\Gamma(d/2)^rJ_{dr}(\upsilon)|\mathrm{Li}|_{dr/2+2}^{dr/2}(-e^\upsilon)|\mathrm{Li}|_{d/2+1}^{r(d/2+2)}(-e^\upsilon)}}^{4/(dr+4)}.
\end{align*}

Now, the efficiencies of the $L_\mathrm{vMF}^S$ and $L_\mathrm{Epa}^P$ kernels are
\begin{align*}
    \mathrm{eff}_{d,r}(L_\mathrm{vMF}^S)%
    &=\lrc{\frac{4\Gamma(dr/2+2)}{(2\pi)^{dr/2}(dr+4)^{dr/2+1}}} \big[(8\pi)^{-dr/2}\big]^{-1}
    = \frac{2^{dr+2} \Gamma(dr/2+2)}{(dr+4)^{dr/2+1}},\\
    \mathrm{eff}_{d,r}(L_\mathrm{Epa}^P)%
    &=\lrc{\frac{4\Gamma(dr/2+2)}{(2\pi)^{dr/2}(dr+4)^{dr/2+1}}} \lrc{\frac{4^r \Gamma(d/2+2)^r}{(2\pi)^{dr/2}(d+4)^{r(d/2+1)}}}^{-1}\\
    &= \frac{4^{1-r}\Gamma(dr/2+2)(d+4)^{r(d/2+1)}}{\Gamma(d/2+2)^r(dr+4)^{dr/2+1}}.
\end{align*}

The efficiencies for $L_{\mathrm{sfp}_\upsilon}^S$ and $L_{\mathrm{sfp}_\upsilon}^P$ are
\begin{align*}
    \mathrm{eff}_{d,r}(L_{\mathrm{sfp}_\upsilon}^S)%
    &=\lrc{\frac{4\Gamma(dr/2+2)}{(2\pi)^{dr/2}(dr+4)^{dr/2+1}}}\lrc{\frac{\upsilon^{dr/2}J_{dr}(\upsilon) \,|\mathrm{Li}|_{dr/2+2}^{dr/2}(-e^\upsilon)}{(4\pi)^{dr/2}\Gamma(dr/2) \,|\mathrm{Li}|^{dr/2+2}_{dr/2+1}(-e^\upsilon)}}^{-1}\\
    &=\frac{2^{dr/2+2}\Gamma(dr/2+2)\Gamma(dr/2)|\mathrm{Li}|^{dr/2+2}_{dr/2+1}(-e^\upsilon)}{(dr+4)^{dr/2+1} \upsilon^{dr/2}J_{dr}(\upsilon) \,|\mathrm{Li}|_{dr/2+2}^{dr/2}(-e^\upsilon)},\\
    \mathrm{eff}_{d,r}(L_{\mathrm{sfp}_\upsilon}^P)%
    &= \mathrm{eff}_{d,r}(L_{\mathrm{sfp}_\upsilon}^S) \times\frac{\Gamma(d/2)^rJ_{dr}(\upsilon)|\mathrm{Li}|_{dr/2+2}^{dr/2}(-e^\upsilon)|\mathrm{Li}|_{d/2+1}^{r(d/2+2)}(-e^\upsilon)}{\Gamma(dr/2)J_d^r(\upsilon)|\mathrm{Li}|_{dr/2+1}^{dr/2+2}(-e^\upsilon)|\mathrm{Li}|_{d/2+2}^{dr/2}(-e^\upsilon)}.
\end{align*}
\end{proof}

\begin{proof}[Proof of Proposition \ref{prop:nabla2}]
If $\bar{f}$ is twice continuously differentiable on $\Sdr$, then from Proposition 1 in \cite{Garcia-Portugues2023}, it can be obtained that
\begin{align*}
    \tr{\bHcal_{jj}\bar{f}(\bx)}=&\;\mbox{tr}\, [(\bI_{d+1}-\bx_j\bx_j')\bHcal_{jj} f(\bx)(\bI_{d+1}-\bx_j\bx_j')\nonumber\\
    &-(\bnab_j f(\bx)\bx_j)(\bI_{d+1}-\bx_j\bx_j')-\big[\bx_j\bnab_j f(\bx)+(\bx_j\bnab_j f(\bx))'-2(\bnab_j f(\bx)\bx_j)\bx_j\bx_j'\big]]\nonumber\\
    =&\;\tr{\bHcal_{jj}f(\bx)}-\bx_j'\bHcal_{jj} f(\bx)\bx_j-d_j\bnab f_j(\bx)\bx_j.
\end{align*}

Note that $f_{\mathrm{PvMF}}(\bx;\bmu,\bkappa)=L_{1/\sqrt{\bkappa}}(\bx,\bmu)$ if $L_\mathrm{vMF}^P$ is considered in \eqref{eq:estimator}. Thus,
\begin{align*}
   \bnabla_j f_{\mathrm{PvMF}}(\bx;\bmu,\bkappa) &= \kappa_j f_{\mathrm{PvMF}}(\bx;\bmu,\bkappa) \bmu_{j}',\\
   \bHcal_{jj} f_{\mathrm{PvMF}}(\bx;\bmu,\bkappa) &= \kappa_j^2 f_{\mathrm{PvMF}}(\bx;\bmu,\bkappa) \bmu_{j}\bmu_{j}',
\end{align*}
and, then $\tr{\bHcal_{jj}f_{\mathrm{PvMF}}(\bx;\bmu,\bkappa)}= \kappa_j^2 f_{\mathrm{PvMF}}(\bx;\bmu,\bkappa)$ and $\bx_j'\bHcal_{jj} f_{\mathrm{PvMF}}(\bx;\bmu,\bkappa)\bx_j =\allowbreak [\kappa_j\bmu_{j}'\bx_j]^2 f_{\mathrm{PvMF}}(\bx;\bmu,\bkappa)$.

As a consequence, it follows that
\begin{align*}
    \tr{\bHcal_{jj}\bar{f}_{\mathrm{PvMF}}(\bx;\bmu,\bkappa)}=&\;\psi_j(\bx_j'\bmu_j,\kappa_j)f_{\mathrm{PvMF}}(\bx;\bmu,\bkappa),
\end{align*}
with $\psi_j(t,\kappa)=\kappa[\kappa(1-t^2)-d_jt]$ and, hence, that
\begin{align*}
    \bT_{\bar{f}_{\mathrm{PvMF}}}(\bx)=f_{\mathrm{PvMF}}(\bx;\bmu,\bkappa)\diag{\psi_1(\bx_1'\bmu_1,\kappa_1),\ldots,\psi_r(\bx_r'\bmu_r,\kappa_r)}.
\end{align*}
\end{proof}

\begin{proof}[Proof of Proposition \ref{prop:plugin}]
In order to compute $\mathrm{AMISE}[\hat{f}(\cdot;\bh_n)]$ and the corresponding AMISE optimal bandwidth for $\mathrm{PvMF}$, we first compute $\int_{\Sr}((\bh_n^{\odot 2})'\bT_{\bar{f}_{\mathrm{PvMF}}}(\bx)\bb_{\bd}(L))^2\,\sigmar(\rd\bx)$. From Proposition \ref{prop:nabla2}, we have
\begin{align*}
    (\bh_n^{\odot 2})'\bT_{\bar{f}_{\mathrm{PvMF}}}(\bx)\bb_{\bd}(L)=f_{\mathrm{PvMF}}(\bx;\bmu,\bkappa)\sum_{j=1}^rh_{j,n}^2b_{\bd,j}(L)\psi_j(\bx_j'\bmu_j,\kappa_j).
\end{align*}

For $m=0,1,2$, let us define
\begin{align*}
    R_m(\kappa)\defin\int_{\Sd} f_{\mathrm{vMF}}(\bx;\bmu,\kappa)^2 \psi(\bx'\bmu,\kappa)^m\,\sigmad(\rd\bx).
\end{align*}
It is obtained that
\begin{align*}
   \int_{\Sr}&((\bh_n^{\odot 2})'\bT\bar{f}(\bx)\bb_{\bd}(L))^2\,\sigmar(\rd\bx)\\
    =&\;\sum_{j,k=1}^r\int_{\Sr} f_{\mathrm{PvMF}}(\bx;\bmu,\bkappa)^2h_{j,n}^2h_{k,n}^2b_{\bd,j}(L)b_{\bd,k}(L) \psi(\bx_j'\bmu_j,\kappa_j)\psi(\bx_k'\bmu_k,\kappa_k)\,\sigmar(\rd\bx)\\
    =&\;\sum_{j=1}^rh_{j,n}^4b^2_{\bd,j}(L)\int_{\Sj} f_{\mathrm{vMF}}(\bx_j;\bmu_j,\kappa_j)^2 \psi(\bx_j'\bmu_j,\kappa_j)^2\,\sigmaj(\rd\bx_j)\times \prod_{\substack{\ell=1\\\ell\neq j}}^r \int_{\Sell} f_{\mathrm{vMF}}(\bx_\ell;\bmu_\ell,\kappa_\ell)^2 \,\sigmaell(\rd\bx_\ell)\\
    &+\sum_{\substack{j,k=1\\j\neq k}}^rh_{j,n}^2h_{k,n}^2b_{\bd,j}(L)b_{\bd,k}(L)\int_{\Sj} f_{\mathrm{vMF}}(\bx_j;\bmu_j,\kappa_j)^2 \psi(\bx_j'\bmu_j,\kappa_j)\, \sigmaj(\rd\bx_j)\\
    &\times\int_{\Sk} f_{\mathrm{vMF}}(\bx_k;\bmu_k,\kappa_k)^2 \psi(\bx_k'\bmu_k,\kappa_k)\, \sigmak(\rd\bx_k)
    \times\prod_{\substack{\ell=1\\\ell\neq j\\\ell\neq k}}^{r}\int_{\Sell} f_{\mathrm{vMF}}(\bx_\ell;\bmu_\ell,\kappa_\ell)^2\,\sigmaell(\rd\bx_\ell)\\
    =&\;\sum_{j=1}^rh_{j,n}^4b^2_{\bd,j}(L)\frac{R_2(\kappa_j)}{R_0(\kappa_j)}\prod_{k=1}^r R_0(\kappa_k)+\sum_{\substack{j,k=1\\j\neq k}}^rh_{j,n}^2h_{k,n}^2b_{\bd,j}(L)b_{\bd,k}(L)\frac{R_1(\kappa_j) R_1(\kappa_k)}{R_0(\kappa_j) R_0(\kappa_k)}\prod_{\ell=1}^{r} R_0(\kappa_\ell)\\
    =&\;(\bh_n^{\odot 2}\odot\bb_{\bd}(L))'\bR(\bkappa)(\bh_n^{\odot 2}\odot\bb_{\bd}(L)),
\end{align*}
where $\bR(\bkappa)$ is the symmetric $r\times r$ matrix with entries
\begin{align}
    \bR(\bkappa)_{jj}&=\frac{R_2(\kappa_j)}{R_0(\kappa_j)}\times\prod_{\ell=1}^{r} R_0(\kappa_\ell),\quad j=1,\ldots,r,\label{eq:rkk}\\
    \bR(\bkappa)_{jk}&=\frac{R_1(\kappa_j) R_1(\kappa_k)}{R_0(\kappa_j) R_0(\kappa_k)}\times\prod_{\ell=1}^{r} R_0(\kappa_\ell),\quad j,k=1,\ldots,r,j\neq k\label{eq:rjk}.
\end{align}

Now, since
\begin{align*}
    R_m(\kappa)=\int_{\Sd} f_{\mathrm{vMF}}(\bx;\bmu,\kappa)^2 \psi(\bx'\bmu,\kappa)^m\,\sigmad(\rd\bx)=c_d^{\mathrm{vMF}}(\kappa)^2\int_{\Sd} e^{2\kappa\bx'\bmu}\psi(\bx'\bmu,\kappa)^m\,\sigmad(\rd\bx),
\end{align*}
it follows that
\begin{align}
    R_0(\kappa)&=\frac{c_d^{\mathrm{vMF}}(\kappa)^2}{c_d^{\mathrm{vMF}}(2\kappa)}%
    =\frac{\kappa^{(d-1)/2}\mathcal{I}_{(d-1)/2}(2\kappa)}{2^d \pi^{(d+1)/2}\mathcal{I}_{(d-1)/2}(\kappa)^2},\label{eq:R0}\\
    R_1(\kappa)%
    &=c_d^{\mathrm{vMF}}(\kappa)^2\omega_{d-1}\int_{-1}^1 e^{2\kappa t} \kappa[\kappa(1-t^2)-dt] (1-t^2)^{d/2-1}\,\rd t\nonumber\\
    &=c_d^{\mathrm{vMF}}(\kappa)^2\omega_{d-1}\lrb{-\sqrt{\pi}\Gamma(d/2+1) \kappa^{(3-d)/2} \mathcal{I}_{(d+1)/2}(2\kappa)}\nonumber\\
    &=\frac{\kappa^{d-1}}{(2\pi)^{d+1}\mathcal{I}_{(d-1)/2}(\kappa)^2} \lrb{-\pi^{(d+1)/2} d \kappa^{(3-d)/2} \mathcal{I}_{(d+1)/2}(2\kappa)}\nonumber\\
    &=-\frac{d\kappa}{2}\, \frac{\mathcal{I}_{(d+1)/2}(2\kappa)}{\mathcal{I}_{(d-1)/2}(2\kappa)} \times R_0(\kappa),\label{eq:R1}\\
    R_2(\kappa)%
    &=c_d^{\mathrm{vMF}}(\kappa)^2\omega_{d-1}\int_{-1}^1 e^{2\kappa t} \kappa^2[\kappa(1-t^2)-dt]^2 (1-t^2)^{d/2-1}\,\rd t\nonumber\\
    &=\frac{\kappa^{d-1}}{(2\pi)^{d+1}\mathcal{I}_{(d-1)/2}(\kappa)^2} \lrb{d\pi^{(d+1)/2}/2 \kappa^{(3-d)/2}\lrc{2d\mathcal{I}_{(d+1)/2}(2\kappa)+(d+2)\kappa \mathcal{I}_{(d+3)/2}(2\kappa)}}\nonumber\\
    &=\frac{d\kappa^{(d+1)/2}}{2^{d+2}\pi^{(d+1)/2}\mathcal{I}_{(d-1)/2}(\kappa)^2} \lrc{2d\mathcal{I}_{(d+1)/2}(2\kappa)+(d+2)\kappa \mathcal{I}_{(d+3)/2}(2\kappa)}\label{eq:connect}\\
    &=\frac{d\kappa}{4}\lrc{2d+(d+2)\kappa \frac{\mathcal{I}_{(d+3)/2}(2\kappa)}{\mathcal{I}_{(d+1)/2}(2\kappa)}}\times\frac{\kappa^{(d-1)/2} \mathcal{I}_{(d+1)/2}(2\kappa)}{2^{d}\pi^{(d+1)/2}\mathcal{I}_{(d-1)/2}(\kappa)^2}\nonumber\\
    &= \frac{d^2\kappa}{2}\, \frac{\mathcal{I}_{(d+1)/2}(2\kappa)}{\mathcal{I}_{(d-1)/2}(2\kappa)} \lrc{1+\frac{(d+2)\kappa}{2d} \frac{\mathcal{I}_{(d-1)/2}(2\kappa)-((d+1)/(2\kappa)) \mathcal{I}_{(d+1)/2}(2\kappa)}{\mathcal{I}_{(d+1)/2}(2\kappa)}}\times R_0(\kappa)\nonumber\\
    &= \frac{d\kappa}{8}\, \lrc{2(2+d)\kappa-(d^2-d+2)\frac{\mathcal{I}_{(d+1)/2}(2\kappa)}{\mathcal{I}_{(d-1)/2}(2\kappa)}}\times R_0(\kappa).\label{eq:R2}
\end{align}

Note that the term $R_2(\kappa)$ as given in \eqref{eq:connect} coincides with $d^2 R(\Psi(f_{\mathrm{vMF}}(\cdot ;\bmu, \kappa), \cdot))$ from the proof of Proposition 2 in \cite{Garcia-Portugues2013a}:
\begin{align*}
    d^2 R(\Psi(f_{\mathrm{vMF}}(\cdot ;\bmu, \kappa), \cdot))&=d^2 \frac{\kappa^{(d+1)/2}}{2^{d+2} \pi^{(d+1)/2} \mathcal{I}_{(d-1)/2}(\kappa)^2 d}\left[2 d \mathcal{I}_{(d+1)/2}(2 \kappa)+(d+2) \kappa \mathcal{I}_{(d+3)/2}(2 \kappa)\right].%
\end{align*}

Using \eqref{eq:R0}--\eqref{eq:R2}, it can be obtained that
\begin{align*}
    \frac{R_1(\kappa)}{R_0(\kappa)}&=-\frac{d\kappa}{2}\, \frac{\mathcal{I}_{(d+1)/2}(2\kappa)}{\mathcal{I}_{(d-1)/2}(2\kappa)}\quad\text{and}\quad\frac{R_2(\kappa)}{R_0(\kappa)}= \frac{d\kappa}{8}\, \lrc{2(2+d)\kappa-(d^2-d+2)\frac{\mathcal{I}_{(d+1)/2}(2\kappa)}{\mathcal{I}_{(d-1)/2}(2\kappa)}},
\end{align*}
and, consequently, the matrix $\bR(\bkappa)$ whose entries are specified in \eqref{eq:rkk}--\eqref{eq:rjk} can be expressed in a more straightforward form:
\begin{align*}
    \bR(\bkappa)&=\frac{1}{4}\lrb{\frac{1}{2}\mathrm{diag}(\bv(\bkappa))+(\bu(\bkappa)\bu(\bkappa)')^\circ}\times\prod_{j=1}^{r} R_0(\kappa_j),
\end{align*}
where $\bv(\bkappa)= \bd\odot\bkappa\odot\lrc{2(2+\bd)\odot\bkappa-(\bd^{\odot 2}-\bd+2)\odot\mathbfcal{I}}$ and $\bu(\bkappa)=\bd\odot\bkappa\odot\mathbfcal{I}$, being $\mathbfcal{I}=\big(\mathcal{I}_{(d_1+1)/2}(2\kappa_1)\big/\mathcal{I}_{(d_1-1)/2}(2\kappa_1),\ldots,\mathcal{I}_{(d_r+1)/2}(2\kappa_r)\big/\mathcal{I}_{(d_r-1)/2}(2\kappa_r)\big)'$, and $\bA^\circ$ stands for a matrix $\bA$ whose diagonal is set to zero.

Once $\int_{\Sr}((\bh_n^{\odot 2})'\bT\bar{f}(\bx)\bb_{\bd}(L))^2\,\sigmar(\rd\bx)$ is computed, it directly follows that
\begin{align}
    \mathrm{AMISE}[\hat{f}(\cdot;\bh_n)]=&\;(\bh_n^{\odot 2}\odot\bb_{\bd}(L))'\bR(\bkappa)(\bh_n^{\odot 2}\odot\bb_{\bd}(L))+\frac{v_{\bd}(L)}{n\rho(\bh_n)}.\label{eq:amisevmf}
\end{align}
Differentiating, it follows that $\bh_{\mathrm{AMISE}}$ is the solution to
\begin{align*}
    4 \bR(\bkappa)({\bh}_{\mathrm{AMISE}}^{\odot 2}\odot\bb_{\bd}(L))\odot {\bh}_{\mathrm{AMISE}}\odot\bb_{\bd}(L)-\frac{v_{\bd}(L)}{n\rho(\bh_{\mathrm{AMISE}})}\big(\bd\odot{\bh}_{\mathrm{AMISE}}^{\odot(-1)}\big)=\zero_r.
\end{align*}

For the special case of equal bandwidth $h_n$, \eqref{eq:amisevmf}
becomes
\begin{align*}
    \mathrm{AMISE}[\hat{f}(\cdot;h_n\one_r)]=h_n^4 \bb_{\bd}(L)'\bR(\bkappa)\bb_{\bd}(L)+\frac{v_{\bd}(L)}{nh_n^{\tilde{d}}}
\end{align*}
and hence its minimization is the same as \eqref{eq:amise1d_proof}.
\end{proof}

\begin{proof}[Proof of Proposition \ref{prop:LSCV}] For a general kernel $L$, it follows that
\begin{align*}
\mathrm{LSCV}(\bh)=&\;\int_{\Sr}\hat{f}(\bx;\bh)^2 \,\sigmar (\rd\bx)-2n^{-1}\sum_{i=1}^n\hat{f}_{-i}(\bX_i;\bh)\\
    =&\;\frac{1}{n^2}\int_{\Sr}\Bigg[\sum_{i=1}^n L^2_{\bh}(\bx,\bX_i)+2\sum_{i=1}^n\sum_{j>i}^n L_{\bh}(\bx,\bX_i)L_{\bh}(\bx,\bX_j)\Bigg] \,\sigmar (\rd\bx)\\
    &-\frac{4}{n(n-1)}\sum_{i=1}^n\sum_{j>i}^n L_{\bh}(\bX_i,\bX_j).%
\end{align*}

In the particular case in which the kernel is $L_\mathrm{vMF}^P$,
\begin{align*}
\mathrm{LSCV}(\bh)
 =&\;\frac{c_{\bd}^{\mathrm{vMF}}\big(\bh^{\odot(-2)}\big)^2}{n^2}\int_{\Sr}\Bigg[\sum_{i=1}^n\lrc{\prod_{\ell=1}^r \exp\{2\bx_{\ell}'\bX_{i\ell}/h_\ell^2\}}\\
    &+2\sum_{i=1}^n\sum_{j>i}^n \Bigg[\prod_{\ell=1}^r\exp\{\bx_{\ell}'\bX_{i\ell}/h_\ell^2+\bx_{\ell}'\bX_{j\ell}/h_\ell^2\}\Bigg] \,\sigmar (\rd\bx)\\
    &-\frac{4 c_{\bd}^{\mathrm{vMF}}\big(\bh^{\odot(-2)}\big)}{n(n-1)}\sum_{i=1}^n\sum_{j>i}^n \Bigg[\prod_{\ell=1}^r \exp\{\bX_{i\ell}'\bX_{j\ell}/h_\ell^2\}\Bigg]\\
    =&\;\frac{c_{\bd}^{\mathrm{vMF}}\big(\bh^{\odot(-2)}\big)^2}{n^2}\lrc{\frac{n}{ c_{\bd}^{\mathrm{vMF}}\big(2 \bh^{\odot(-2)}\big)}+2\sum_{i=1}^n\sum_{j>i}^n\frac{1}{ c_{\bd}^{\mathrm{vMF}}\big(\|\bX_i+\bX_j\|  \bh^{\odot(-2)}\big)}}\\
    &-\frac{4 c_{\bd}^{\mathrm{vMF}}\big(\bh^{\odot(-2)}\big)}{n(n-1)}\sum_{i=1}^n\sum_{j>i}^n \Bigg[\prod_{\ell=1}^r \exp\{\bX_{i\ell}'\bX_{j\ell}/h_\ell^2\}\Bigg]
\end{align*}
using Proposition 6 in \cite{Garcia-Portugues2013b} for the second equality.
\end{proof}

\begin{proof}[Proof of Proposition \ref{prop:hmin}]
For each $i=1,\ldots,n$, $\hat{f}_{-i}(\bX_i;\bh)>0$ if and only if
\begin{align}
    \text{there exists at least one }j\neq i\text{ such that }1-\bX_{ik}'\bX_{jk}<h_{k}^2\text{ for all }k=1,\ldots,r \label{eq:iff}
\end{align}
(i.e., the $j$th leave-one-out kernel evaluated at $\bX_i$ does not vanish in any $k$th sphere). When $\bh=h\one_r$, \eqref{eq:iff} is equivalent to $\min_{j\neq i} \max_{k=1,\ldots,r}\{1-\bX_{ik}'\bX_{jk}\}<h^2$. Since it is required that $\hat{f}_{-i}(\bX_i;\bh)>0$ for all $i=1,\ldots,n$ to yield $\mathrm{LCV}(\bh)>-\infty$, then the previous condition is equivalent to $h_{\min,\mathrm{Epa}}^2=\max_{i=1,\ldots,n}\min_{j\neq i}\max_{k=1,\ldots,r}\{1-\bX_{ik}'\bX_{jk}\}< h^2$.

For a general bandwidth $\bh$, $\bh>h_{\min,\mathrm{Epa}}\one_r$ componentwisely implies that \eqref{eq:iff} is satisfied for all $i=1,\ldots,n$.
\end{proof}

%-------------------------------%
\section{Numerical experiments}
\label{sec:num}
%-------------------------------%

%-------------------------------%
\subsection{Asymptotic normality validity}
%-------------------------------%

This section is devoted to a numerical experiment that illustrates and empirically validates Theorems \ref{thm:biasvar}--\ref{thm:distr}. Specifically, the objective of the experiment is to verify that, for a fixed $\bx\in\Sr$ and a sequence of increasing sample sizes $n$, the asymptotic distributions in \eqref{eq:asymp1} and \eqref{eq:asymp2} are satisfied.

We consider $r=2$ and $d=d_1=d_2=2$, so the sample $\bX_1,\ldots,\bX_n$ lies on the polysphere $(\mathbb{S}^2)^2$. The data generating model is the PvMF distribution given in \eqref{eq:pmvf} with mean vector $\bmu=(1,0,0,1,0,0)'$ and concentration vector $\bkappa=(5,5)'$. In this setting, we generate samples of sizes $n=2^\ell$, $\ell=7,8,\ldots,17$ (i.e., ranging from $n=128$ to $n=131,\!072$). For these samples, we compute $\hat{f}(\bx;h_{n,\delta}\one_r )$ for $\bx=\bmu$ and the product kernel $L_\mathrm{vMF}^P(\bs)= L_\mathrm{vMF}(s_1)L_\mathrm{vMF}(s_2)$. We use sequences of bandwidths $h_n$ based on \eqref{eq:hamise} and satisfying assumption \ref{A3}:
\begin{align*}
    h_{n,\delta}\defin C\times n^{-1/(dr+4+\delta)},\quad C=\lrc{\frac{v^r_d(L_{\mathrm{vMF}})}{b^{2}_d(L_{\mathrm{vMF}})}\times\frac{dr}{4R\lrpbig{\nabla^2 \bar{f}_{\mathrm{PvMF}}(\cdot;\bmu,\bkappa)}}}^{1/(dr+4)},
\end{align*}
with $\delta=-2,-1,0,1,2,4$. Hence, the specification $\delta=0$ yields the AMISE-optimal rate, while $\delta<0$ or $\delta>0$ give sequences that decay faster or slower with respect to the AMISE order, respectively (i.e., they prioritize the reduction of bias or variance, respectively). Explicit expressions for the kernel moments $b_d(L_{\mathrm{vMF}})$ and $v_d(L_{\mathrm{vMF}})$, and the curvature term $R\lrpbig{\nabla^2 \bar{f}_{\mathrm{PvMF}}(\cdot;\bmu,\bkappa)}$ are given in subsequent Propositions \ref{prop:moments} and \ref{prop:plugin}, respectively.

\begin{figure}[t!]
    \centering
    \begin{subfigure}{0.33\textwidth}
        \centering
        \includegraphics[width=\textwidth]{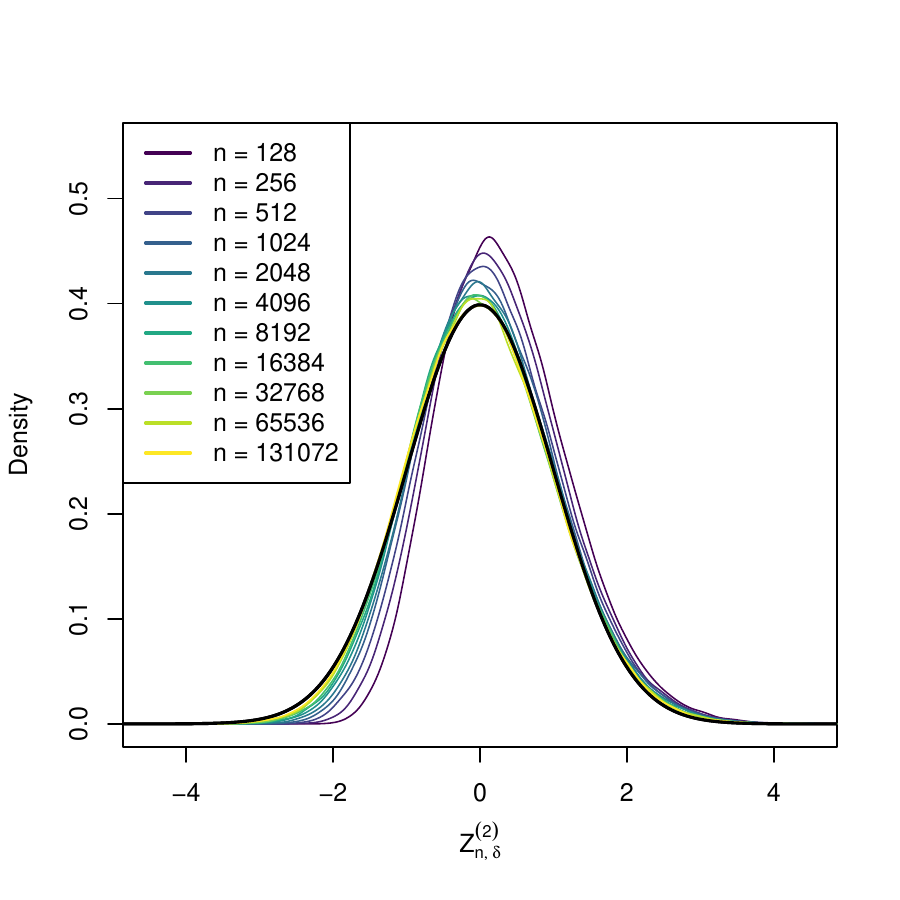}%
        \caption{\small $\delta=-2$}
    \end{subfigure}%
    \begin{subfigure}{0.33\textwidth}
        \centering
        \includegraphics[width=\textwidth]{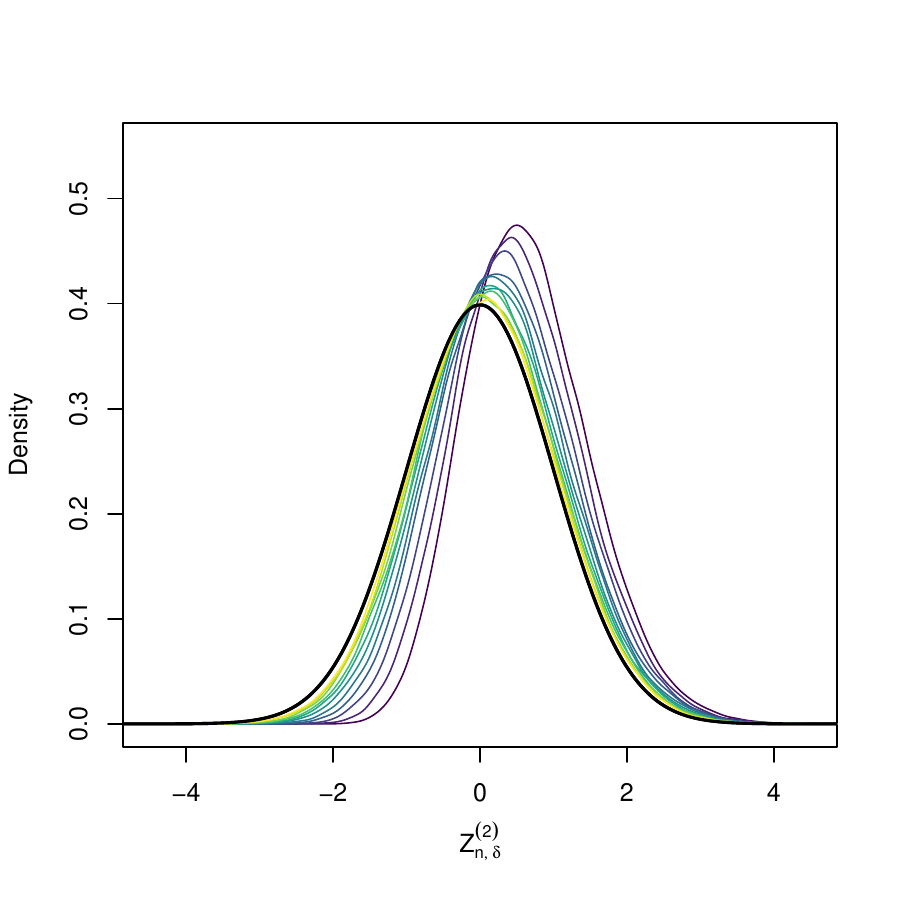}%
        \caption{\small $\delta=-1$}
    \end{subfigure}%
    \begin{subfigure}{0.33\textwidth}
        \centering
        \includegraphics[width=\textwidth]{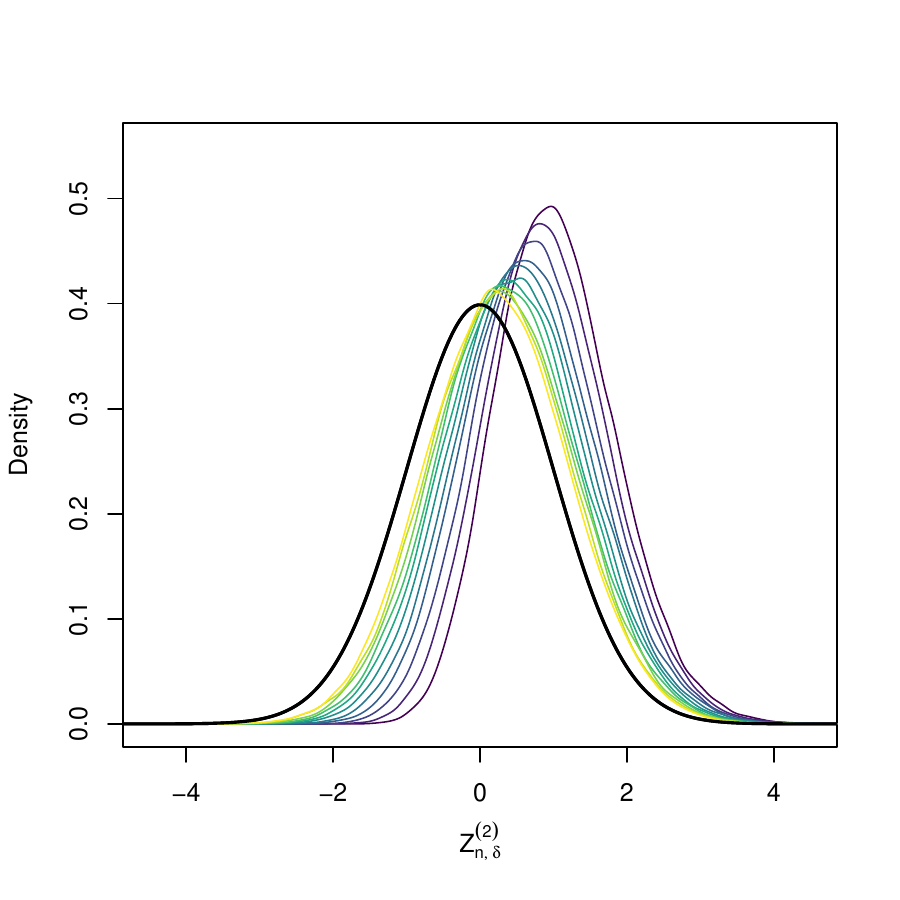}%
        \caption{\small $\delta=0$}
    \end{subfigure}\\
    \begin{subfigure}{0.33\textwidth}
        \centering
        \includegraphics[width=\textwidth]{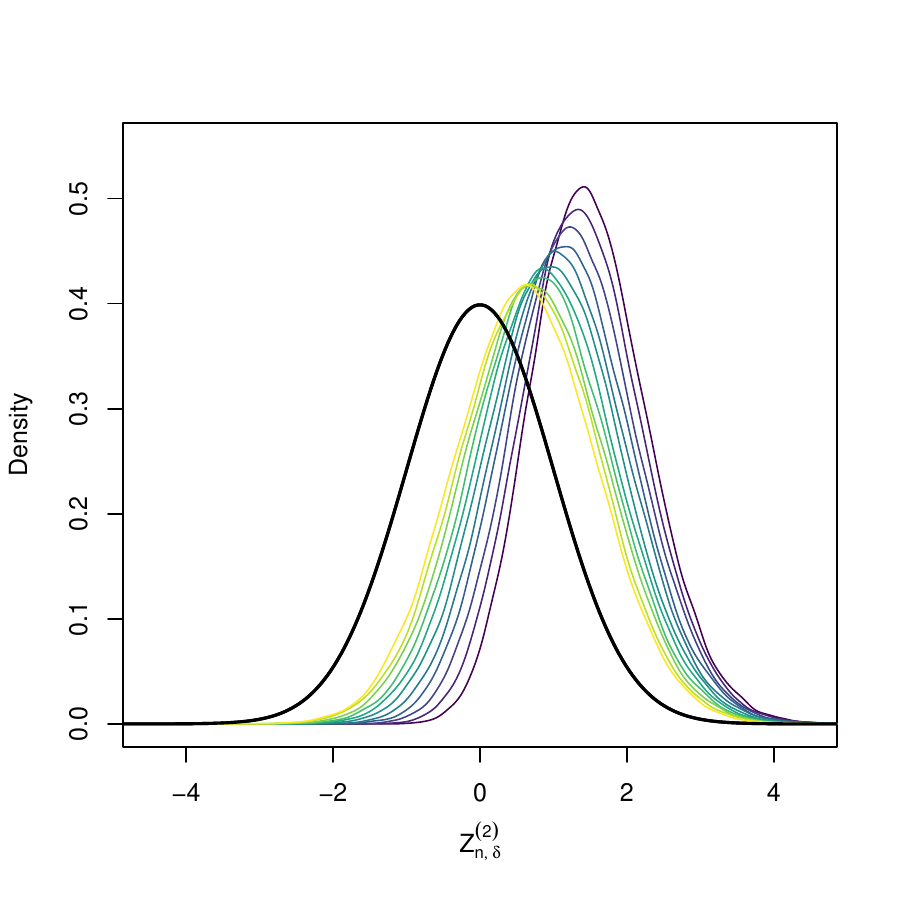}%
        \caption{\small $\delta=1$}
    \end{subfigure}%
    \begin{subfigure}{0.33\textwidth}
        \centering
        \includegraphics[width=\textwidth]{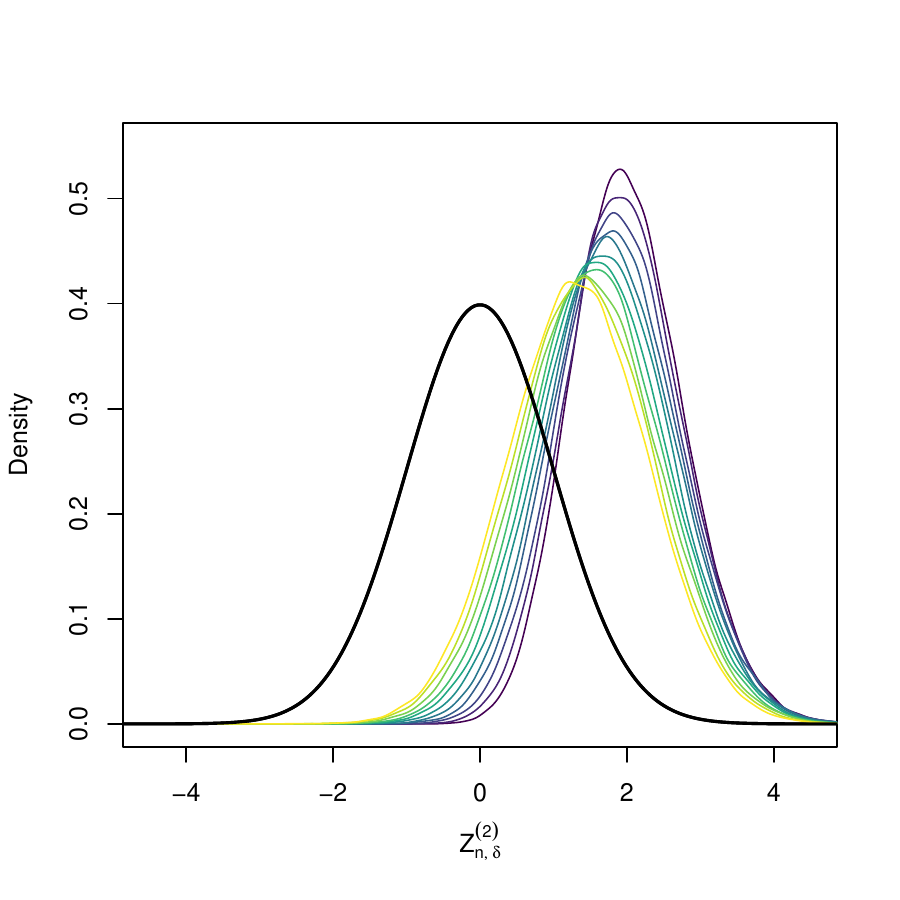}%
        \caption{\small $\delta=2$}
    \end{subfigure}%
    \begin{subfigure}{0.33\textwidth}
        \centering
        \includegraphics[width=\textwidth]{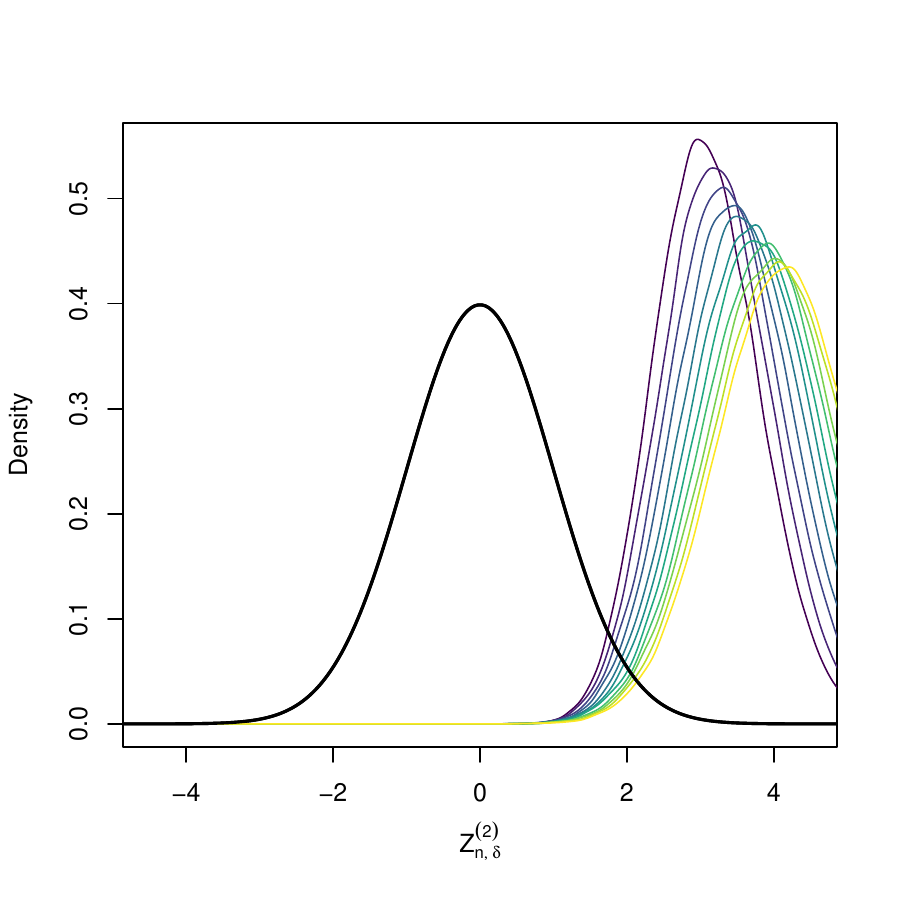}%
        \caption{\small $\delta=4$}
    \end{subfigure}
    \caption{\small Density of $\mathcal{N}(0,1)$ (black curves) and kdes of $M=10^5$ observations of $\smash{Z_{n,\delta}^{(2)}}$, for $n=2^\ell$, $\ell=7,8,\ldots,17$ (curves in color gradient from dark violet to yellow) and $\delta=-2,-1,0,1,2,4$.}
    \label{fig:zn2}
\end{figure}

For each combination of $(n,\delta)$, $M=10^5$ samples were computed from the statistics
\begin{align*}
    \smash{Z_{n,\delta}^{(1)}}\defin&\;\sqrt{\frac{n h_{n,\delta}^{dr}}{v_{d}^r(L)f(\bx)}}\lrp{\hat{f}(\bx;h_{n,\delta} \one_r)-\mathbb{E}[\hat{f}(\bx;h_{n,\delta} \one_r)]},\\
    \smash{Z_{n,\delta}^{(2)}}\defin&\;\sqrt{\frac{n h_{n,\delta}^{dr}}{v_{d}^r(L)f(\bx)}}\lrp{\hat{f}(\bx;h_{n,\delta} \one_r)-f(\bx)-b_{d}(L)\nabla^2 \bar{f}(\bx)h_{n,\delta}^2}
\end{align*}
that, according to Theorem \ref{thm:distr}, are asymptotically $\mathcal{N}(0,1)$-distributed, the latter if $nh_{n,\delta}^{dr+4}=O(1)$. In the current setting, this restriction implies that the asymptotic standard normality of $\smash{Z_{n,\delta}^{(2)}}$ is only guaranteed for $\delta\leq0$. The expectation in $\smash{Z_{n,\delta}^{(1)}}$ was estimated empirically.

\begin{figure}[htpb!]
    \centering
    \begin{subfigure}{0.33\textwidth}
        \centering
        \includegraphics[width=\textwidth,clip,trim={0cm 0.5cm 0cm 1.5cm}]{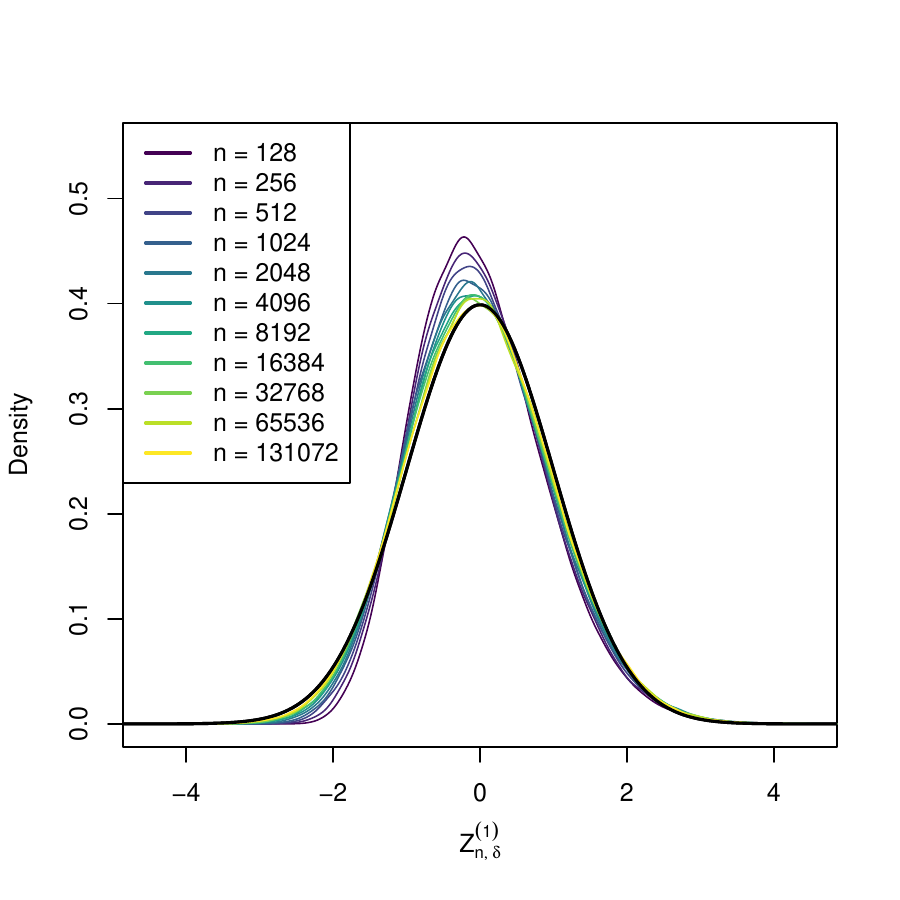}%
        \caption{\small $\delta=-2$}
    \end{subfigure}%
    \begin{subfigure}{0.33\textwidth}
        \centering
        \includegraphics[width=\textwidth,clip,trim={0cm 0.5cm 0cm 1.5cm}]{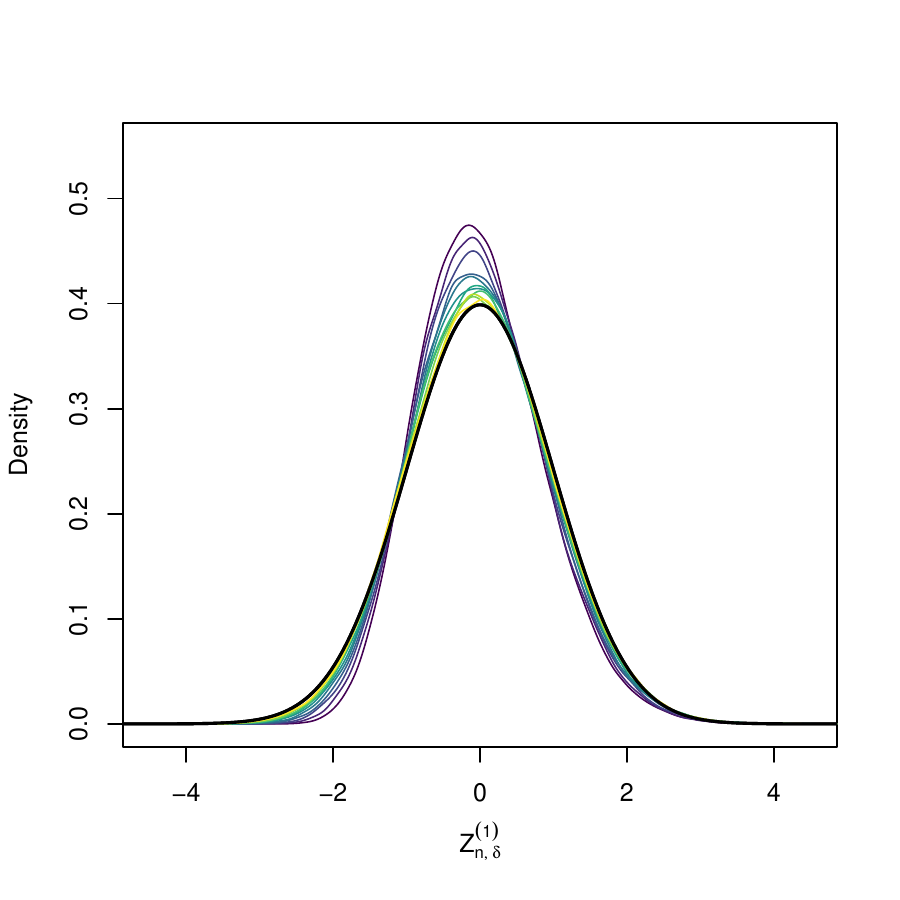}%
        \caption{\small $\delta=-1$}
    \end{subfigure}%
    \begin{subfigure}{0.33\textwidth}
        \centering
        \includegraphics[width=\textwidth,clip,trim={0cm 0.5cm 0cm 1.5cm}]{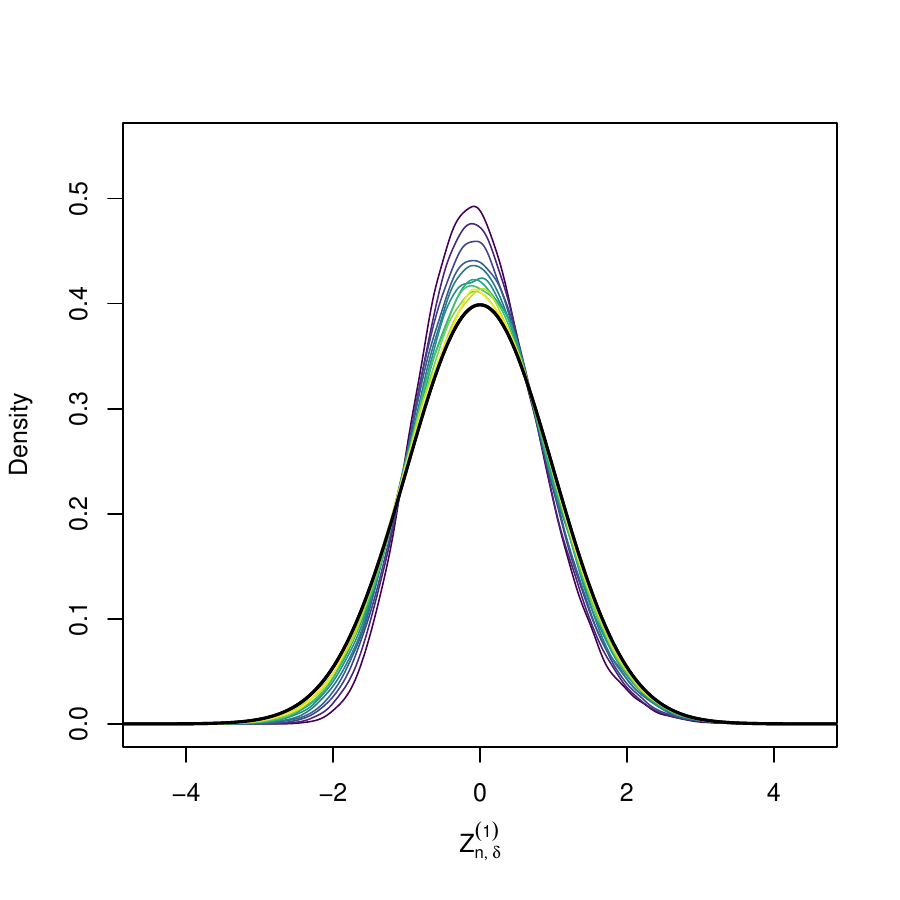}%
        \caption{\small $\delta=0$}
    \end{subfigure}\\
    \begin{subfigure}{0.33\textwidth}
        \centering
        \includegraphics[width=\textwidth,clip,trim={0cm 0.5cm 0cm 1.5cm}]{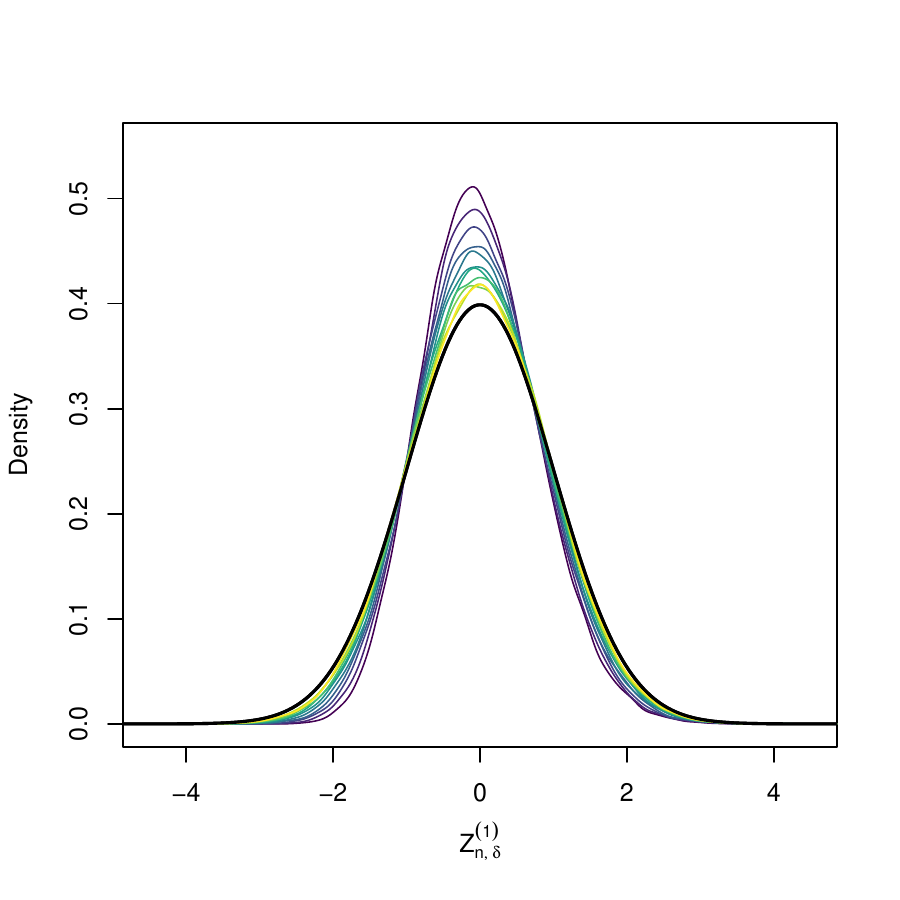}%
        \caption{\small $\delta=1$}
    \end{subfigure}%
    \begin{subfigure}{0.33\textwidth}
        \centering
        \includegraphics[width=\textwidth,clip,trim={0cm 0.5cm 0cm 1.5cm}]{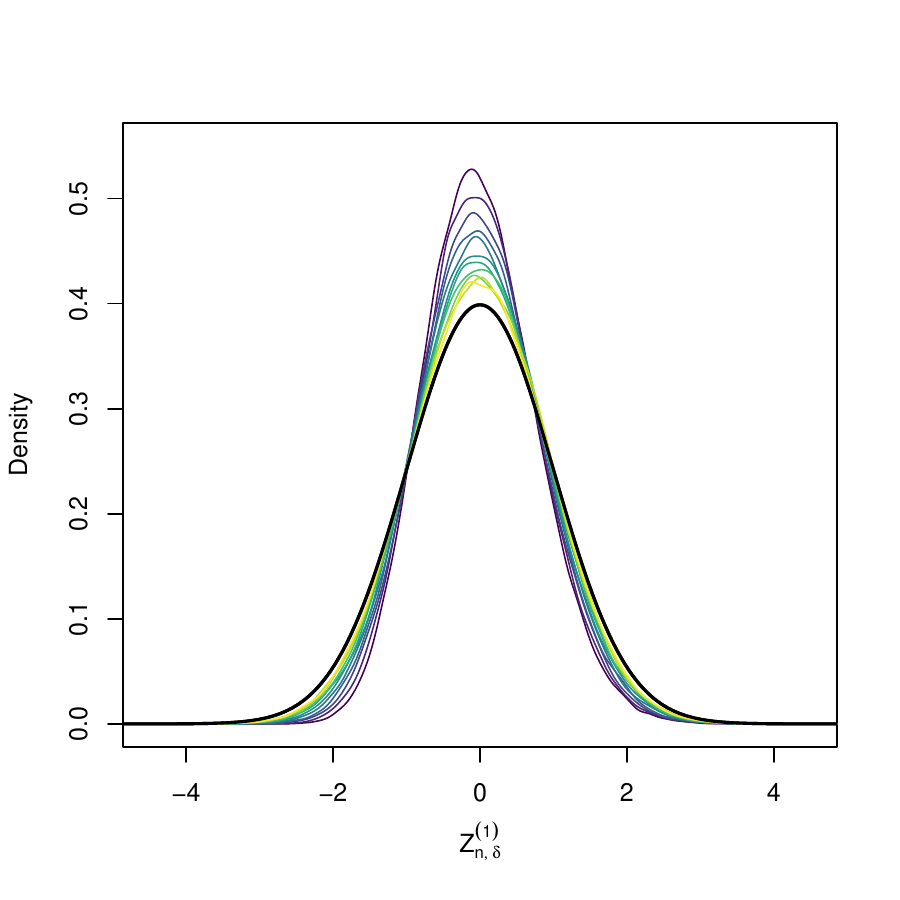}%
        \caption{\small $\delta=2$}
    \end{subfigure}%
        \begin{subfigure}{0.33\textwidth}
        \centering
        \includegraphics[width=\textwidth,clip,trim={0cm 0.5cm 0cm 1.5cm}]{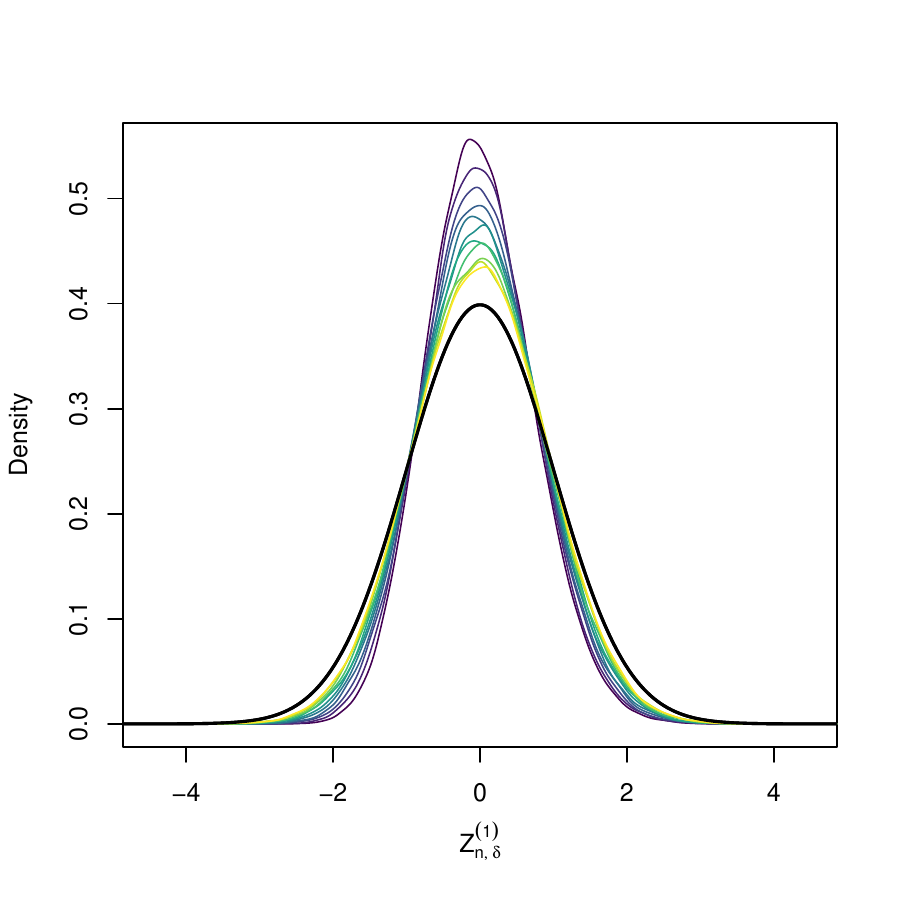}%
        \caption{\small $\delta=4$}
    \end{subfigure}
    \caption{\small Density of $\mathcal{N}(0,1)$ (black curves) and kdes of $M=10^5$ observations of $\smash{Z_{n,\delta}^{(1)}}$, for $n=2^\ell$, $\ell=7,8,\ldots,17$ (curves in color gradient from dark violet to yellow) and $\delta=-2,-1,0,1,2,4$.}
    \label{fig:zn1}
\end{figure}

Figure \ref{fig:zn2} shows the evolution of the standard kdes for $M$-size samples of $\smash{Z_{n,\delta}^{(2)}}$, with increasing values of $n$ and $\delta$. The rule-of-thumb bandwidth selector was used in all kdes. Several interesting facts are revealed from this display. First, when the sample size $n$ increases, the convergence of the empirical densities of $\smash{Z_{n,\delta}^{(2)}}$ toward the standard normal density is evidenced in all cases, except for $\delta=4$. The speed of this convergence is not homogeneous: bandwidth sequences with $\delta<0$ boost this convergence by minimizing the bias, while those with $\delta\geq0$ slow the convergence (observe the slow attraction at $\delta=2$), eventually reaching a divergence for $\delta=4$. In this divergent case, $\smash{h_n=C\times n^{-1/(dr+8)}}$, and hence the restriction of bandwidths stated in Remark \ref{rem:orders_norm2} is not satisfied. The convergence in cases $\delta=1,2$ also reveals that the remaining order in \eqref{eq:estimator:bias} is actually smaller than $o(h_n^2)$ (the PvMF pdf satisfies \ref{A1*}) and therefore the restriction $nh_{n,\delta}^{dr+4}=O(1)$ in Theorem \ref{thm:distr} can be weakened. The corresponding plots for $\smash{Z_{n,\delta}^{(1)}}$ are in Figure \ref{fig:zn1}, where it can be seen that convergence occurs for any of the bandwidth rates explored. Both graphics corroborate a common folklore on kernel smoothing: the effect of smoothing is more delicate on the bias than on the variance.

Further insight into the asymptotic convergence of $\smash{Z_{n,\delta}^{(1)}}$ and $\smash{Z_{n,\delta}^{(2)}}$ can be extracted from Figure \ref{fig:zn2evol}. It shows the evolution of several summaries of the empirical distributions of $\smash{Z_{n,\delta}^{(1)}}$ and $\smash{Z_{n,\delta}^{(2)}}$ as a function of $n$. On the left, it displays the evolution of the expectation of $\smash{Z_{n,\delta}^{(2)}}$, convergent to zero for $\delta<4$ and divergent for $\delta=4$. In the center, it shows the evolution of the standard deviation of $\smash{Z_{n,\delta}^{(1)}}$ and $\smash{Z_{n,\delta}^{(1)}}$ towards one, happening for all $\delta$'s. Lastly, on the right, it plots the Kolmogorov--Smirnov statistics for testing $\smash{Z_{n,\delta}^{(1)}},\smash{Z_{n,\delta}^{(2)}}\sim\mathcal{N}(0,1)$, with much faster convergence to zero for $\smash{Z_{n,\delta}^{(1)}}$ and varying convergence rate for ${Z_{n,\delta}^{(2)}}$, again with $\delta=4$ showing divergence.

\begin{figure}[htpb!]
    \centering
    \includegraphics[width=0.33\textwidth,clip,trim={0cm 0.5cm 0cm 1.5cm}]{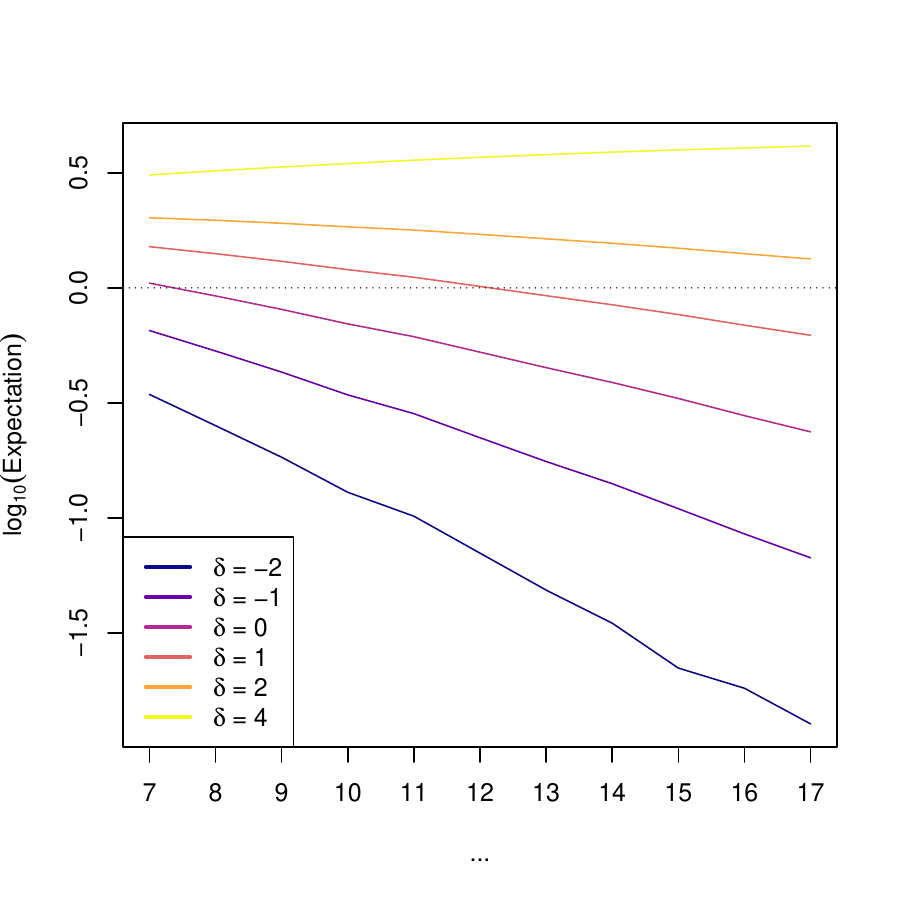}%
    \includegraphics[width=0.33\textwidth,clip,trim={0cm 0.5cm 0cm 1.5cm}]{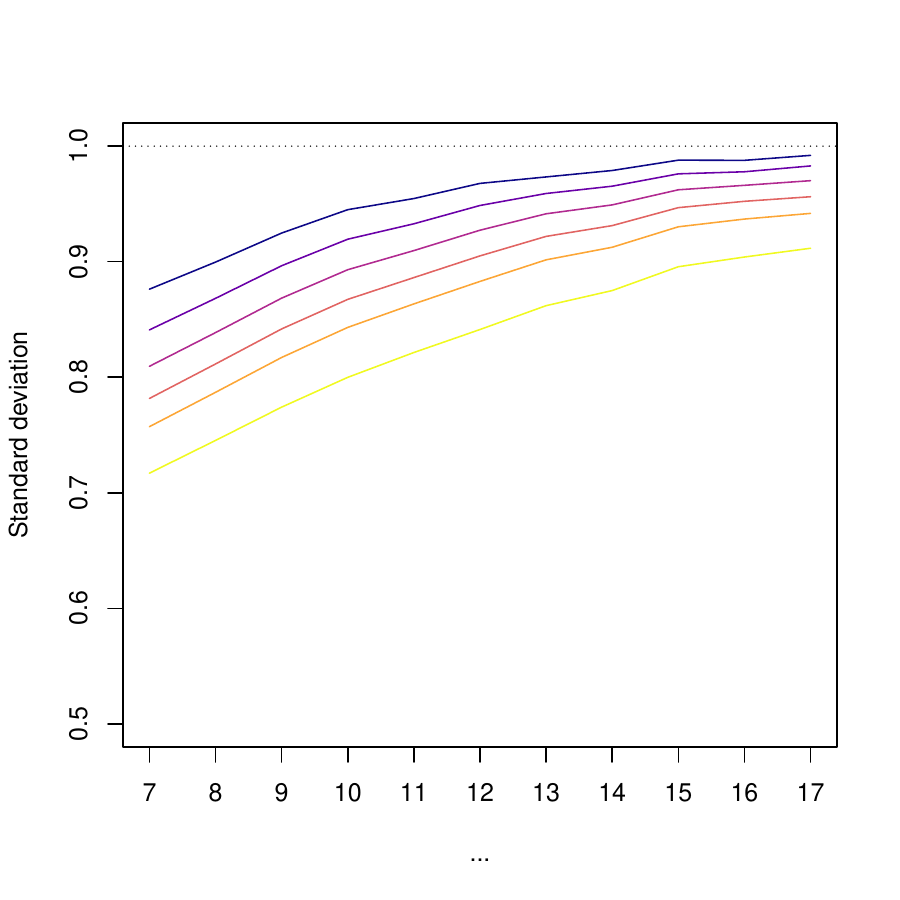}%
    \includegraphics[width=0.33\textwidth,clip,trim={0cm 0.5cm 0cm 1.5cm}]{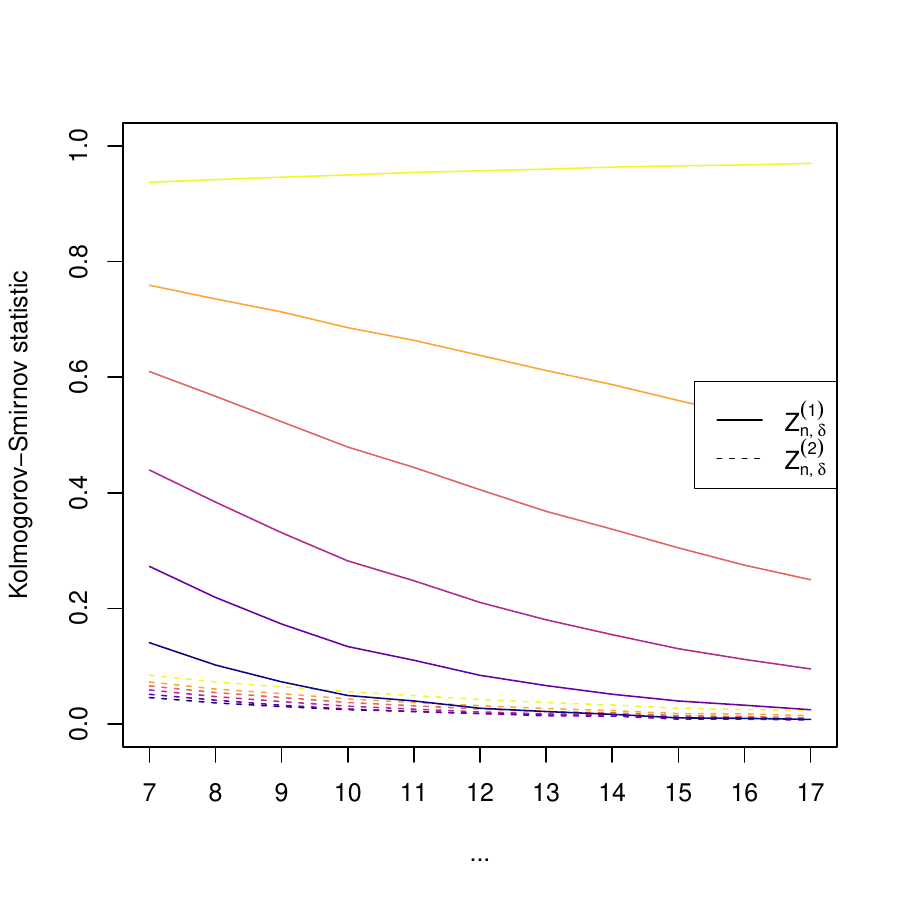}
    \caption{\small Evolution of several summaries for the $\smash{Z_{n,\delta}^{(1)}}$ and $\smash{Z_{n,\delta}^{(2)}}$, for $\delta=-2,-1,0,1,2,4$ as a function of $n=2^\ell$, $\ell=7,8,\ldots,17$. Left: estimated $\log_{10}$-expectation of $\smash{Z_{n,\delta}^{(2)}}$. Center: estimated standard deviation of $\smash{Z_{n,\delta}^{(1)}}$ and $\smash{Z_{n,\delta}^{(2)}}$ (both equal). Right: Kolmogorov--Smirnov statistics measuring the distance of the empirical distributions of $\smash{Z_{n,\delta}^{(1)}}$ and $\smash{Z_{n,\delta}^{(2)}}$ with respect to $\mathcal{N}(0,1)$.}
    \label{fig:zn2evol}
\end{figure}

Finally, Figure \ref{fig:biasvar} corroborates the accuracy of the asymptotic expansions \eqref{eq:estimator:bias}--\eqref{eq:estimator:variance}. For large $n$, it can be seen that the asymptotic expansions become closer to the exact forms, with several insights. The matches between asymptotic and exact forms for expectation and bias (first two leftmost plots) strongly depend on $\delta$, even for large $n$'s: the difference for $\delta=-2$ and $\delta=2$ is notable. For the standard deviation (rightmost plot), the accuracy of the matches depends much less on $\delta$, and the accuracy for large $n$'s is approximately uniform.

\begin{figure}[htpb!]
    \centering
    \includegraphics[width=0.33\textwidth,clip,trim={0cm 0.5cm 0cm 1.5cm}]{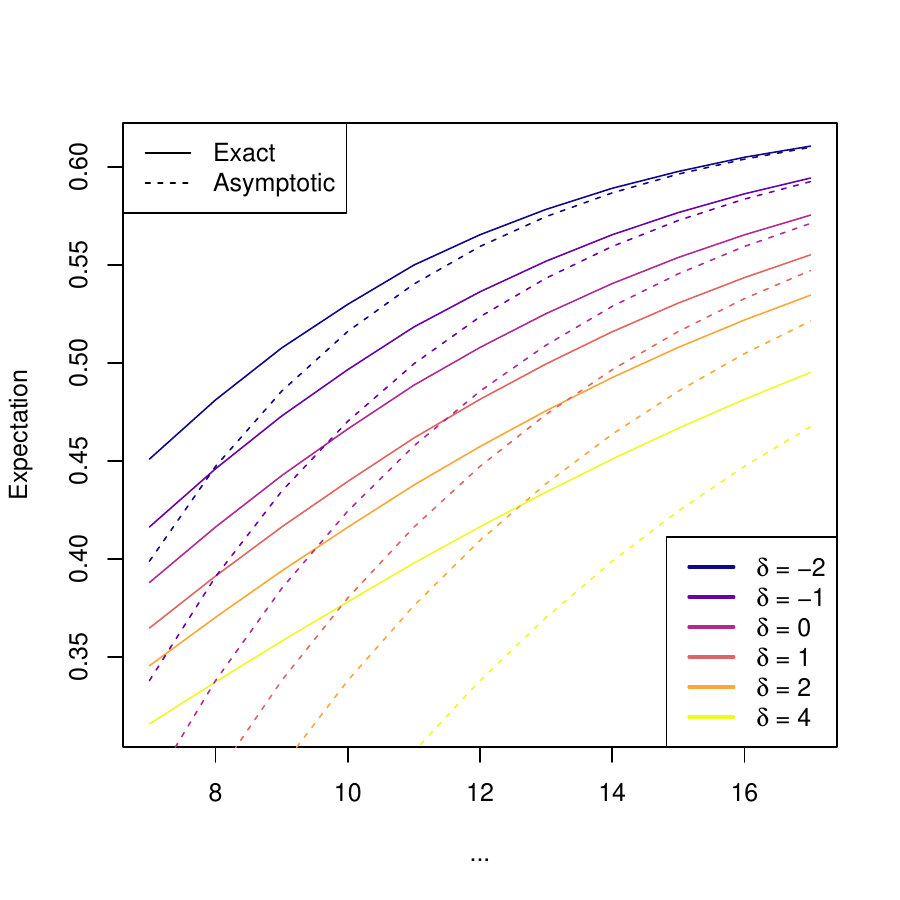}%
    \includegraphics[width=0.33\textwidth,clip,trim={0cm 0.5cm 0cm 1.5cm}]{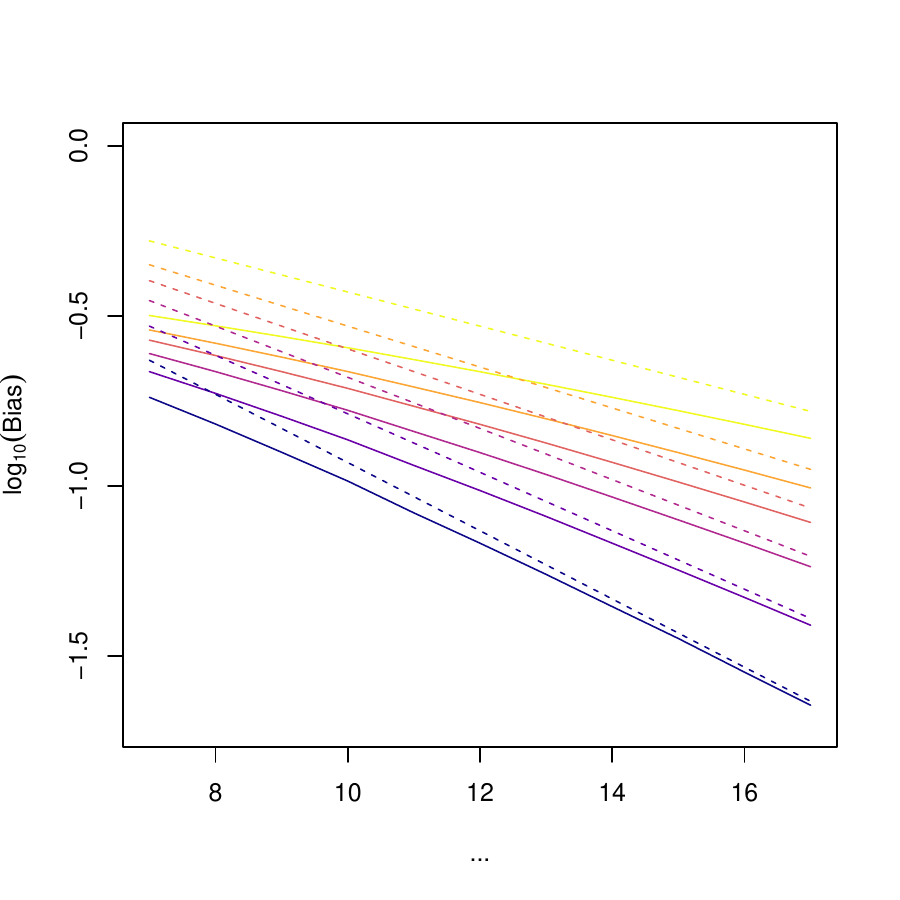}%
    \includegraphics[width=0.33\textwidth,clip,trim={0cm 0.5cm 0cm 1.5cm}]{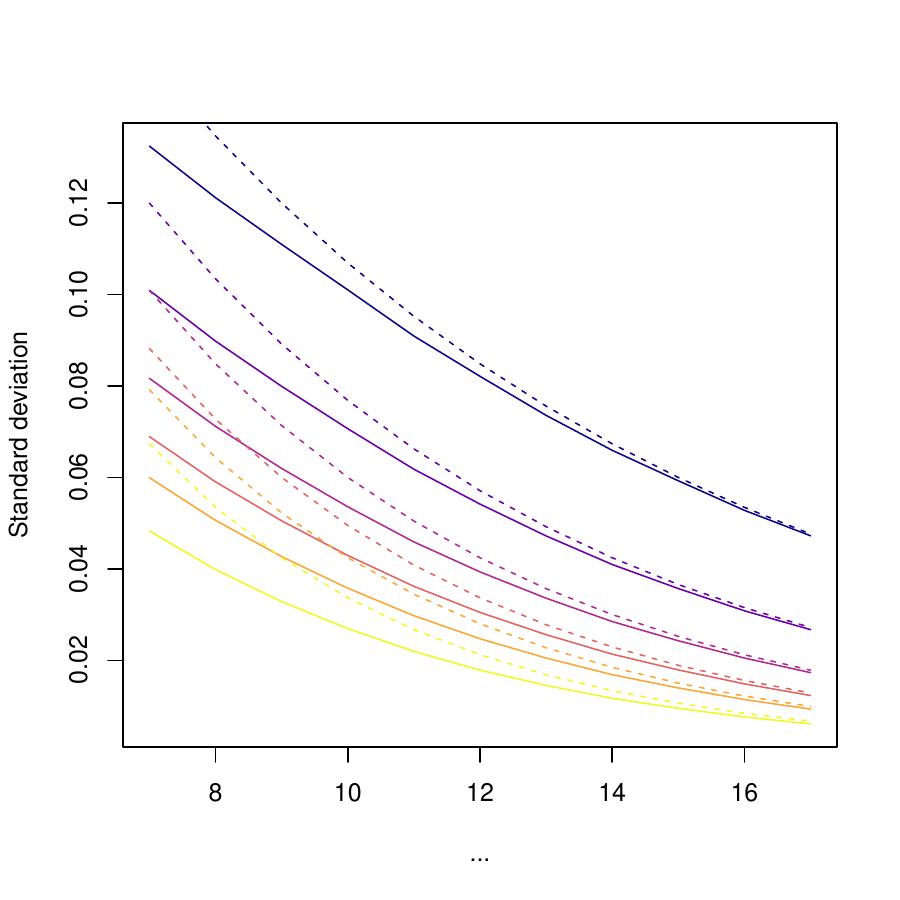}
    \caption{\small Exact and asymptotic versions of the expectation, bias, and standard deviation, for $\delta=-2,-1,0,1,2,4$ as a function of $n=2^\ell$, $\ell=7,8,\ldots,17$. Left: (estimated) $\mathbb{E}[\hat{f}(\bx;h_{n,\delta} \one_r)]$ vs. $f(\bx)+b_{d}(L)\nabla^2 \bar{f}(\bx)h_{n,\delta}^2$. Center: $\log_{10}(f(\bx)-\mathbb{E}[\hat{f}(\bx;h_{n,\delta} \one_r)])$ vs. $\log_{10}(-b_{d}(L)\nabla^2 \bar{f}(\bx)h_{n,\delta}^2)$. Right: (estimated) $\{\mathbb{V}\mathrm{ar}[\hat{f}(\bx;h_{n,\delta} \one_r)]\}^{1/2}$ vs. $\{(v_d^r(L)f(\bx))/(nh_{n,\delta})^{dr}\}^{1/2}$.}
    \label{fig:biasvar}
\end{figure}

The whole experiment was replicated with $(r,d)\in\{(1,1),(1,2),(2,1)\}$ and the Epa product kernel. Analogous results, yielding the same qualitative conclusions, were obtained.

%-------------------------------%
\subsection{Kernel efficiency}
%-------------------------------%

The efficiencies of the vMF, sfp, and Epa kernels are listed in Table \ref{tab:effic} for $r,d=1,2,3,5,10$. Several comments are in order. First, when the dimension $dr$ increases, the efficiencies of all the kernels drop, as expected. This drop is the same on $(r,d)$ for spherically symmetric kernels. Second, the $\mathrm{Epa}^{P}$ kernel dominates in efficiency the vMF kernel, for all $(d,r)$, and notably for large dimensions. Third, the sfp kernels also improve the efficiency of the vMF kernel, at least for $\upsilon\geq1$, for both its spherically symmetric and product versions. Fourth, the efficiency of the $\mathrm{sfp}^{S}_{\upsilon}$ (respectively, $\mathrm{sfp}^{P}_{\upsilon}$) kernel converges to that of the $\mathrm{Epa}^{S}$ ($\mathrm{Epa}^{P}$) kernel as $\upsilon\to\infty$, with a slower convergence as $dr$ increases. Fifth, for low or intermediate $\upsilon$'s ($\upsilon=1,10$), the product version $\mathrm{sfp}^{P}_{\upsilon}$ is more efficient than the spherically symmetric version $\mathrm{sfp}^{S}_{\upsilon}$, i.e., the second factor in \eqref{eq:eff:sfpprod} is larger than one. This phenomenon can be partly attributed to the hybrid nature of $\mathrm{sfp}^{S}_{\upsilon}$ for $r,d>1$ (the kernel is spherically symmetric on each component $\Sd$).

\begin{table}[htpb!]
    \centering
    \begin{tabular}{ll|rrrr|rrrr}
    \toprule
    $r$ & $d$ & vMF & $\mathrm{sfp}^{S}_{1}$ & $\mathrm{sfp}^{S}_{10}$ & $\mathrm{sfp}^{S}_{100}$ & $\mathrm{Epa}^{P}$ & $\mathrm{sfp}^{P}_{1}$ & $\mathrm{sfp}^{P}_{10}$ & $\mathrm{sfp}^{P}_{100}$ \\
    \midrule
    1 & 1 & 95.12 & 97.20 & 99.86 & 100.00 & 100.00 & 97.20 & 99.86 & 100.00\\
     & 2 & 88.89 & 93.05 & 99.58 & 100.00 & 100.00 & 93.05 & 99.58 & 100.00\\
     & 3 & 82.03 & 87.85 & 99.13 & 100.00 & 100.00 & 87.85 & 99.13 & 100.00\\
     & 5 & 68.08 & 75.40 & 97.59 & 100.00 & 100.00 & 75.40 & 97.59 & 100.00\\
     & 10 & 39.17 & 43.77 & 88.92 & 99.98 & 100.00 & 43.77 & 88.92 & 99.98\\
    \midrule
    2 & 1 & 88.89 & 93.05 & 99.58 & 100.00 & 98.24 & 92.82 & 97.97 & 98.24\\
     & 2 & 75.00 & 81.87 & 98.47 & 100.00 & 94.92 & 82.19 & 94.13 & 94.92\\
     & 3 & 61.44 & 68.72 & 96.45 & 99.99 & 91.30 & 70.46 & 89.72 & 91.30\\
     & 5 & 39.17 & 43.77 & 88.92 & 99.98 & 84.51 & 48.05 & 80.49 & 84.50\\
     & 10 & 11.00 & 11.38 & 48.39 & 99.86 & 71.72 & 13.74 & 56.71 & 71.69\\
    \midrule
    3 & 1 & 82.03 & 87.85 & 99.13 & 100.00 & 95.32 & 87.54 & 94.92 & 95.32\\
     & 2 & 61.44 & 68.72 & 96.45 & 99.99 & 87.48 & 70.49 & 86.38 & 87.48\\
     & 3 & 44.04 & 49.46 & 91.29 & 99.98 & 79.78 & 54.09 & 77.71 & 79.78\\
     & 5 & 21.12 & 22.64 & 71.89 & 99.93 & 66.94 & 28.70 & 62.22 & 66.93\\
     & 10 & 2.82 & 2.83 & 10.82 & 99.56 & 46.90 & 3.93 & 32.97 & 46.86\\
    \midrule
    5 & 1 & 68.08 & 75.40 & 97.59 & 100.00 & 87.43 & 75.86 & 86.82 & 87.43\\
     & 2 & 39.17 & 43.77 & 88.92 & 99.98 & 70.58 & 49.24 & 69.11 & 70.58\\
     & 3 & 21.12 & 22.64 & 71.89 & 99.93 & 56.86 & 29.75 & 54.42 & 56.86\\
     & 5 & 5.61 & 5.69 & 25.70 & 99.74 & 38.36 & 9.35 & 33.96 & 38.35\\
     & 10 & 0.16 & 0.16 & 0.21 & 98.11 & 17.88 & 0.29 & 9.94 & 17.86\\
    \midrule
    10 & 1 & 39.17 & 43.77 & 88.92 & 99.98 & 64.60 & 48.63 & 63.71 & 64.60\\
     & 2 & 11.00 & 11.38 & 48.39 & 99.86 & 35.73 & 17.39 & 34.26 & 35.73\\
     & 3 & 2.82 & 2.83 & 10.82 & 99.56 & 20.42 & 5.59 & 18.71 & 20.42\\
     & 5 & 0.16 & 0.16 & 0.21 & 98.11 & 7.71 & 0.46 & 6.04 & 7.70\\
     & 10 & 0.00 & 0.00 & 0.00 & 82.94 & 1.25 & 0.00 & 0.39 & 1.25\\
    \bottomrule
    \end{tabular}\caption{\small Efficiencies of the vMF, $\mathrm{sfp}_{\upsilon}^S$, $\mathrm{Epa}^P$, and $\mathrm{sfp}_{\upsilon}^P$ kernels with respect to the $\mathrm{Epa}^S$ kernel. The $100\times \mathrm{eff}_{d,r}(L)$ efficiencies are reported for $d,r=1,2,3,5,10$ and $\upsilon=1,10,100$.}
    \label{tab:effic}
\end{table}

%-------------------------------%
\subsection{\texorpdfstring{$k$-sample test validity and relevance}{k-sample test validity and relevance}}
%-------------------------------%

To verify by simulations the adequate performance of the $k$-sample test, we propose three experiments featuring different Data Generating Processes (DGPs) for each sample. In these DGPs, $a\in[0,1]$ controls the strength of the deviation from the null hypothesis of homogeneity ($a=0$) to the strongest non-homogeneous case with $a=1$. Specifically, with $a=1$ each sample has a different DGP, while for $a=0$ these DGPs are completely mixed into a common one.

\begin{enumerate}[label=\textbf{E\arabic{*}},ref=\textbf{E\arabic{*}}]
    \item The setup of the first experiment is a two-sample test (that is, $k=2$) on $\mathbb{S}^2$, and $n_1=n_2=100$. The DGPs of the two samples are given by
    \begin{align}
        \bX_{1,j},\ldots,\bX_{n_j,j}\sim f_{j,a} = \frac{1 + (-1)^j a}{2} f_1 + \frac{1 - (-1)^j a}{2} f_2,\quad j=1,2, \label{eq:mixdgp}
    \end{align}
    where $f_1=\frac{1}{4} \{f_\mathrm{vMF}(\cdot;\be_1,50)+f_\mathrm{vMF}(\cdot;-\be_1,50)+f_\mathrm{vMF}(\cdot;\be_2,50)+\allowbreak f_\mathrm{vMF}(\cdot;-\be_2,50)\}$ and $f_2=f_\mathrm{SC}(\cdot;\be_3,25,0)$, with $f_\mathrm{SC}(\bx;\bmu,\eta,\tau)\propto \exp(-\eta(\bx'\bmu-\tau)^2)$ being the Small Circle pdf with $\eta\in\R$, $\tau\in[-1,1]$, and $\bx,\bmu\in\mathbb{S}^2$. The data pattern generated by $f_1$ is four equally-separated and equally-concentrated clusters on the equator, while $f_2$ generates a girdle-like sample concentrated along the equator. Figure \ref{fig:mixdgp} shows two samples from \eqref{eq:mixdgp}, illustrating the mixing process. In this experiment, the tests based on $T_{n,\mathrm{JSD}}(\bh)$, $T_{n,\mathrm{loc}}$, and $T_{n,\mathrm{sc}}$ are compared. \label{E1}

    \item The second experiment is set with $k=2$ on $(\mathbb{S}^2)^{168}$, the framework of the hippocampi data set discussed in Section \ref{sec:data}, and $n_1=n_2=45$. The design of the DGPs in \eqref{eq:mixdgp} is replicated, but with different densities. We use the ranking of hippocampi given in Section \ref{sec:inout} to define two groups: ``inward'' and ``outward'' hippocampi. These are defined as the top-$25\pct$ and bottom-$25\pct$ hippocampi in ranking \eqref{eq:rank} applied with $\hat{\bh}_\mathrm{ROT}$ and the vMF kernel. The densities $f_1$ and $f_2$ now correspond to the kdes $\hat{f}_\mathrm{inw}(\cdot;\hat{\bh}_\mathrm{ROT})$ and $\hat{f}_\mathrm{out}(\cdot;\hat{\bh}_\mathrm{ROT})$ of each group, computed with the vMF kernel (so that simulation is direct from the kdes). The experiment evaluates the performance of the test based on $T_{n,\mathrm{JSD}}(\bh)$ in a setup akin to which it will be run later.\label{E2}

    \item Finally, the third experiment considers $k=3$ on $(\mathbb{S}^{10})^2$, with $n_1=100$, $n_2=75$, and $n_3=50$. The DGPs are
    \begin{align*}
    	\bX_{1,j},\ldots,\bX_{n_j,j}&\sim f_{j,a} = \frac{1}{3}\Big\{(1 + 2a) f_\mathrm{vMF}(\cdot;\be_j,10) + (1 - a)\sum_{\substack{\ell=1\\\ell \neq j}}^3 f_\mathrm{vMF}(\cdot;\be_\ell,10)\Big\},\quad j=1,2,3.
    \end{align*}
    The experiment investigates the performance of the $k$-sample test based on \eqref{eq:TnJSD} when $k>2$. \label{E3}
\end{enumerate}

\begin{figure}
    \centering
    \begin{subfigure}{0.33\textwidth}
        \centering
        \includegraphics[width=\textwidth]{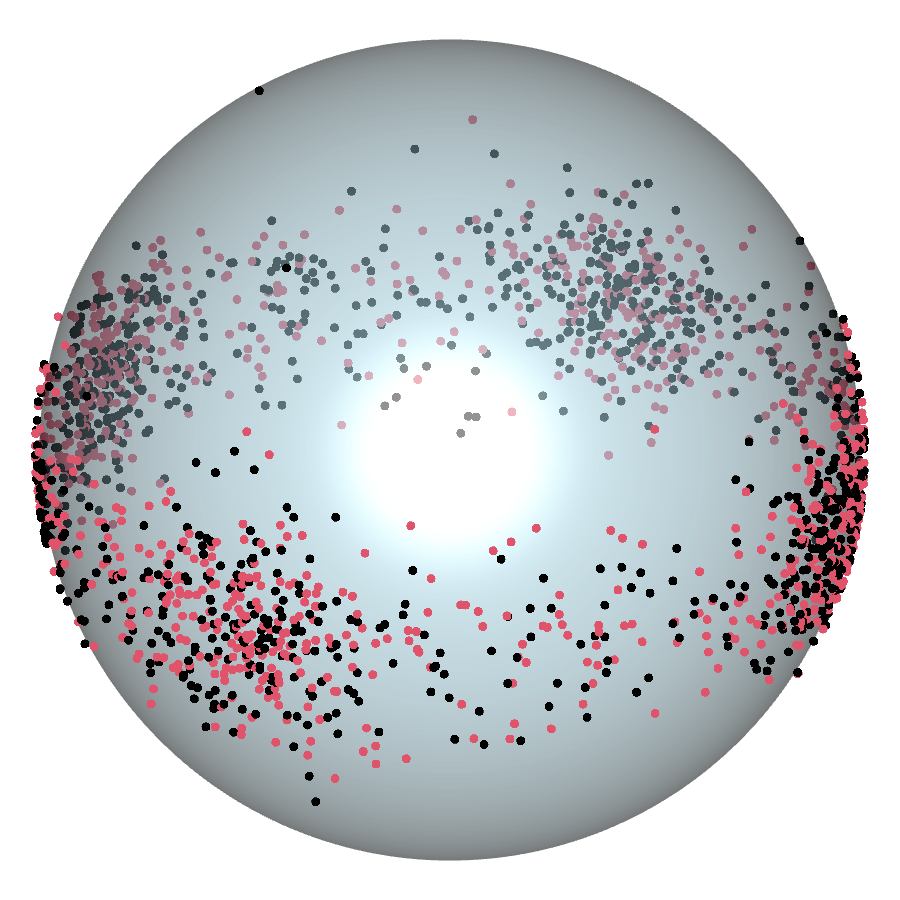}%
        \caption{\small $a=0$} \label{fig:mixdgp:a}
    \end{subfigure}%
    \begin{subfigure}{0.33\textwidth}
        \centering
        \includegraphics[width=\textwidth]{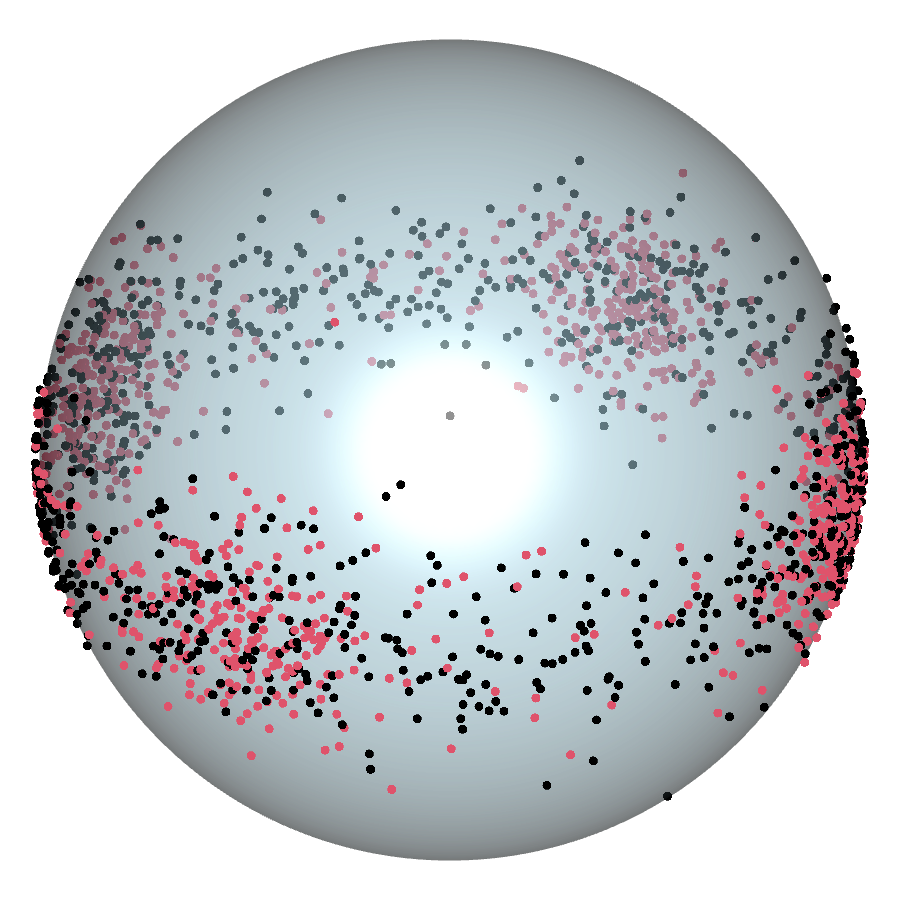}%
        \caption{\small $a=0.5$} \label{fig:mixdgp:b}
    \end{subfigure}%
    \begin{subfigure}{0.33\textwidth}
        \centering
        \includegraphics[width=\textwidth]{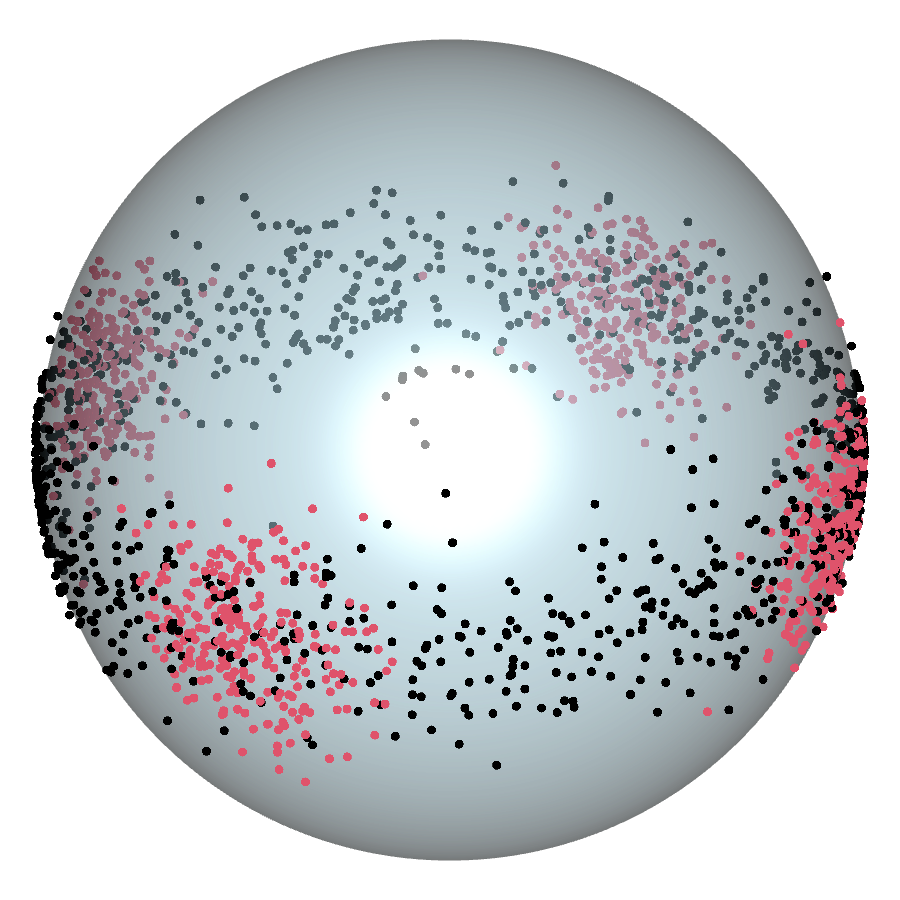}%
        \caption{\small $a=1$} \label{fig:mixdgp:c}
    \end{subfigure}%
    \caption{\small Effect of $a$ in the DGP \ref{E1}. Two samples of $n_1=1,\!000$ (red) and $n_2=1,\!000$ (black) points are sampled from the pdfs $f_{1,a}$ and $f_{2,a}$ given in \eqref{eq:mixdgp}.}
    \label{fig:mixdgp}
\end{figure}

For the three experiments, we considered $M=10^4$ Monte Carlo samples and conducted the simulations adapting the warp-speed bootstrap \citep{Giacomini2013} for hypothesis testing. We ran the test based on $T_{n,\mathrm{JSD}}(\bh)$ for a grid of bandwidths $c\times \bh_\mathrm{med}$, $c>0$, where $\bh_\mathrm{med}$ is the bandwidth vector of the marginal medians of $\hat{\bh}_\mathrm{ROT}$ for \ref{E2}--\ref{E3}, and marginal medians of $\hat{\bh}_\mathrm{LCV}$ for \ref{E1} sampled for $a=0$. The bandwidths $\hat{\bh}_\mathrm{ROT}$ and $\hat{\bh}_\mathrm{LCV}$ are computed on the pooled sample. In the three scenarios, $\bh_\mathrm{med}$ represents a sensible estimation bandwidth. The faster median ROT bandwidths are considered in \ref{E2}--\ref{E3}; they perform very similarly to the median LCV bandwidths (mean ratios close to one %
across coordinates, respectively for \ref{E2}--\ref{E3}). The situation is different in \ref{E1} (ratio: $6.34$): the non-vMFness of the underlying distributions makes the ROT selector output bandwidths that are too large. Hence, in \ref{E1} the median LCV bandwidth is more sensible for estimation. In all scenarios, the sfp kernel with $\upsilon\in\{10,100\}$ is used to approximate the Epa kernel in a smooth form. For the low-dimensional scenarios \ref{E1} and \ref{E3}, $\upsilon=10$ suffices, while in the higher-dimensional case \ref{E2}, $\upsilon=100$ is considered.

The results are given in Figures \ref{fig:E1}--\ref{fig:E3} in the form of ROC curves showing the empirical rejection rate (power, vertical axis) for varying significance level ($\alpha$, horizontal axis), and for varying $a\in\{0,0.25,0.5,0.75,1\}$ and $c=2^\ell$ with $\ell\in\{-3,-2,\ldots,5\}$. The ROC curves are on a logarithmic scale to improve visualization at lower significance levels.

\begin{figure}[htpb!]
    \centering
    \begin{subfigure}{0.25\textwidth}
        \centering
        \includegraphics[width=\textwidth,clip,trim={0.75cm 1.5cm 1cm 1cm}]{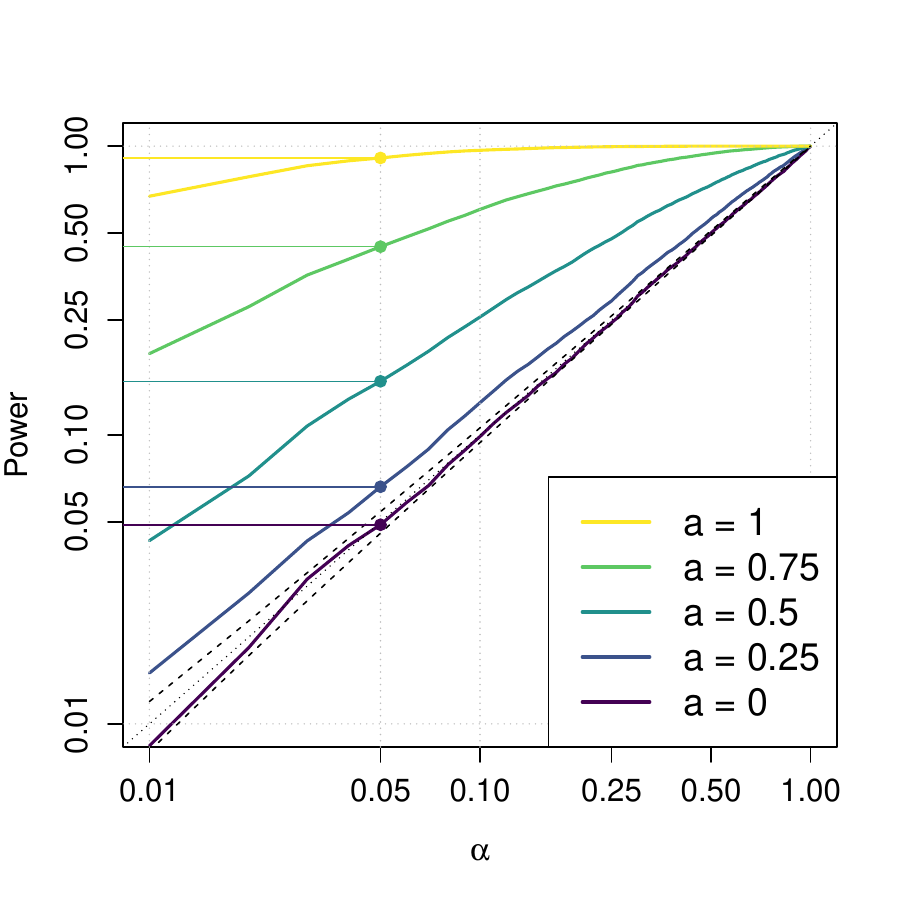}%
        \caption{\small JSD, $c=0.125$} \label{fig:E1:a0}
    \end{subfigure}%
    \begin{subfigure}{0.25\textwidth}
        \centering
        \includegraphics[width=\textwidth,clip,trim={0.75cm 1.5cm 1cm 1cm}]{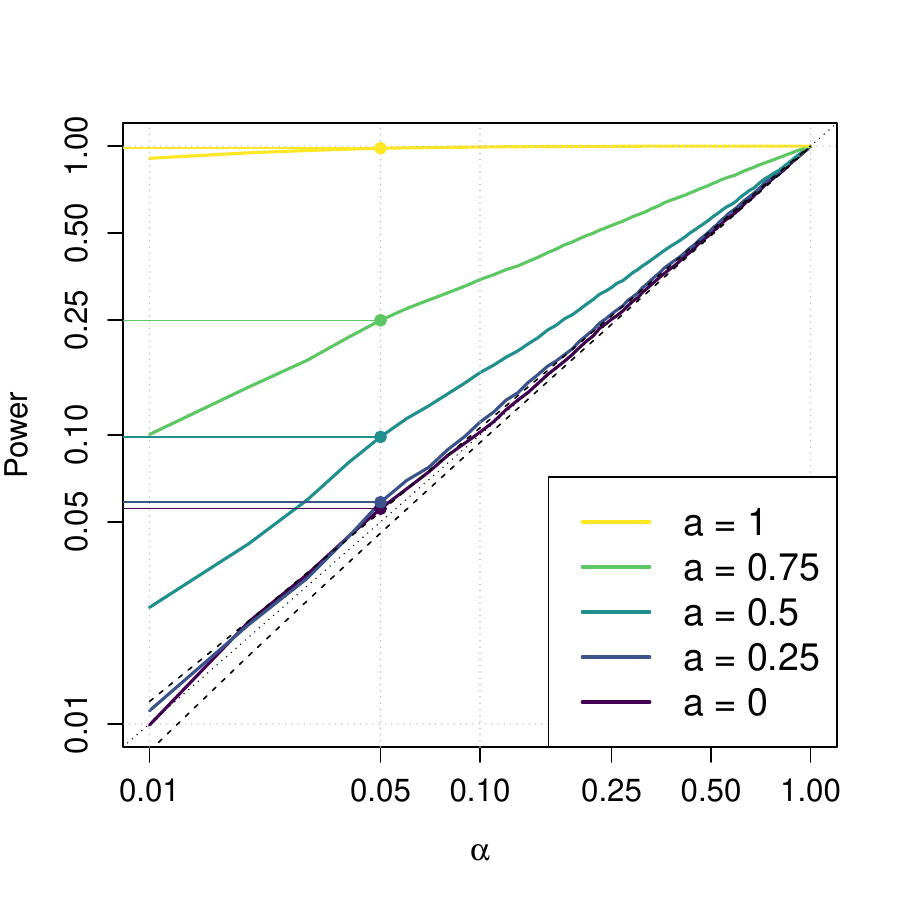}%
        \caption{\small JSD, $c=0.25$} \label{fig:E1:a}
    \end{subfigure}%
    \begin{subfigure}{0.25\textwidth}
        \centering
        \includegraphics[width=\textwidth,clip,trim={0.75cm 1.5cm 1cm 1cm}]{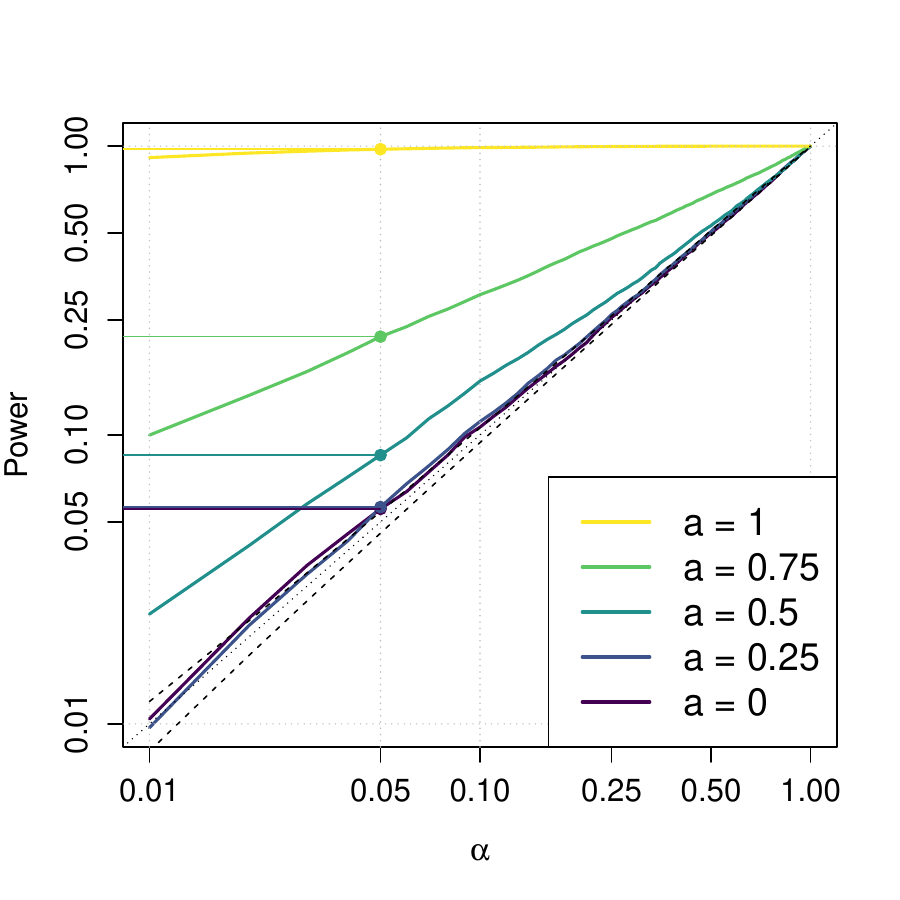}%
        \caption{\small JSD, $c=0.50$} \label{fig:E1:b}
    \end{subfigure}%
    \begin{subfigure}{0.25\textwidth}
        \centering
        \includegraphics[width=\textwidth,clip,trim={0.75cm 1.5cm 1cm 1cm}]{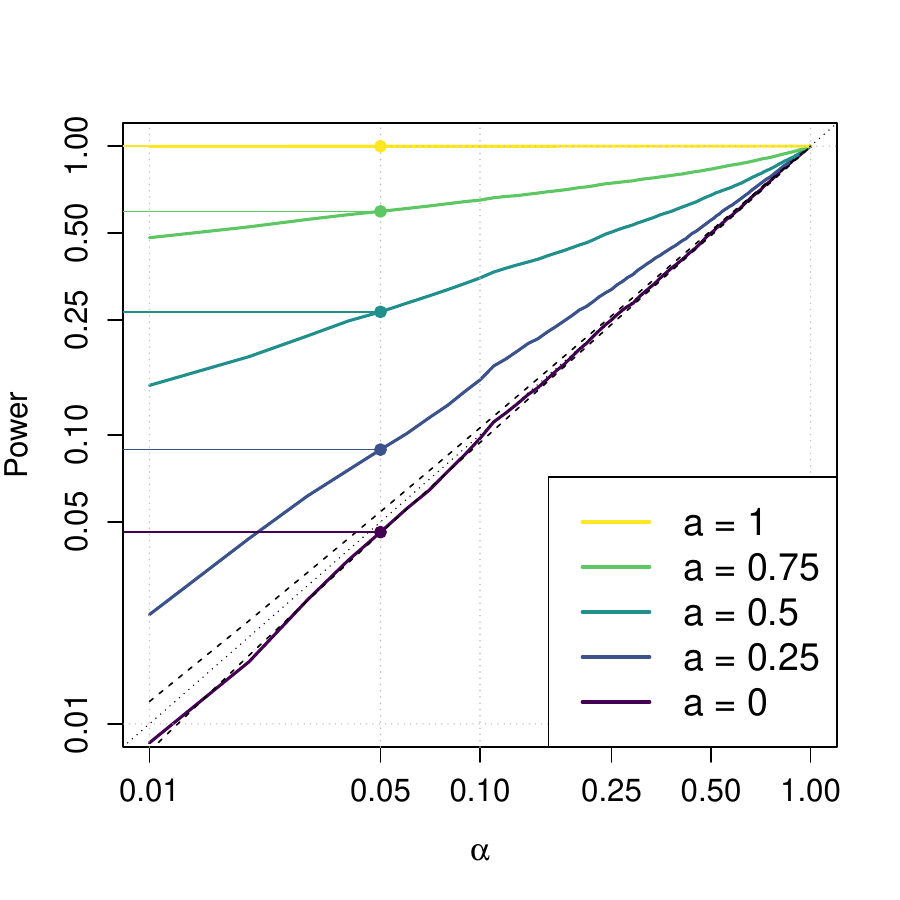}%
        \caption{\small JSD, $c=1$} \label{fig:E1:c}
    \end{subfigure}\\
    \begin{subfigure}{0.25\textwidth}
        \centering
        \includegraphics[width=\textwidth,clip,trim={0.75cm 1.5cm 1cm 1cm}]{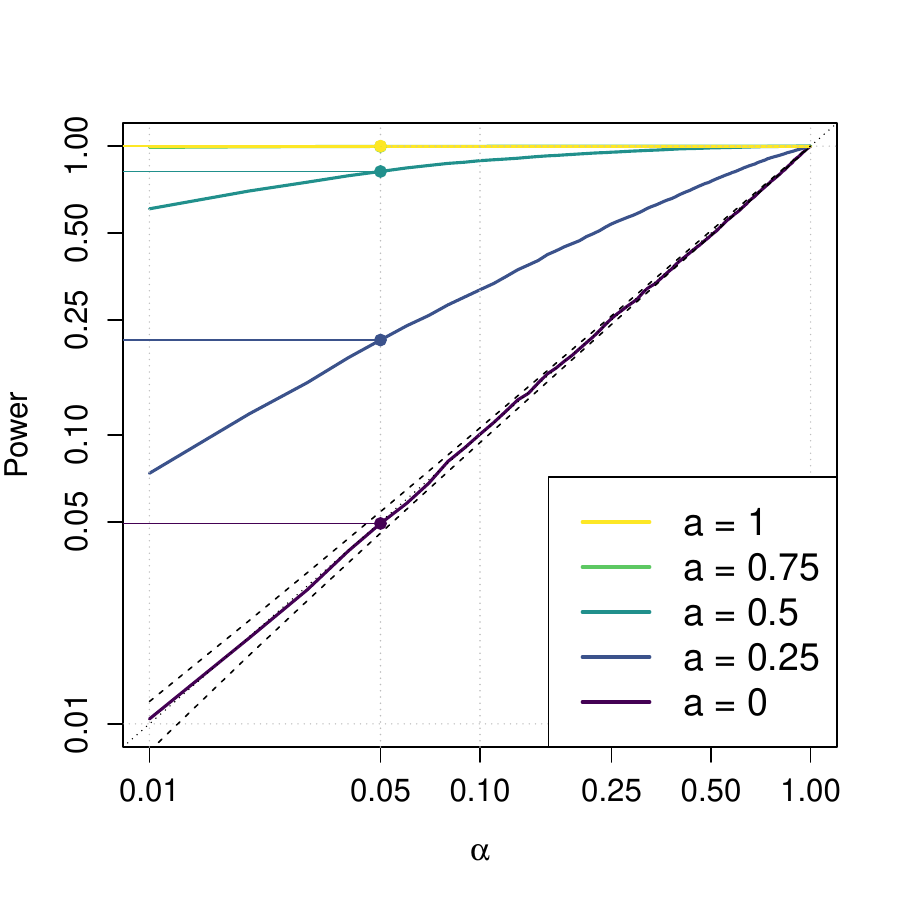}%
        \caption{\small JSD, $c=2$} \label{fig:E1:d}
    \end{subfigure}%
    \begin{subfigure}{0.25\textwidth}
        \centering
        \includegraphics[width=\textwidth,clip,trim={0.75cm 1.5cm 1cm 1cm}]{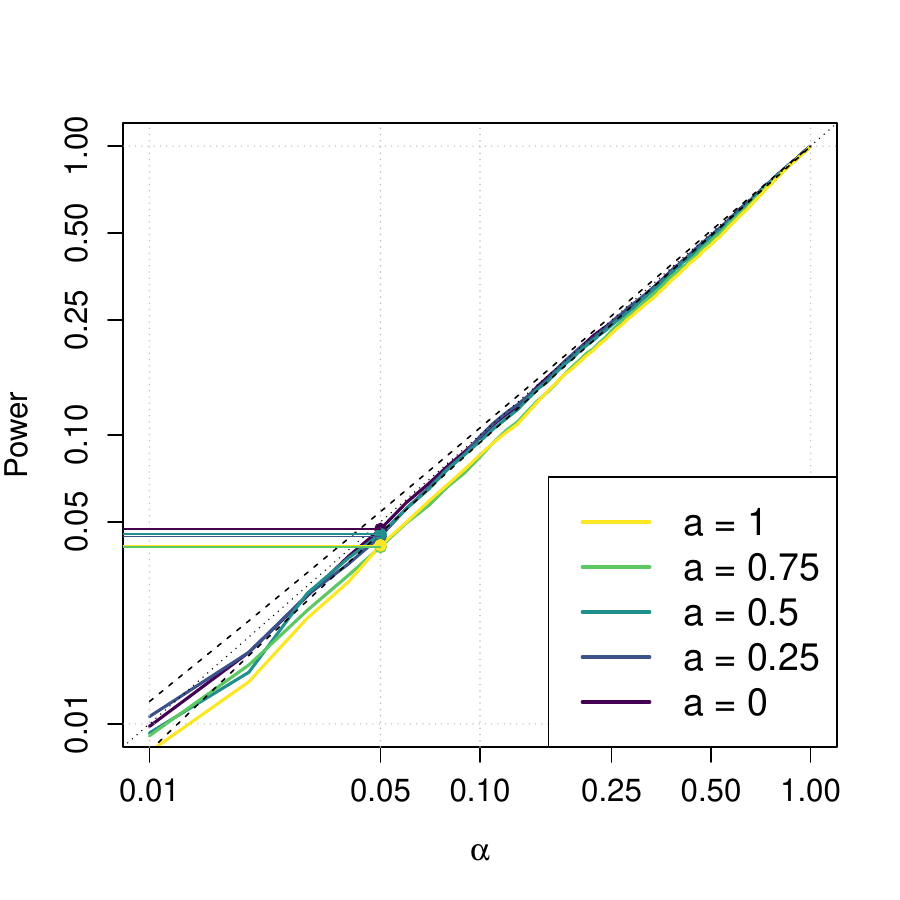}%
        \caption{\small JSD, $c=4$} \label{fig:E1:e}
    \end{subfigure}%
    \begin{subfigure}{0.25\textwidth}
        \centering
        \includegraphics[width=\textwidth,clip,trim={0.75cm 1.5cm 1cm 1cm}]{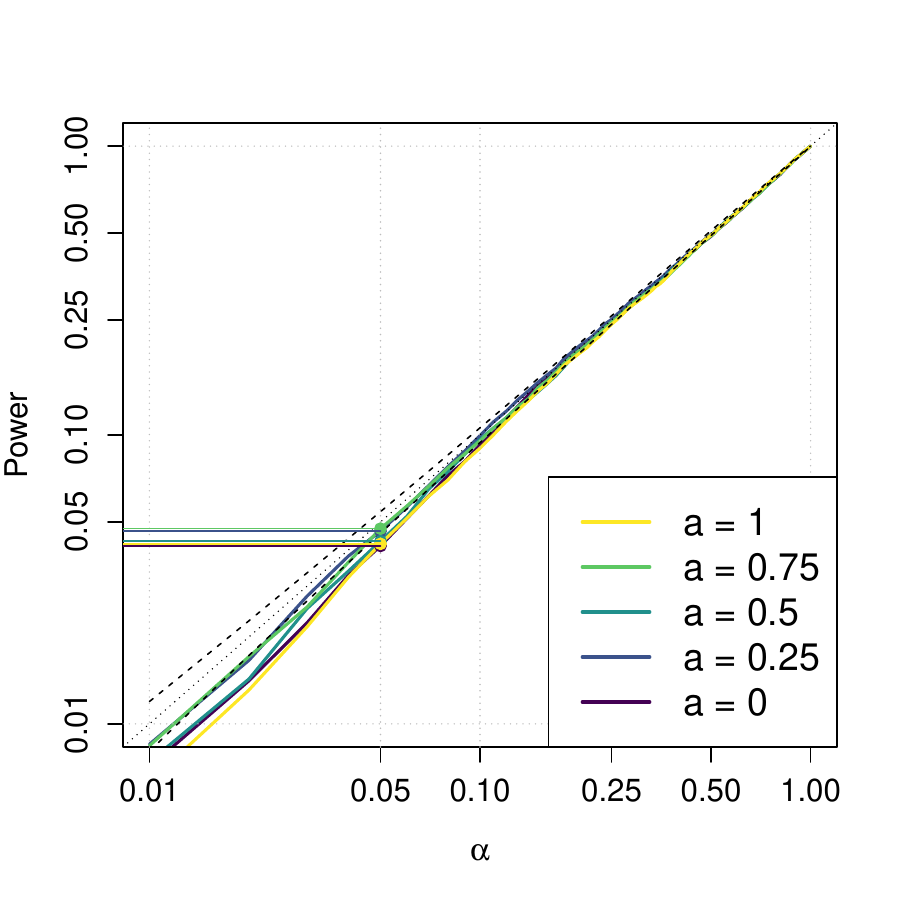}%
        \caption{\small Location} \label{fig:E1:g}
    \end{subfigure}%
    \begin{subfigure}{0.25\textwidth}
        \centering
        \includegraphics[width=\textwidth,clip,trim={0.75cm 1.5cm 1cm 1cm}]{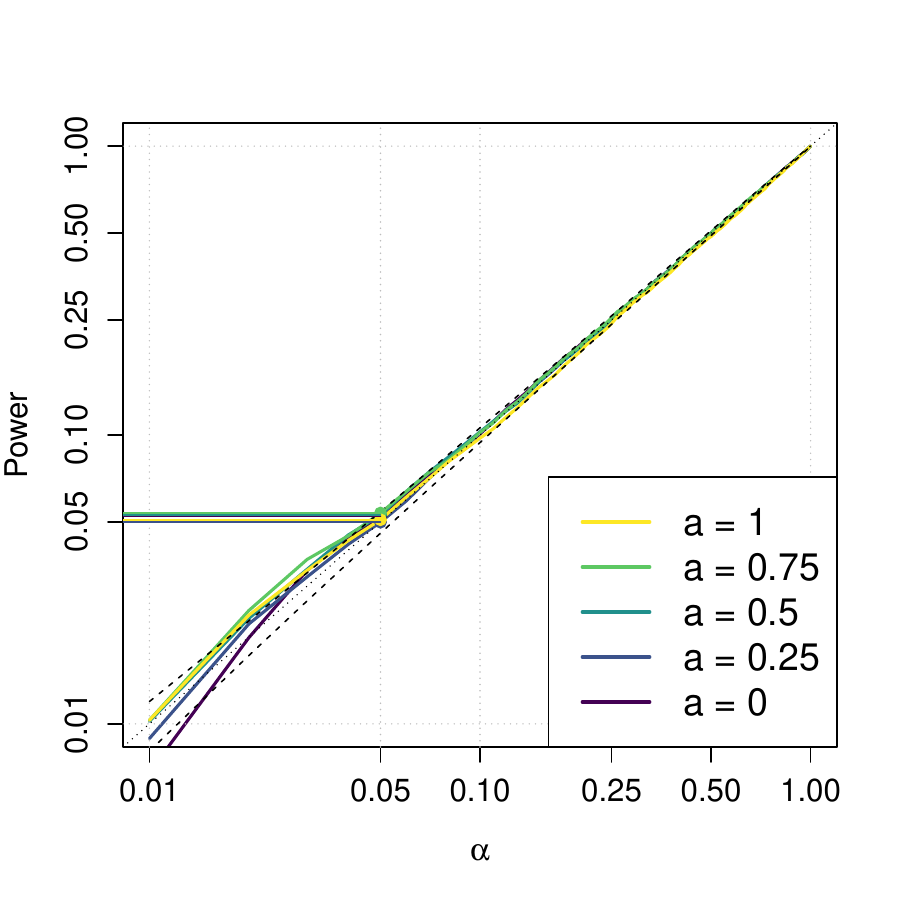}%
        \caption{\small Scatter} \label{fig:E1:h}
    \end{subfigure}
    \caption{\small ROC curves for the tests evaluated in \ref{E1}, with varying color indicating the value of $a$. The JSD, location, and scatter tests are based on the statistics $T_{n,\mathrm{JSD}}(c\times\bh_\mathrm{med})$ (for several $c$'s), $T_{n,\mathrm{loc}}$, and $T_{n,\mathrm{sc}}$, respectively. The horizontal axis gives the significance of the tests $\alpha$, while the empirical power is shown in the vertical axis. Both axes use a logarithmic scale. The colored horizontal segments signal the empirical powers achieved for $\alpha=0.05$. The black dashed lines about the dotted diagonal line indicate the asymptotic $95\pct$-confidence intervals for the significance level of the test.}
    \label{fig:E1}
\end{figure}

\begin{figure}[htpb!]
    \centering
    \begin{subfigure}{0.25\textwidth}
        \centering
        \includegraphics[width=\textwidth,clip,trim={0.75cm 1.5cm 1cm 1cm}]{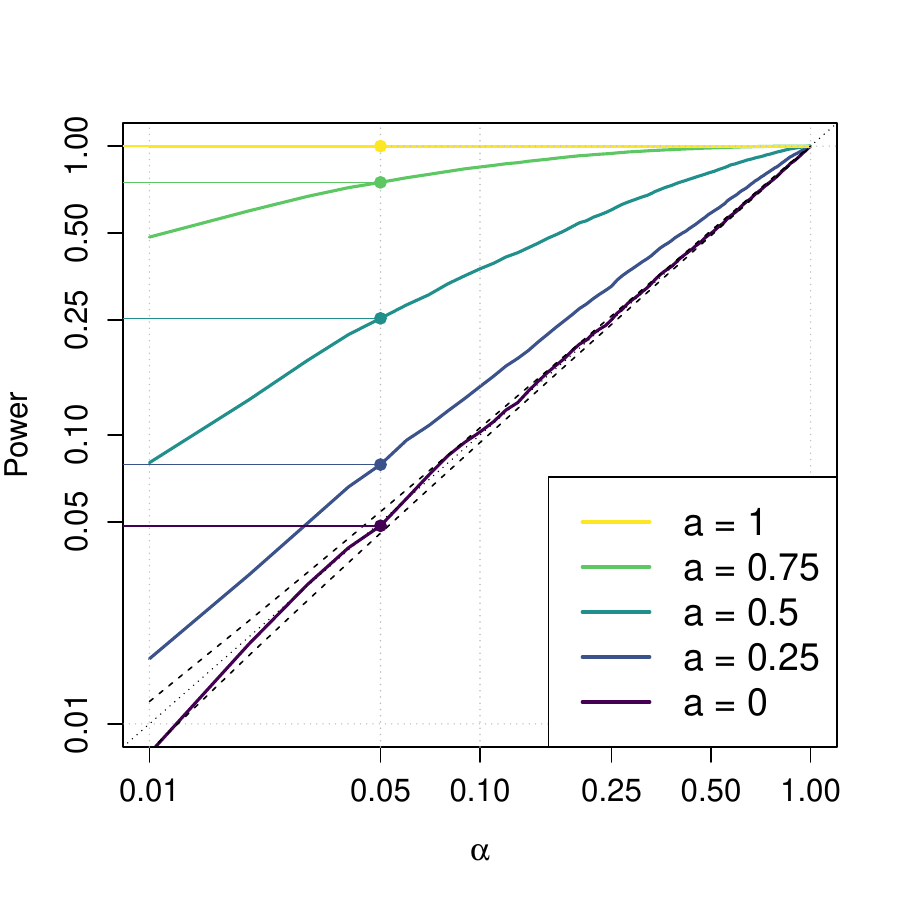}%
        \caption{\small JSD, $c=0.25$} \label{fig:E2:a}
    \end{subfigure}%
    \begin{subfigure}{0.25\textwidth}
        \centering
        \includegraphics[width=\textwidth,clip,trim={0.75cm 1.5cm 1cm 1cm}]{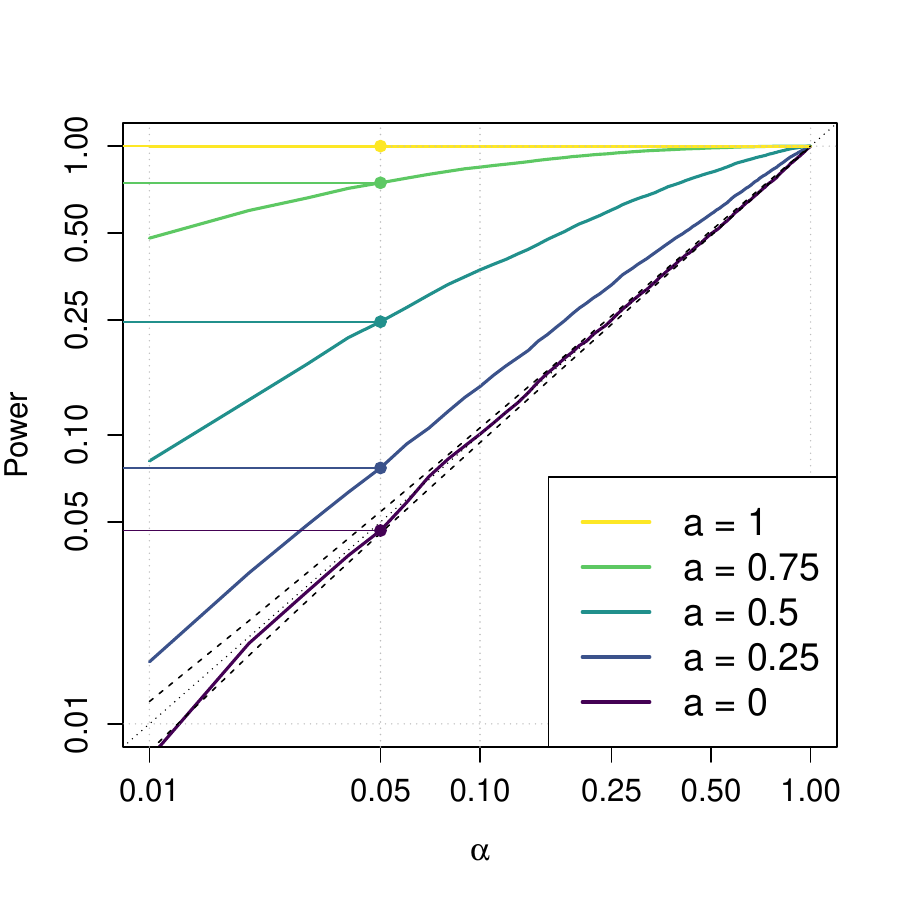}%
        \caption{\small JSD, $c=1$} \label{fig:E2:b}
    \end{subfigure}%
    \begin{subfigure}{0.25\textwidth}
        \centering
        \includegraphics[width=\textwidth,clip,trim={0.75cm 1.5cm 1cm 1cm}]{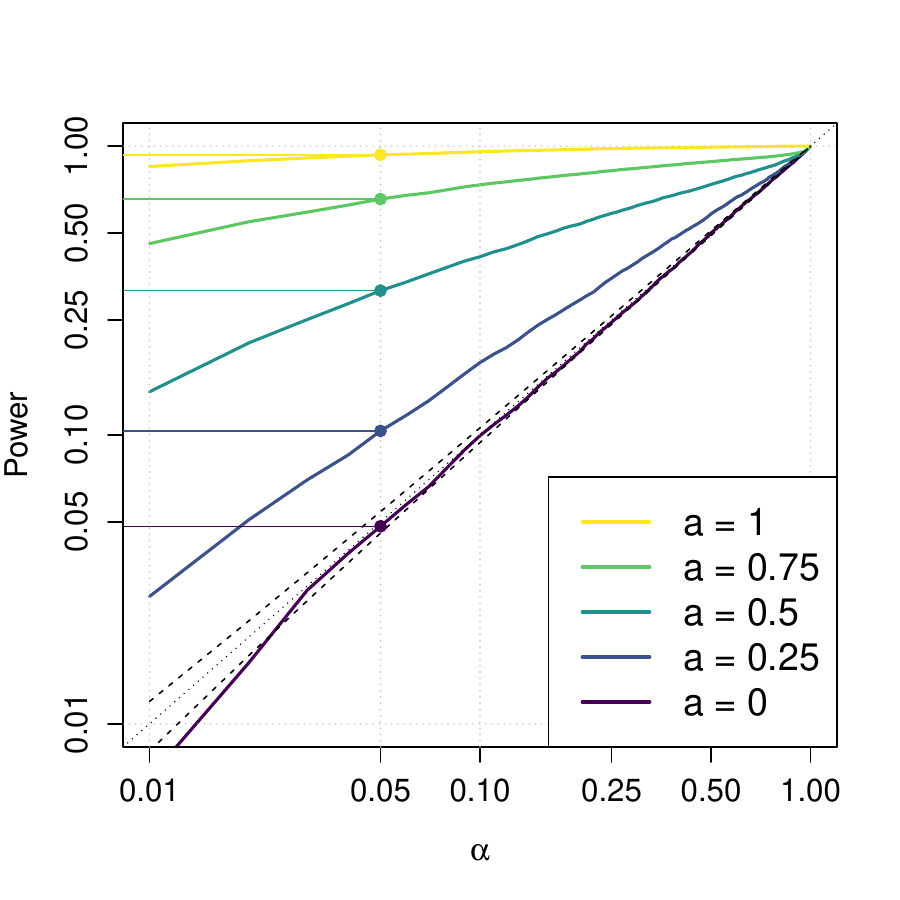}%
        \caption{\small JSD, $c=2$} \label{fig:E2:b2}
    \end{subfigure}%
    \begin{subfigure}{0.25\textwidth}
        \centering
        \includegraphics[width=\textwidth,clip,trim={0.75cm 1.5cm 1cm 1cm}]{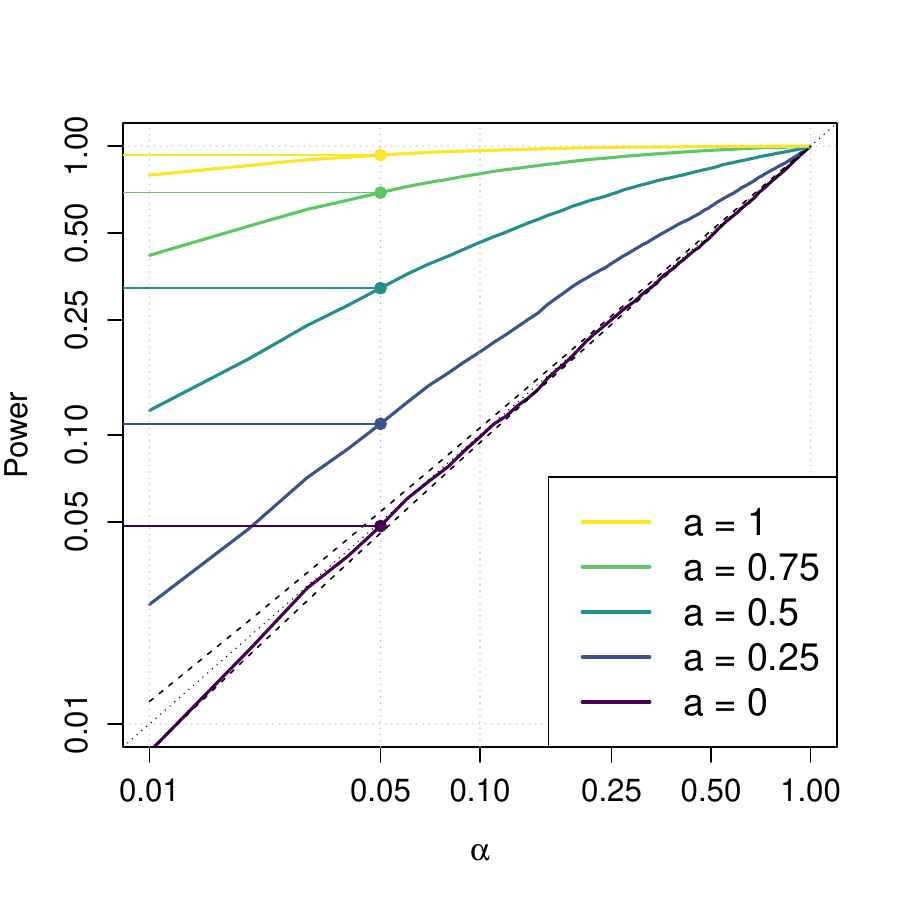}%
        \caption{\small JSD, $c=4$} \label{fig:E2:c}
    \end{subfigure}\\pct
    \begin{subfigure}{0.25\textwidth}
        \centering
        \includegraphics[width=\textwidth,clip,trim={0.75cm 1.5cm 1cm 1cm}]{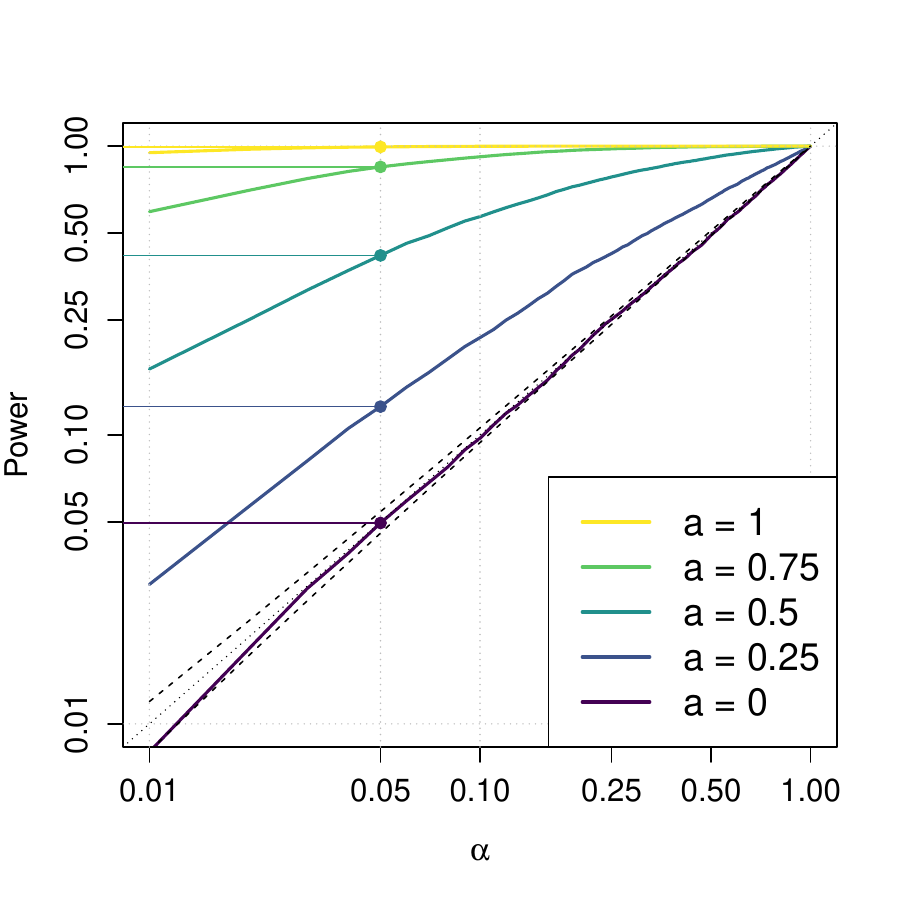}%
        \caption{\small JSD, $c=8$} \label{fig:E2:d}
    \end{subfigure}%
    \begin{subfigure}{0.25\textwidth}
        \centering
        \includegraphics[width=\textwidth,clip,trim={0.75cm 1.5cm 1cm 1cm}]{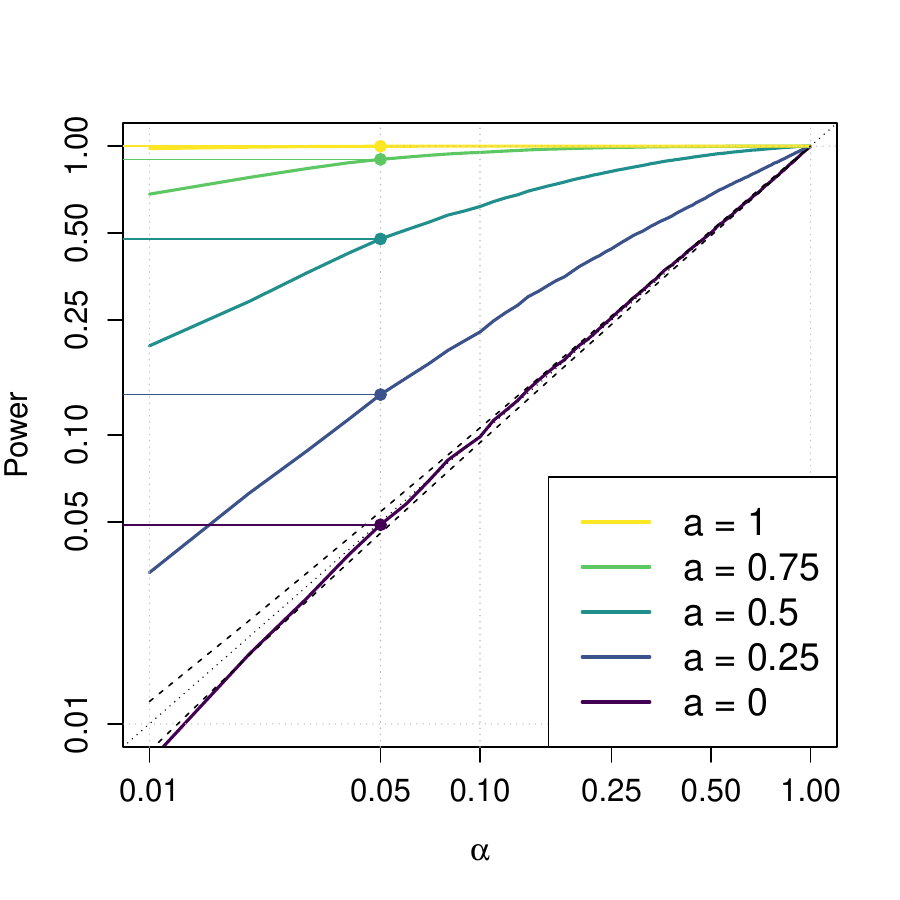}%
        \caption{\small JSD, $c=32$} \label{fig:E2:f}
    \end{subfigure}%
    \begin{subfigure}{0.25\textwidth}
        \centering
        \includegraphics[width=\textwidth,clip,trim={0.75cm 1.5cm 1cm 1cm}]{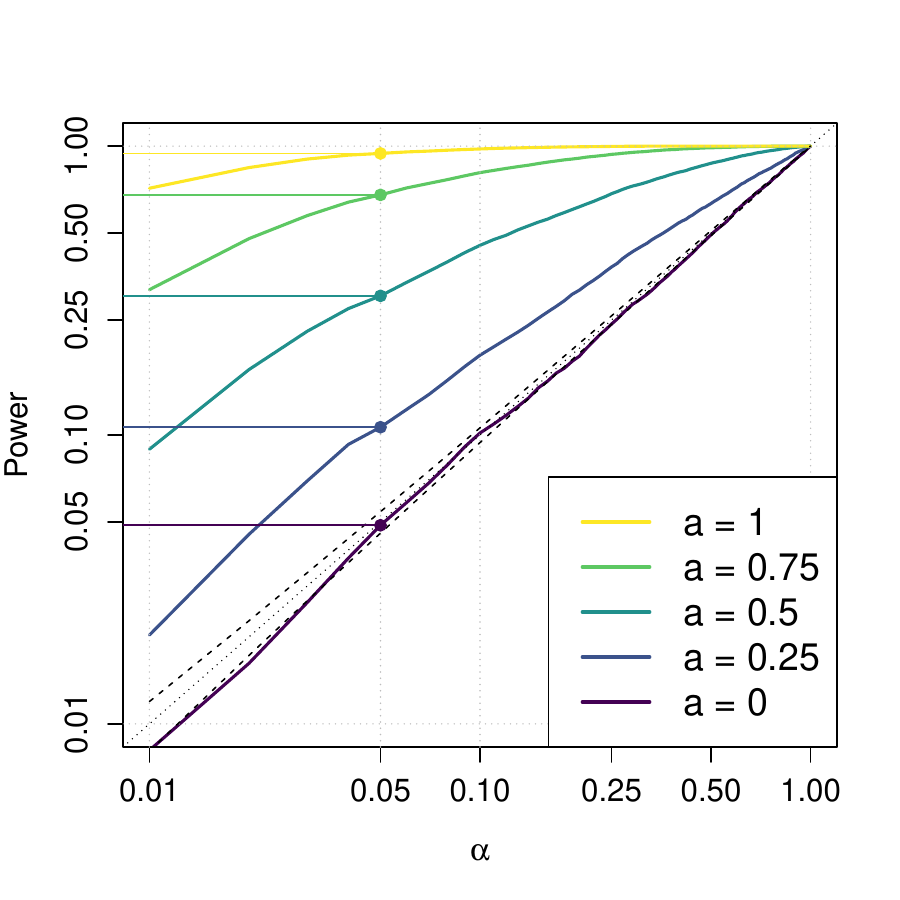}%
        \caption{\small Location} \label{fig:E2:g}
    \end{subfigure}%
    \begin{subfigure}{0.25\textwidth}
        \centering
        \includegraphics[width=\textwidth,clip,trim={0.75cm 1.5cm 1cm 1cm}]{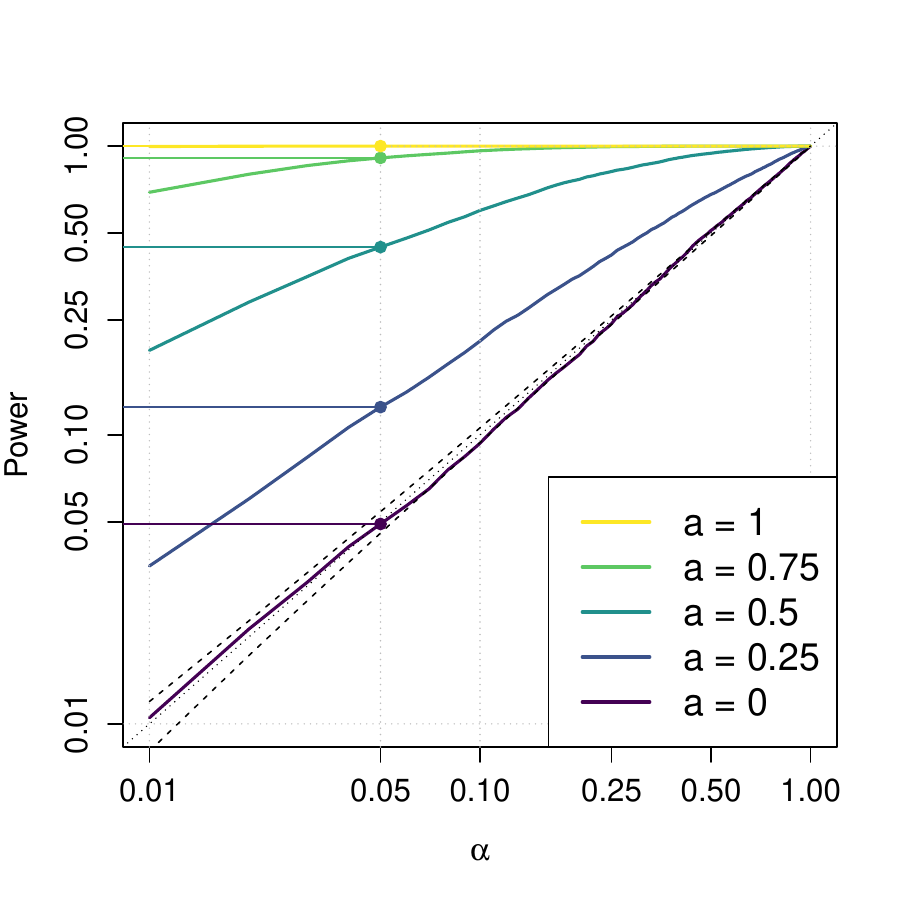}%
        \caption{\small Scatter} \label{fig:E2:h}
    \end{subfigure}
    \caption{\small ROC curves for the tests evaluated in \ref{E2}. The same description of Figure \ref{fig:E1} applies.}
    \label{fig:E2}
\end{figure}

\begin{figure}[htpb!]
    \centering
    \begin{subfigure}{0.25\textwidth}
        \centering
        \includegraphics[width=\textwidth,clip,trim={0.75cm 1.5cm 1cm 1cm}]{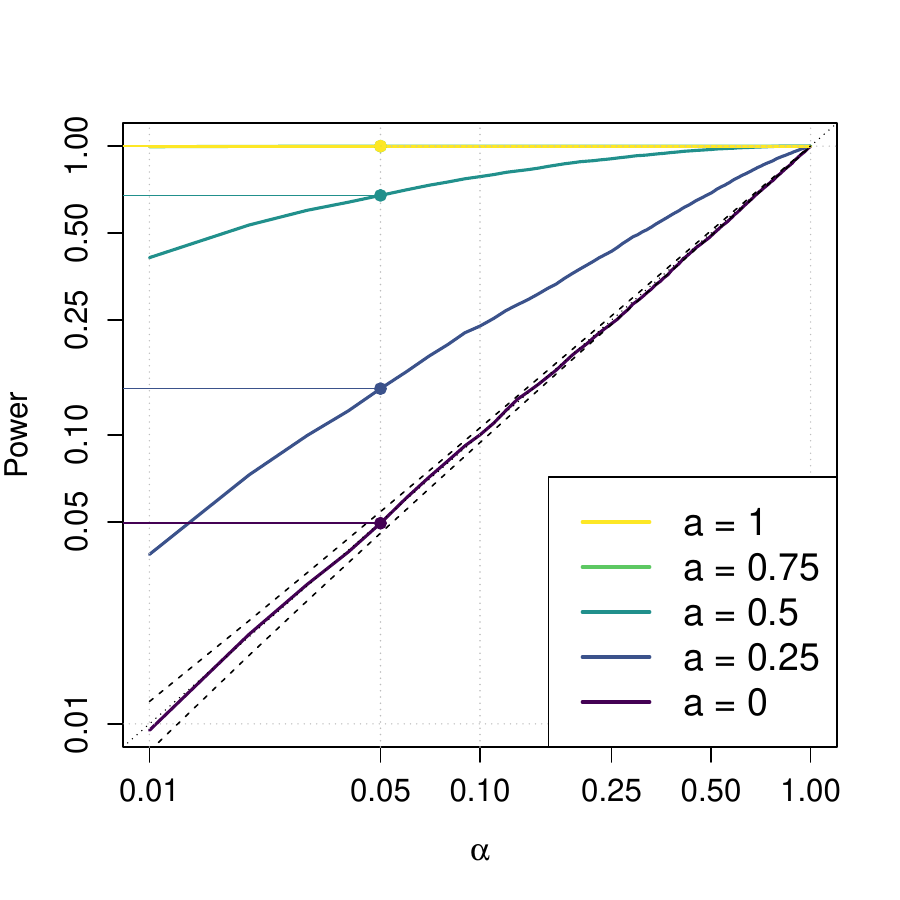}%
        \caption{\small JSD, $c=0.5$} \label{fig:E3:a}
    \end{subfigure}%
    \begin{subfigure}{0.25\textwidth}
        \centering
        \includegraphics[width=\textwidth,clip,trim={0.75cm 1.5cm 1cm 1cm}]{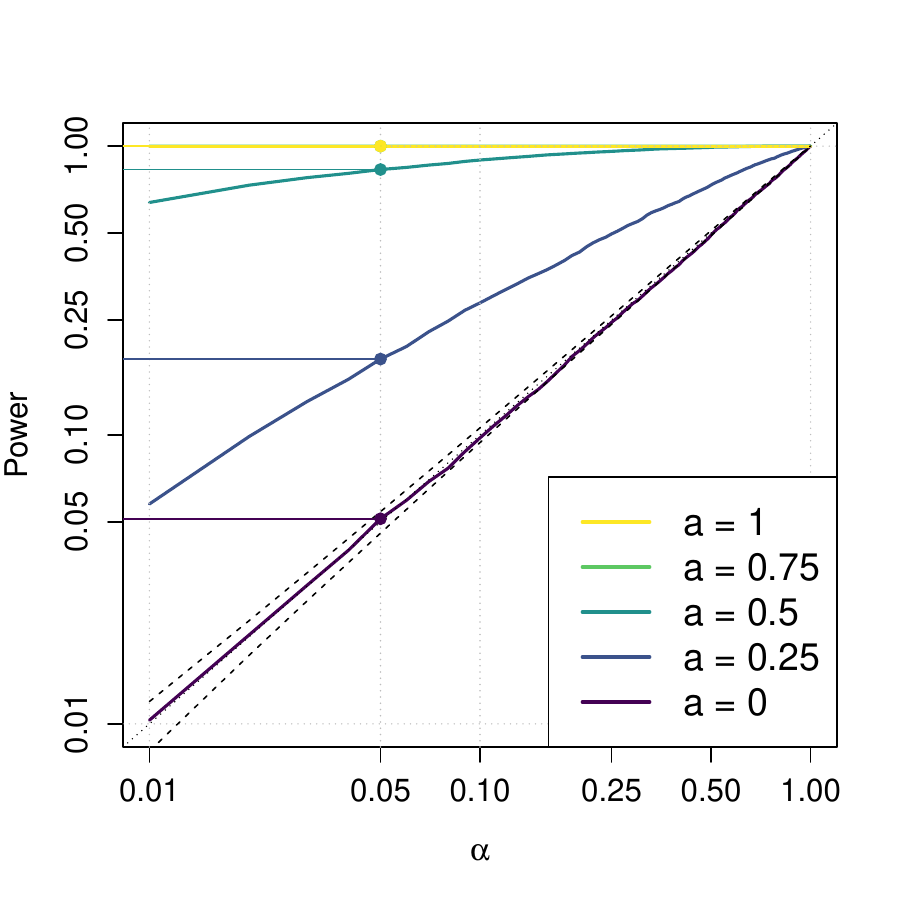}%
        \caption{\small JSD, $c=1$} \label{fig:E3:b}
    \end{subfigure}%
        \begin{subfigure}{0.25\textwidth}
        \centering
        \includegraphics[width=\textwidth,clip,trim={0.75cm 1.5cm 1cm 1cm}]{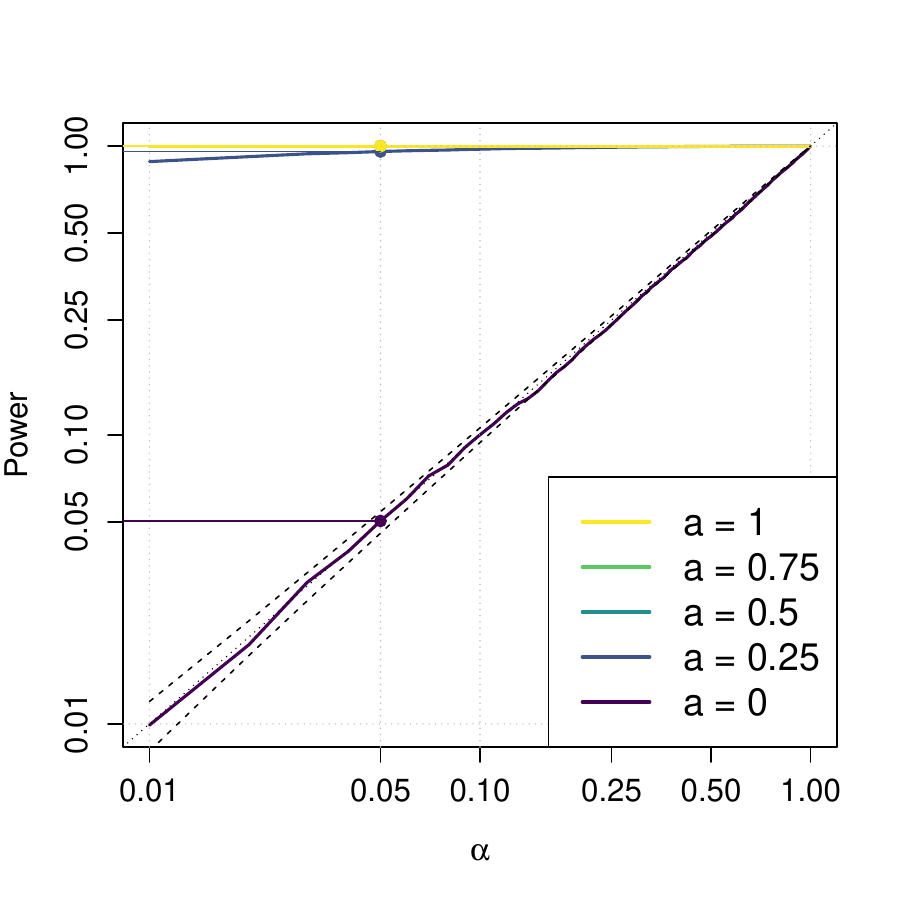}%
        \caption{\small JSD, $c=2$} \label{fig:E3:c}
    \end{subfigure}%
        \begin{subfigure}{0.25\textwidth}
        \centering
        \includegraphics[width=\textwidth,clip,trim={0.75cm 1.5cm 1cm 1cm}]{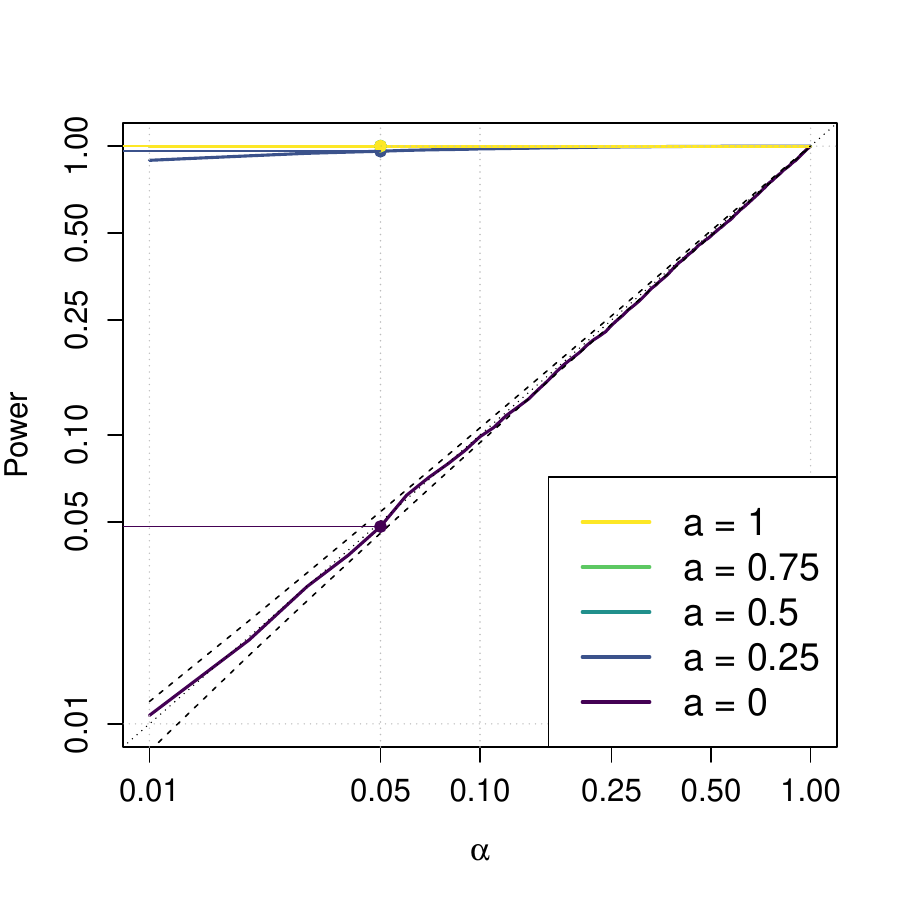}%
        \caption{\small JSD, $c=8$} \label{fig:E3:d}
    \end{subfigure}
    \caption{\small ROC curves for the test evaluated in \ref{E3}. The same description of Figure \ref{fig:E1} applies.}
    \label{fig:E3}
\end{figure}

As shown in Figures \ref{fig:E1}--\ref{fig:E3}, the permutation-based JSD respects the significance level for the three experiments and the bandwidth range considered: the ROC power curves are within the asymptotic confidence intervals for the rejection level when $a=0$ (dark violet). This corroborates that the calibration of the null distribution of $T_{n,\mathrm{JSD}}(\bh)$ is correct for the whole distribution and not just for its upper tail. The bandwidths reported in the figures are representative of the larger grid of bandwidths explored, with figures that were too similar omitted.

The effect of the bandwidths in the JSD test is clear: a larger bandwidth than the sensible estimation bandwidth $\bh_\mathrm{med}$ is preferable for a larger power of the test, up to a certain threshold in which smoothing is excessive. Indeed, in Figures \ref{fig:E1}--\ref{fig:E3} some of the power curves for $c>1$ (precisely, Figures \ref{fig:E1:d}, \ref{fig:E2:c}--\ref{fig:E2:f}, and \ref{fig:E3:c}--\ref{fig:E3:d}) clearly dominate those for $c=1$. Consequently, the power curves for $c<1$ are smaller or equal to those for $c=1$. An excessive amount of smoothing can lead to a complete loss of power (Figures \ref{fig:E1:e}) or, less problematically, to a stabilization of power with respect to smaller bandwidths (Figures \ref{fig:E2:f} and \ref{fig:E3:d}). Another relevant observation is that the factor $c$ that maximizes the power appears to depend on the dimension, with a much higher magnification required in \ref{E2} compared to \ref{E1} and \ref{E3}.

In the two-sample experiments \ref{E1} and \ref{E2}, the JSD test is shown to be very competitive with respect to the location and scatter tests. In \ref{E1}, both the location and scatter tests are ``blind'' to the alternative hypothesis (Figures \ref{fig:E1:g}--\ref{fig:E1:h}), failing to reject it due to the high symmetry of the distributions for $a>0$. In \ref{E2}, the JSD test outperforms for $c=8,16,32$ the location test, while it has comparable or slightly larger power than the scatter test for $c=16,32$. Therefore, the JSD test exhibits competitive power for substantially larger bandwidths than the average estimation bandwidth.

A practical recommendation deduced from the outcomes of the experiments is to take bandwidths in the JSD test a factor $c>1$ (possibly significantly) larger than a sensible estimation bandwidth.

%-------------------------------%
\section{Central and extreme hippocampi}
\label{sec:extremehippo}
%-------------------------------%

Figure \ref{fig:in_out} extracts the nine most central/extreme hippocampi shapes, in terms of their density within the sample, from Figure \ref{fig:subjects_sorted}. These hippocampi can be regarded as the most prototypical/outlying, respectively.

\begin{figure}[htpb!]
    \centering
    \begin{subfigure}[t]{0.5\textwidth}
        \centering
        \includegraphics[width=\textwidth]{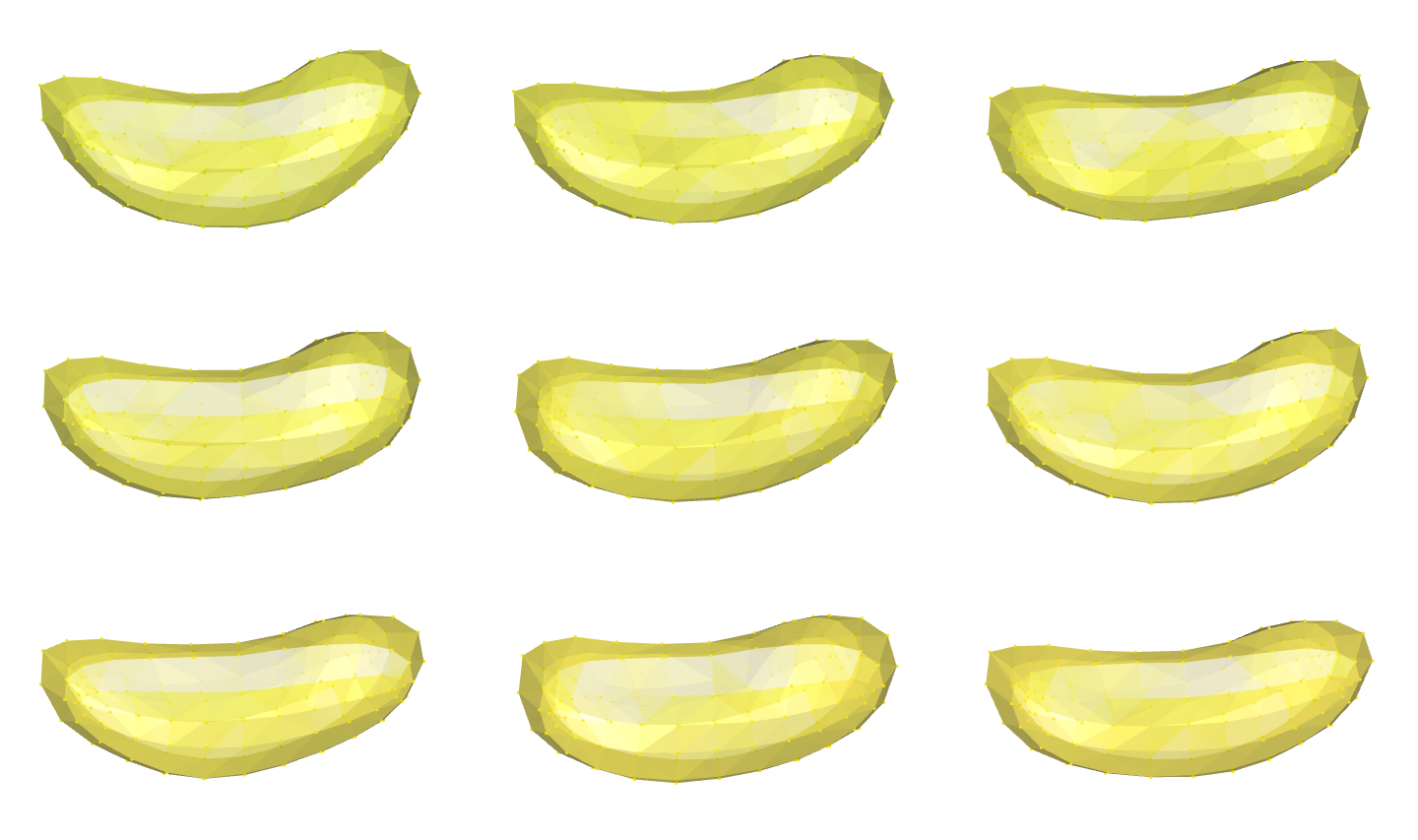}%
        \caption{\small Most central hippocampi}
    \end{subfigure}%
    \begin{subfigure}[t]{0.5\textwidth}
        \centering
        \includegraphics[width=\textwidth]{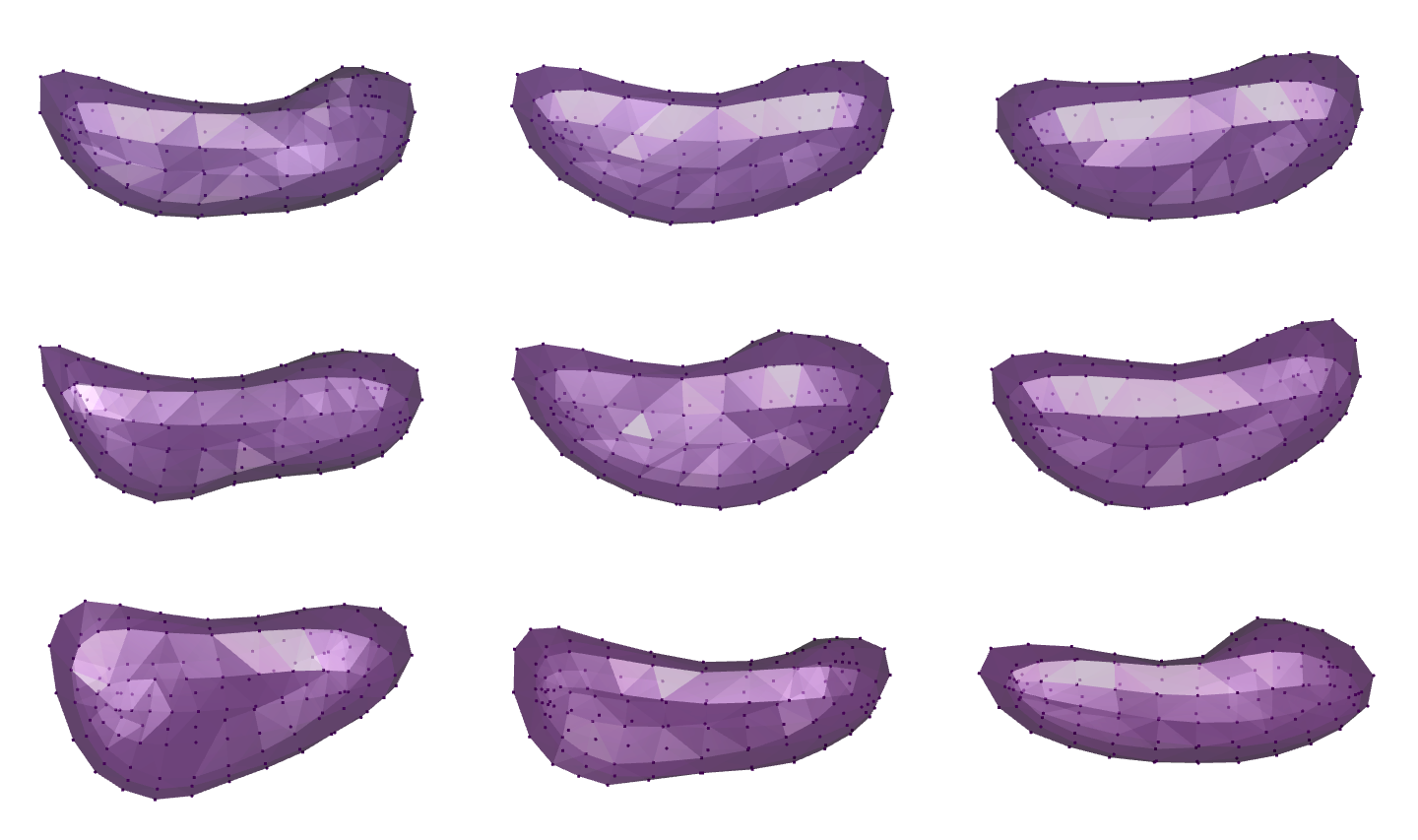}
        \caption{\small Most extreme hippocampi}
    \end{subfigure}%
    \caption{\small Most central and extreme hippocampi within the analyzed dataset. The most central hippocampi are markedly more regular, rounded, and symmetric than the extreme hippocampi.}
    \label{fig:in_out}
\end{figure}

\fi

\end{document}